\documentclass[twocolumn]{aastex63}
\usepackage{amsmath}
\usepackage{multirow}
\usepackage{graphicx}
\usepackage[caption=false]{subfig}
\usepackage{subfig}
\usepackage[figuresright]{rotating}
\DeclareCaptionFormat{cont}{#1 (cont.)#2#3\par}

\shortauthors{Liu et al.}
\begin{document}

\title{SiO Outflows as Tracers of Massive Star Formation in Infrared Dark Clouds}

\author{Mengyao Liu}
\affil{Dept. of Astronomy, University of Virginia, Charlottesville, Virginia 22904, USA}
\author{Jonathan C. Tan}
\affil{Dept. of Space, Earth \& Environment, Chalmers University of Technology, 412 93 Gothenburg, Sweden}
\affil{Dept. of Astronomy, University of Virginia, Charlottesville, Virginia 22904, USA}
\author{Joshua Marvil}
\affil{National Radio Astronomy Observatory, 1003 L{\'o}pezville Rd., Socorro, NM 87801, USA}
\author{Shuo Kong}
\affil{Steward Observatory, University of Arizona, Tucson, AZ 85719, USA}
\author{Viviana Rosero}
\affil{National Radio Astronomy Observatory, 1003 L{\'o}pezville Rd., Socorro, NM 87801, USA}
\author{Paola Caselli}
\affil{Centre for Astrochemical Studies, Max-Planck-Institut f{\"u}r extraterrestrische Physik, Gie{\ss}enbachstra{\ss}e 1, D-85749 Garching bei M{\"u}nchen, Germany}
\author{Giuliana Cosentino}
\affil{Dept. of Space, Earth \& Environment, Chalmers University of Technology, 412 93 Gothenburg, Sweden}

\begin{abstract}
To study the early phases of massive star formation, we present ALMA
observations of SiO(5-4) emission and VLA observations of 6\,cm continuum
emission towards 32 Infrared Dark Cloud (IRDC) clumps, spatially resolved down to $\lesssim 0.05$\,pc. Out of the 32 clumps, we detect SiO emission in 20 clumps, and in 11 of them the SiO emission is relatively strong and likely tracing protostellar outflows. Some SiO outflows are collimated, while others are less ordered. For the six strongest SiO outflows, we estimate basic outflow properties. In our entire sample, where there is SiO emission, we find 1.3\,mm continuum and infrared emission nearby, but not vice versa. We build the spectral energy distributions (SEDs) of cores with 1.3\,mm continuum emission and fit them with radiative transfer (RT) models. The low luminosities and stellar masses returned by SED fitting suggest these are early stage protostars. We see a slight trend of increasing SiO line luminosity with bolometric luminosity, which suggests more powerful shocks in the vicinity of more massive YSOs. We do not see a clear relation between the SiO luminosity and the evolutionary stage indicated by $L/M$. We conclude that as a protostar approaches a bolometric luminosity of $\sim 10^2 \: L_{\odot}$, the shocks in the outflow are generally strong enough to form SiO emission. The VLA 6\,cm observations toward the 15 clumps with the strongest SiO emission detect emission in four clumps, which is likely shock
ionized jets associated with the more massive ones of these protostellar cores.
\end{abstract}


\section{Introduction}\label{sec:intro}

Massive stars play a key role in the regulation of galactic
environments and galaxy evolution, yet there is no consensus on even
the basics of their formation mechanism. Theories range from models
based on Core Accretion, i.e., formation from massive self-gravitating
cores (e.g., McKee \& Tan 2003), to Competitive Accretion, i.e.,
chaotic clump-fed accretion concurrent with star cluster formation
(e.g., Bonnell et al. 2001; Wang et al. 2010), to Protostellar
Collisions (e.g., Bonnell et al. 1998; Bally \& Zinnecker 2005).

Magneto-centrifugally driven protostellar outflows are thought to be
an ubiquitous feature of ongoing formation of all masses of stars
(e.g., Arce et al. 2007; Beltr\'an \& de Wit 2016) and are likely to
be essential or at least important for removing angular momentum from
the accreting gas. Such outflows involving large scale magnetic fields
threading the disk (i.e., ``disk winds''; e.g., K\"onigl \& Pudritz
2000) and/or stellar magnetic fields (i.e., ``x-winds''; e.g., Shu et
al. 2000) lead to collimated bipolar outflows that can be the most
obvious signpost of early star formation activity. Thus studies of the
morphologies and kinematics of such outflows can help us understand
protostellar activities from inner disk to core envelope scales and
beyond. Whether protostellar outflow properties scale smoothly with
the mass of the driving protostar will also shed light on how massive
star formation differs from low-mass star formation. As one example,
radio surveys have found an association of ``radio jets'', i.e.,
collimated radio emission, and molecular outflows that appears to be a
common phenomenon in both low-mass (Anglada 1996) and high-mass
protostars (Purser et al. 2016; Rosero et al. 2016, 2019b).

With the development of high-resolution, high-sensitivity facilities
like ALMA and VLA, more and more observations have been carried
out towards massive star-forming regions.
While most mm and radio surveys are towards more evolved massive
protostars (e.g., Motte et al. 2018; CORE, Beuther et al. 2018; ATOMS, Liu et al. 2020; Rosero et al. 2016; Purser et al. 2016), the earlier stages remain relatively unexplored with only a few surveys being carried out recently (e.g., ASHES, Sanhueza et al. 2019; CORE-extension, Beuther et al. 2021). To study
such early stages, we focus on protostars in Infrared Dark Clouds
(IRDCs). IRDCs are cold ($T<$20~K), dense ($n_{\rm H}\gtrsim
10^{4}\:{\rm cm}^{-3}$) regions of molecular clouds that are opaque at
wavelengths $\sim$10 $\rm \mu m$ or longer and thus appear dark
against the diffuse Galactic background emission. They are likely to
harbor the earliest stages of star formation (see, e.g., Tan et
al. 2014; Sanhueza et al. 2019; Li et al. 2019b).

We have conducted a survey of 32 IRDC clumps with ALMA (see Table 1 in
Kong et al. 2017). This high mass surface density sample was selected
from 10 IRDCs (A-J) by the mid-infrared (\textit{Spitzer}-IRAC 8
$\mu$m) extinction (MIREX) mapping methods of Butler \& Tan (2009,
2012). The distances range from 2.4 kpc to 5.7 kpc.  The first goal of
this study was to find massive pre-stellar cores. About 100 such core
candidates have been detected via their $\rm N_{2}D^{+}(3-2)$ emission
(Kong et al. 2017).

Next, we identified 1.3~mm continuum cores in the 32 clumps (Liu
et al. 2018). In total, 107 cores were found with a dendrogram
algorithm with a mass range from 0.150 to 178~$M_{\odot}$ assuming a
temperature of 20~K.

The ALMA observations are also sensitive to SiO(5-4) emission. SiO is
believed to form through sputtering and grain-grain collisions of dust grains (e.g., Schilke et al. 1997; Caselli et al. 1997). Unlike
CO, SiO emission does not suffer from confusion with easily excited
ambient material. While SiO emission with a narrow
velocity range may come from large-scale colliding gas flows (e.g.,
Jim{\'e}nez-Serra et al. 2010; Cosentino et al. 2018, 2020),
SiO emission with a broad velocity range is considered to be an
effective tracer of fast shocks from protostellar outflows. In
  low-mass protostars, SiO jets are often detected and the detection
  rate increases with source luminosity (e.g., Podio et al. 2021). Here we
use such SiO emission as a tracer of outflows, which are then
signposts to help identify and characterize the protostars.

We have also carried out VLA follow-up observations towards our
protostar sample to determine the onset of the appearance of radio
continuum emission and thus diagnose when the protostars transition
from a ``radio quiet" to a ``radio loud" phase.

The structure of the paper is as follows. In \S2 we summarize the ALMA
and VLA observations. We present the results of SiO outflows in \S3
and 6\,cm radio emission in \S4. We discuss the implications of the
results in \S5. We summarize the conclusions in \S6.

\section{Observations}\label{sec:obs}

\subsection{ALMA data}

We use data from ALMA Cycle 2 project 2013.1.00806.S (PI: Tan), which
observed 30 positions in IRDCs on 04-Jan-2015, 10-Apr-2015 and
23-Apr-2015, using 29 12 m antennas in the array. Track 1 with central
$v_{\rm LSR}$ = 58 km/s includes clumps A1, A2, A3 ($v_{\rm LSR} \sim$66 km/s), B1,
B2 ($v_{\rm LSR} \sim$26 km/s), C2, C3, C4, C5, C6, C7, C8, C9 ($v_{\rm LSR}
\sim$79 km/s), E1, E2 ($v_{\rm LSR} \sim$80 km/s) with notations following
Butler \& Tan (2012) (see Table 1 in Kong et al. 2017 for a list of
targets). Track 2 with central $v_{\rm LSR}$ = 66 km/s includes D1, D2
(also contains D4), D3, D5 (also contains D7), D6, D8, D9 ($v_{\rm LSR}
\sim$87 km/s), F3, F4 ($v_{\rm LSR} \sim$58 km/s), H1, H2, H3, H4, H5, H6
($v_{\rm LSR} \sim$44 km/s). In Track 1, J1924-2914 was used as the
bandpass calibrator, J1832-1035 was used as the gain calibrator, and
Neptune was the flux calibrator. In Track 2, J1751+0939 was used as
the bandpass calibrator, J1851+0035 was used as the gain calibrator,
and Titan was the flux calibrator. The total observing time
including calibration was 2.4~hr. The actual on-source time was
$\sim$2-3~min for each pointing.
 
The spectral set-up included a continuum band centered at 231.55~GHz
(LSRK frame) with width 1.875~GHz from 230.615~GHz to 232.490~GHz.  At
1.3~mm, the primary beam of ALMA 12m antennas is $\sim 27\arcsec$
(FWHM) and the largest recoverable scale for the array is $\sim
7\arcsec$. The data were processed using NRAO's Common Astronomy Software
Applications (CASA) package (McMullin et al. 2007). For reduction of the data, we used ``Briggs"
cleaning and set {\tt robust = 0.5}. The continuum image reaches a 1$\sigma$ rms noise of
0.2~mJy in a synthesized beam of 1.2$\arcsec$ $\times$ 0.8$\arcsec$.
The spectral line sensitivity per 0.2 $\rm km \: s^{-1}$ channel is
$\sim$0.02~Jy beam$^{-1}$ for Track 1 and $\sim$0.03~Jy beam$^{-1}$
for Track 2. Other basebands were tuned to observe N$_2$D$^{+}$(3-2),
SiO(5-4), $\rm C^{18}O$ (2-1), DCN(3-2), DCO$^{+}$(3-2) and $\rm
CH_{3}OH (5_{1,4}-4_{2,2})$.  

\subsection{VLA data}

The 6 cm (C-band) observations were made towards some of the clumps
with strongest SiO in both C configuration and A configuration. The C
configuration data taken in 2017 are from project code 17A-371 (PI:
Liu) and include A1, A2, A3 in Cloud A (gain calibrator: J1832-1035,
positional accuracy 0.01 - 0.15$\arcsec$), B1, B2 in Cloud B (gain
calibrator: J1832-1035), C1 (RA: 18:42:46.498, DEC: -4:04:15.964), C2,
C4, C5, C6, C9 in Cloud C (gain calibrator: J1832-1035), D1, D8 in
Cloud D (gain calibrator: J1832-1035) and H5, H6 in Cloud H (gain
calibrator: J1824+1044, positional accuracy $<$ 0.002$\arcsec$). The A
configuration data taken in 2018 are from project code 18A-405 (PI:
Liu) and include A1, A2, A3 in Cloud A (gain calibrator: J1832-1035),
C1, C2, C4, C5, C6, C9 in Cloud C (gain calibrator: J1804+0101,
positional accuracy $<$ 0.002$\arcsec$) and H5, H6 in Cloud H (gain
calibrator: J1824+1044). 3C286 was used as flux density and bandpass
calibrator for all the regions with both configurations. Both sets of
data consist of two 2 GHz wide basebands (3 bit samplers) centered at
5.03 and 6.98 GHz, where the first baseband was divided into 16
spectral windows (SPWs), each with a bandwidth of 128 MHz and the
second baseband was divided into 15 SPWs with 14 SPWs 128 MHz wide
each and 1 SPW 2 MHz wide for the 6.7 GHz methanol maser. The data
were recorded in 31 unique SPWs, 30 comprised of 64 channels with each
channel being 2 MHz wide and 1 comprised of 512 channels with each
channel being 3.906 kHz wide, resulting in a total bandwidth of 3842
MHz (before ``flagging'').  For sources in Cloud A and C the
observations were made alternating on a target source for 9.5 minutes
and a phase calibrator for 1 minute, for a total on-source time of
47.5 minutes. For sources in Cloud B, D and H the observations were
made alternating on a target source for 8.5 minutes and a phase
calibrator for 1 minute, for a total on-source time of 42.5 minutes.

The data were processed using NRAO's CASA package (McMullin et al. 2007). We used
the VLA pipeline CASA v5.1 for the A configuration and pipeline CASA
v4.7.2 for the C configuration for calibration. We ran one additional
pass of CASA's {\tt flagdata} task using the {\tt rflag} algorithm on
each target field to flag low-level interference. For Cloud A, C and
H, for which we have data observed with both configurations, we used
CASA's {\tt concat} task to combine the A configuration and C
configuration data giving a factor of 3 times more weight to the A
configuration data, allowing us to obtain a more Gaussian-like PSF at
a similar resolution as the A-configuration data. We made mosaic
images combining all the fields in the same cloud since there was
substantial overlap in their primary beams to improve sensitivity and
uv-coverage. For Cloud C we used ``mosaic'' gridder in CASA's {\tt
  tclean} task to jointly deconvolve all 6 fields. The continuum
images were made using 29 of the 30 wide-band spectral windows, since
one SPW was excluded due to possible contamination by maser emission.
All images were deconvolved using the {\tt mtmfs} mode of CASA's {\tt
  tclean} task and used {\tt nterms=2} to model the sources' frequency
dependencies. For Cloud A, B, D, H, we tried different approaches and
concluded that the CASA {\tt linearmosaic} tool gave us the optimal
results. Note that mosaics of B and D clouds only include C
configuration data. A list of the beam sizes and rms noise levels are
shown in Table~\ref{tab:vla}.

\begin{deluxetable}{ccccc}
\setlength{\tabcolsep}{0.5pt}
\tabletypesize{\scriptsize}
\tablecaption{VLA 6~cm Observations} \label{tab:vla}
\tablewidth{0pt}
\tablehead{
\colhead{Source} & \colhead{Configuration} & \colhead{Beam} & \colhead{Continuum} & \colhead{Continuum 3$\sigma$ rms} \\
\colhead{} & \colhead{} & \colhead{Size} & \colhead{Detection}& \colhead{($\mu$Jy beam$^{-1}$)} 
}
\startdata
A1 & A, C & 0.659$\arcsec$ $\times$ 0.364$\arcsec$ & Y & 9.9 \\
A2 & A, C & 0.659$\arcsec$ $\times$ 0.364$\arcsec$ & N & 15.0 \\
A3 &  A, C & 0.659$\arcsec$ $\times$ 0.364$\arcsec$ & N & 17.4 \\
B1 & C & 4.685$\arcsec$ $\times$ 2.576$\arcsec$ & N & 18.6 \\
B2 & C & 4.685$\arcsec$ $\times$ 2.576$\arcsec$ & N & 20.4  \\
C1 &  A, C & 0.481$\arcsec$ $\times$ 0.396$\arcsec$ & N & 15.0 \\
C2 &  A, C & 0.481$\arcsec$ $\times$ 0.396$\arcsec$ & Y & 9.0 \\
C4 & A, C & 0.481$\arcsec$ $\times$ 0.396$\arcsec$ & Y & 8.1 \\
C5 & A, C & 0.481$\arcsec$ $\times$ 0.396$\arcsec$ & N & 8.7 \\
C6 & A, C & 0.481$\arcsec$ $\times$ 0.396$\arcsec$ & N & 12.6 \\
C9 & A, C & 0.481$\arcsec$ $\times$ 0.396$\arcsec$ & Y & 13.8 \\
D1 & C & 3.332$\arcsec$ $\times$ 2.701$\arcsec$ & N & 20.7 \\
D8 & C & 3.332$\arcsec$ $\times$ 2.701$\arcsec$ & N &  25.2 \\
H5 & A, C & 0.407$\arcsec$ $\times$ 0.326$\arcsec$ & N & 9.3 \\
H6 & A, C & 0.407$\arcsec$ $\times$ 0.326$\arcsec$ & N & 9.0 \\
\enddata
\end{deluxetable}

\section{SiO outflows}

\subsection{SiO detection}\label{sec:sio detection}

\begin{figure*}[htbp]
\centering{\includegraphics[width=18cm]{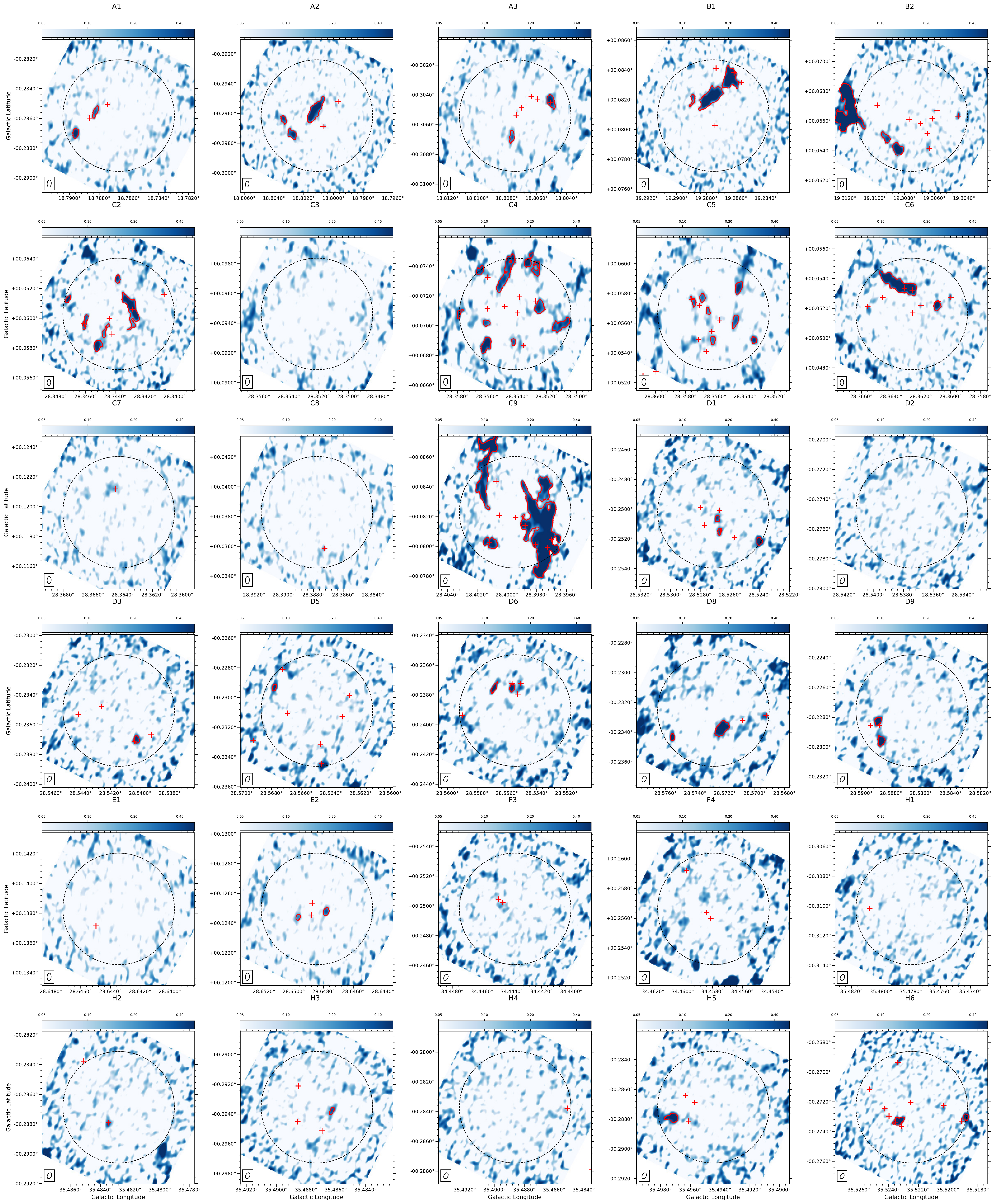}}
\caption{
SiO(5-4) integrated intensity maps within $\pm$ 15 km $\rm s^{-1}$
relative to the cloud velocity of the 30 IRDC positions. The red
contours show the \textit{trunk} structures characterized by the
dendrogram algorithm. The minimum threshold intensity required to
identify a \textit{trunk} structure is 3 $\sigma$. In the images prior
to primary beam correction $\sigma \sim 50 \: {\rm mJy \: km \:
  s}^{-1} \: {\rm beam}^{-1}$ for Clouds A, C, E; $\sigma \sim 40 \:
{\rm mJy \: km \: s^{-1} \: beam}^{-1}$ for Cloud B; and $\sigma \sim
74 \: {\rm mJy \: km \: s^{-1} \: beam}^{-1}$ for Clouds D, F and
H. The minimum area is one synthesized beam size. The black plus sign denotes the center of the field. The red plus signs denote the 1.3 mm continuum cores detected in Liu et al. (2018). The dotted circle in each panel denotes the primary beam. The synthesized beam is shown in the bottom left corner of each panel. }\label{fig:trunk}
\end{figure*}

\begin{figure*}[htbp]
\centering{\includegraphics[width=18cm]{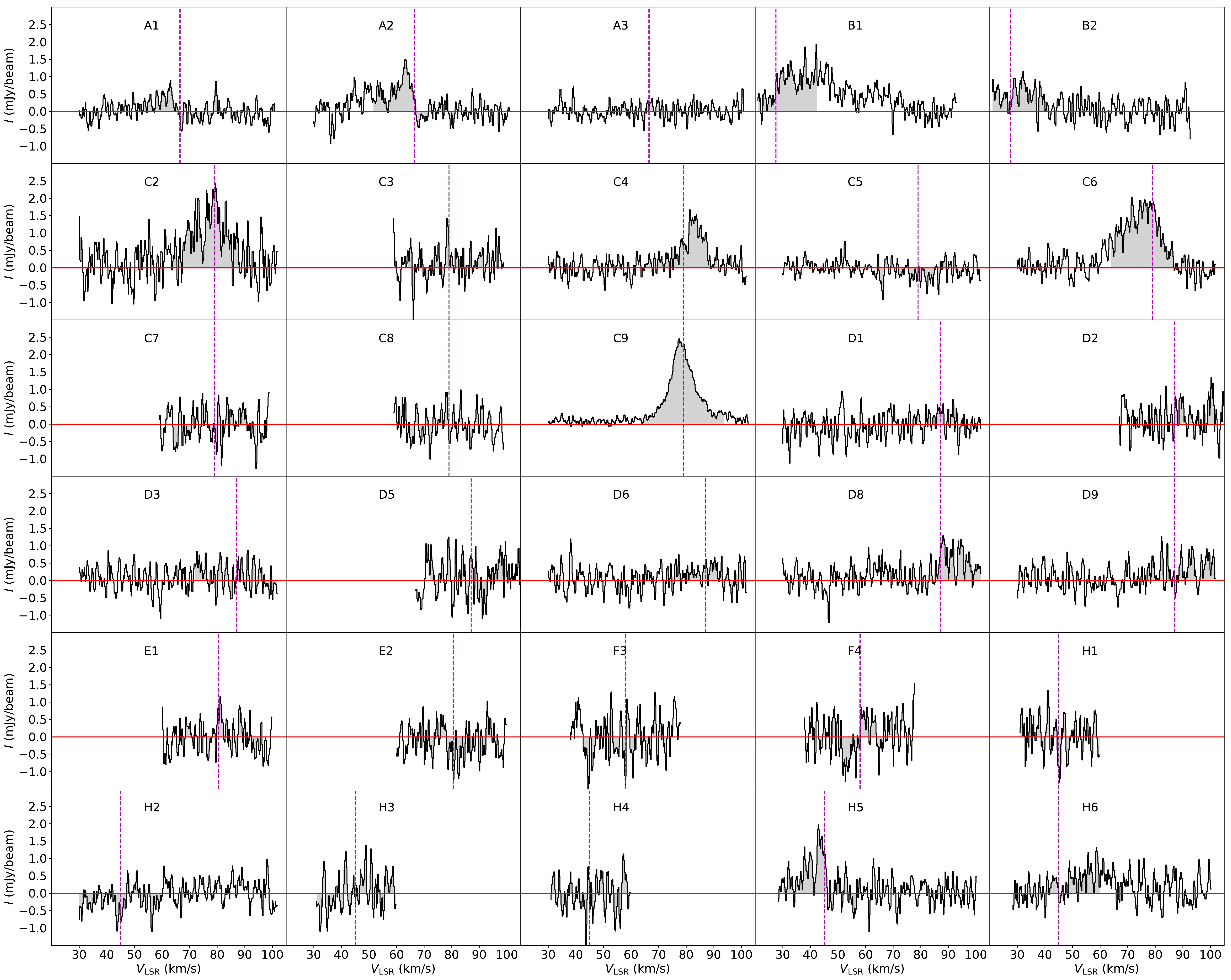}}
\caption{
Spectra of SiO(5-4) emission of the 30 IRDC positions averaged within
the primary beam. The flux of C9 has been reduced by a scale factor of
0.25 for ease of viewing. The dashed lines denote the estimates of the
cloud LSR velocities. The grey area denotes emission within
$\pm$15~km~$\rm s^{-1}$ relative to the cloud velocity, which is used
for identifying structures above 3 $\sigma$ in the integrated
intensity maps of SiO(5-4).}\label{fig:spec30}
\end{figure*}

Out of the 32 clumps observed, we have detected SiO(5-4) emission in
20 sources: A1, A2, A3, B1, B2, C2, C4, C5, C6, C9, D1, D3, D5, D6, D8,
D9, E2, H3, H5, H6.
We define a detection of SiO(5-4) emission with two methods. Method 1
is if there are at least 3 consecutive pixels with the peak emission
greater than $3\sigma$ noise level in their spectra and the projected
area of such pixels is larger than one beam size. Method 2 is if the
integrated intensity of SiO(5-4) within $\pm$15~km~$\rm s^{-1}$
relative to the cloud velocity is higher than the $3\sigma$ noise
level and the area of such pixels is larger than one beam size (see
Figure~\ref{fig:trunk}). We only consider emission inside the primary
beam (except B2, see \S\ref{sec:strong}). When applying Method 2, the
emission features were identified by the dendrogram algorithm
(Rosolowsky et al. 2008), which is a code for identifying hierarchical
structures. We set the minimum threshold intensity required to
identify a parent tree structure ({\it trunk}) to be 3$\sigma$ and a
minimum area of the synthesized beam size. We use the images prior to
primary beam correction to identify cores. We use
astrodendro\footnote{\url{http://www.dendrograms.org/}}, a python
package to compute dendrogram statistics. Only A2, B1, B2, C2, C4, C5,
C6, C9 are reported detection with Method 1.  More detections are made
with Method 2. This indicates the sources detected with Method 2,
  but not Method 1, generally have weak emission in each channel and
  their signal only starts to show up after integration. The spectra
of the 30 IRDC positions averaged within the primary beam are shown in
Figure~\ref{fig:spec30}. These figures also show the estimated
systemic velocity of each cloud, adopted from Kong et al. (2017).

To identify the protostellar sources that are responsible for the
generation of the SiO(5-4) emission, we search for 1.3~mm emission
peaks that are in the vicinity.
Although there is at least one continuum peak in every clump with SiO
detection, it is ambiguous in many cases to tell the association of
SiO and continuum peaks based only on their spatial distributions.

We are more confident about the protostellar nature of a
  continuum peak if it is associated with a dense gas tracer $\rm
  C^{18}O$(2-1), DCN(3-2) or DCO$^{+}$(3-2). For each 1.3~mm
continuum source identified in Liu et al. (2018), we define a
3$\arcsec$ $\times$ 3$\arcsec$ region centered at the continuum peak
as the ``core region'' and derive the spectrum of each dense gas
tracer within the core region. If there are more than 3 channels with
signal above 2$\sigma$ noise level, and the peak channel as well as
its neighboring channels have signal above a 2$\sigma$ noise level,
then we consider there is detection of that line associated with the
continuum core. We record the peak velocity of each line and take the
mean value as the systemic velocity of the core.
Generally $\rm C^{18}O$(2-1) is the most common tracer. DCN(3-2) and
DCO$^{+}$(3-2) can be relatively weak. The SiO detections and the
  association with 1.3~mm continuum emission peaks and dense gas
  tracers are summarized in Table~\ref{tab:sum}.

\begin{longrotatetable} % Landscape page
\centering
\setlength{\tabcolsep}{0.5pt}
\renewcommand{\arraystretch}{0.8}
%\vspace{4in}
\begin{deluxetable*}{ccc|cccc|cccccccccc|cccc}
\tabletypesize{\scriptsize}
\tablecaption{Integrated Flux Densities} \label{tab:sum}
\tablewidth{18pt}
\tablehead{\vspace{-0.6cm}}
\startdata
\multirow{3}{*}{Source} & $d$ & $L_{\rm SiO, total}$ & \multicolumn{4}{c}{SiO Detection} &  \multicolumn{10}{c}{Plausible Driving Protostar}  & \multicolumn{4}{c}{Photometry and SED Fitting}\\
& \multirow{2}{*}{(kpc)} & \multirow{2}{*}{(Jy km s$^{-1}$ pc$^{2}$)} & No.& $l$ & $b$ & $S_{\nu}$ & mm Peak& $l$ & $b$ & $M_{c, \rm raw}$ & DCN & DCO$^{+}$ & C$^{18}$O & CH$_{3}$OH & N$_{2}$D$^{+}$ & velocity & $l_{\rm ap}$  & $b_{\rm ap}$  & $R_{\rm ap}$ & $L_{\rm bol}$ \\
&  & & & ($\arcdeg$) & ($\arcdeg$) & (Jy km s$^{-1}$) &  & ($\arcdeg$) & ($\arcdeg$) & ($M_{\odot}$) & & & & & & (km s$-1$) & ($\arcdeg$) & ($\arcdeg$) & ($\arcsec$) & $L_{\odot}$   \\
\hline
A1 & 4.8 & 2.733$\times 10^{8}$ & 1 & 18.78830 & -0.28567 & 0.328 & A1c2 & 18.78864 & -0.28598 & 23.0 & N &N &Y& N &Y & 64.2 & 18.78897 & -0.28477 & 16 & 5.771$\times 10^{2}$\\ 
& & & 2 & 18.78963 & -0.28711 & 0.616 & A1c2 & 18.78864 & -0.28598 & 23.0 &  N &N &Y& N &Y & 64.2 & & & & \\
\hline
A2 & 4.8 & 1.504e+09 & 1 & 18.80129 & -0.29592 & 3.826  & A2c2 & 18.80070 & -0.29687 & 3.49 & Y &Y& Y& N &N & 55.2 & 18.80177 & -0.29591 & 16 & 4.132$\times 10^{2}$\\
& & & 2 & 18.80340 & -0.29650 & 0.444 & & & & & & & & & & & & & \\
& & & 3 & 18.80281 & -0.29750 & 0.926 & & & & & & & & & & & & & \\
\hline
A3 & 4.8 & 3.706$\times 10^{8}$ & 1 & 18.80510 & -0.30455 & 0.938 & A3c3 & 18.80509 & -0.30452 & 16.5 & N&Y&Y&N&N & 66.0 & 18.80612 & -0.30392 & 16 & 4.194$\times 10^{2}$ \\ 
& & & 2 & 18.80772 & -0.30693 & 0.342 & A3c5 & 18.80738 & -0.30536 & 0.971 & N& Y& Y& N& Y & 65.3 & & & & \\
\hline
B1 & 2.4 & 9.686$\times 10^{8}$ & 1 & 19.28645 & 0.08332 & 5.959 & B1c2 & 19.28614 & 0.08382 & 4.18 &Y &Y &Y& N&N & 26.8 & 19.28618 & 0.08371 & 11& 8.106$\times 10^{1}$  \\
& & & 2 & 19.28777 & 0.08203 & 7.061 & B1c2 & 19.28614 & 0.08382 & 4.18 &Y &Y &Y& N&N & 26.8   & & & & \\
& & & 3 & 19.28898 & 0.08192 & 0.361 & B1c2 & 19.28614 & 0.08382 & 4.18 &Y &Y &Y& N&N & 26.8  & & & & \\
\hline
B2 & 2.4 & 1.050e+10 & 1 & 19.31186 & 0.06678 & 35.549 & B2c9 &  19.31166&  0.06737 & 14.4 & Y& N& Y&Y & N & 26.0 & 19.31172 & 0.06726 & 15 & 2.976$\times 10^{2}$\\
& & &  &  &  &  & B2c10 &  19.31149  &  0.06633  & 2.31 & Y& N& Y& Y&N & 26.4 & & & & \\
& & & 2 & 19.30930 & 0.06477 & 0.700  & & & & & & & & & & & & & \\
& & & 3 & 19.30850 & 0.06399 & 2.048 & & & & & & & & & & & & & \\
\hline
C2 & 5.0 & 2.350e+09 & 1 & 28.34389 & 0.06251 & 0.457 & & & & & & & & & & &28.34574 & 0.06053 & 16 & 9.451$\times 10^{2}$\\
& & & 2 & 28.34721 & 0.06116 & 0.439 & & & & & & & & & & & & & \\
& & & 3 & 28.34298 & 0.06052 & 4.407 & C2c2 & 28.34284 & 0.06061 & 48.5 &Y & Y & Y & N&N & 79.2  & & & & \\
& & & 4 & 28.34607 & 0.05962 & 0.624 & C2c4 & 28.34610 & 0.05963 &  33.1 &Y &Y &Y & Y&N & 83.7 & & & & \\
& & & 5 & 28.34473 & 0.05901 & 0.372 & & & & & & & & & & & & & \\
& & & 6 & 28.34523 & 0.05804 & 1.181 & & & & & & & & & & & & & \\
\hline
C4 & 5.0 & 2.899e+09 & 1 & 28.35328 & 0.07419 & 0.923 & & & & & & & & & & & 28.35456 & 0.07170 & 13 & 2.103$\times 10^{2}$\\
& & & 2 & 28.35274 & 0.07370 & 0.593 & & & & & & & & & & & & & & \\
& & & 3 & 28.35651 & 0.07358 & 0.523 & C4c2 & 28.35596 & 0.07326 & 9.65 & Y&N & N& Y & N & 94.8 & & & & \\
& & & 4 & 28.35475 & 0.07349 & 2.566 & C4c1 & 28.35446 & 0.07388 & 16.8 & N& N& Y&Y& N & 83.1 & & & & \\
& & & 5 & 28.35254 & 0.07112 & 0.952 & C4c4 & 28.35276 & 0.07166 & 6.87 & N & N& Y & Y & N & 81.3 & & & & \\
& & & 6 & 28.35791 & 0.07067 & 0.336 & & & & & & & & & & & & & \\
& & & 7 & 28.35610 & 0.07006 & 0.284 & C4c6& 28.35599 &0.07114 & 0.509& & & & & & & & & \\
& & & 8 & 28.35110 & 0.06992 & 1.303 & & & & & & & & & & & & & \\
& & & 9 & 28.35427 & 0.06879 & 0.196 & C4c8& 28.35356 & 0.06867 & 3.10 & & & & & & & & & \\
& & & 10 & 28.35612 & 0.06858 & 1.552 & & & & & & & & & & & & & \\
\hline
C5 & 5.0 & 7.864$\times 10^{8}$ & 1 & 28.35447 & 0.05825 & 0.768 & & & & & & & & & & & 28.35620 & 0.05782 & 15 & 2.618$\times 10^{2}$\\
& & & 2 & 28.35690 & 0.05758 & 0.276 & C5c2 & 28.35705 & 0.05718& 1.13 & & & & & & & & & \\
& & & 3 & 28.35751 & 0.05732 & 0.247 & C5c1 & 28.35757& 0.05759 & 1.68 & & & & & & & & & \\
& & & 4 & 28.35653 & 0.05670 & 0.150 & & & & & & & & & & & & & \\
& & & 5 & 28.35470 & 0.05602 & 0.492 & & & & & & & & & & & & & \\
& & & 6 & 28.35621 & 0.05475 & 0.247 & & & & & & & & & & & & & \\
& & & 7 & 28.35344 & 0.05475 & 0.323 & C5c4 & 28.35622 & 0.05544 & 2.88 & & & & & & & & & \\
\hline
C6 & 5.0 & 3.511e+09 & 1 & 28.36350 & 0.05346 & 10.276 & C6c1 & 28.36310 & 0.05336 & 9.30 & N& N& Y & Y & Y & 80.0 & 28.36315 & 0.05323 & 15 & 4.484$\times 10^{2}$ \\  
& & &  &  &  &  & C6c2 & 28.36258 & 0.05322 & 3.01 & N& N& Y & Y & Y & 81.0 & & & & \\
& & & 3 & 28.36096 & 0.05206 & 0.901 & C6c5 & 28.36085 & 0.05246  & 8.77 &N &N &Y & N & N & 79.4 & & & & \\
\hline
C9 & 5.0 & 1.981e+10 & 1 & 28.40145 & 0.08545 & 13.221 & & & & & & & & & & & 28.39708 & 0.08034 & 10 & 6.284$\times 10^{3}$  \\
& & & 2 & 28.39759 & 0.08080 & 63.052 &C9c5 & 28.39701 & 0.08045 & 178 & Y &Y &Y & Y&N & 77.0 & & & & \\
& & &  &  &  &  & C9c7 & 28.39806 & 0.08011 &  25.6  & Y & Y &Y &Y & N & 78.5 & & & & \\
& & &  &  &  &  & C9c8 & 28.39726 & 0.07993 &  39.1  & Y &Y &Y & Y&N & 80.0 & & & & \\
& & & 3 & 28.40113 & 0.08011 & 1.883 & C9c6 & 28.40118 & 0.08028 & 13.9 & N &N&N&N&N &  & & & & \\
\hline
D1 & 5.7 & 5.574$\times 10^{8}$ & 1 & 28.52691 & -0.25075 & 0.324 & & & & & & & & & & & 28.52727 & -0.25093 & 15 & 2.957$\times 10^{2}$\\
& & & 2 & 28.52675 & -0.25162 & 0.288 & D1c4 & 28.52666 & -0.25146 &  5.34 & N &Y&Y&N&Y & 86.6 & & & & \\  
& & & 3 & 28.52409 & -0.25228 & 0.754 & & & & & & & & & & & & & \\
\hline
D3 & 5.7 & 2.144$\times 10^{8}$ & 1 & 28.54031 & -0.23711 & 0.525 & D3c4 & 28.54037 & -0.23710 & 1.01 &Y & Y & N &Y & N & 88.1 & 28.54167 & -0.23443 & 15 & 4.533$\times 10^{2}$\\
\hline
D5 & 5.7 & 1.531$\times 10^{8}$ & 1 & 28.56784 & -0.22944 & 0.375 & D5c1 & 28.56724 & -0.22810 &  2.61 &N &N &Y & N & N & 87.4 & 28.56469 & -0.23413 & 11 & 3.726$\times 10^{2}$ \\  
\hline
D6 & 5.7 & 4.713$\times 10^{8}$ & 1 & 28.55691 & -0.23766 & 0.632 & & & & & & & & & & & 28.55816 & -0.23847 & 15 & 2.758$\times 10^{2}$ \\
& & & 2 & 28.55571 & -0.23766 & 0.522 & D6c1 & 28.55565 & -0.23721 &  8.65 & Y&Y&Y&N&N & 83.0 & & & & \\
\hline
D8 & 5.7 & 1.413e+09 & 1 & 28.57227 & -0.23388 & 2.960 & D8c2 & 28.57080 & -0.23321 &  0.843 & N&N&Y&N&N & 88.9 & 28.56960 & -0.23314 & 11 & 2.330$\times 10^{2}$\\
& & & 2 & 28.57559 & -0.23445 & 0.500 & & & & & & & & & & & & & \\
\hline
D9 & 5.7 & 8.494$\times 10^{8}$ & 1 & 28.58892 & -0.22837 & 0.894 & D9c2 & 28.58877 & -0.22855 &  28.3 & N &Y &Y & N&N & 86.3 & 28.58980 & -0.22818 & 12 & 1.653$\times 10^{2}$ \\
& & & 2 & 28.58871 & -0.22972 & 1.187 & D9c2 & 28.58877 & -0.22855 &  28.3 & N &Y &Y & N&N & 86.3  & & & & \\
\hline
E1 & 5.1 & & 0 & - & - & - & - & - & - & - &- & - & - &-&- &- & 28.64497 & 0.13900 & 15 & 1.743$\times 10^{2}$\\
\hline
E2 & 5.1 & 1.487$\times 10^{8}$ & 1 & 28.64787 & 0.12465 & 0.312 & E2c2 & 28.64883 & 0.12454 &  3.72 & N &N &Y & Y& N & 78.7 & 28.64893 & 0.12560 & 15 & 2.083$\times 10^{2}$ \\
& & & 2 & 28.64975 & 0.12427 & 0.143 & E2c2 & 28.64883 & 0.12454 &  3.72 & N &N &Y & Y& N & 78.7 & & & & \\
\hline
F3 & 3.7 & & 0 & - & - & - & - & - & - & - &- & - & - &-&- &- & 34.44478 & 0.25019 & 6 & 2.398$\times 10^{1}$\\
\hline
F4 & 3.7 & & 0 & - & - & - & - & - & - & - &- & - & - &-&- &- & 34.45981 & 0.25909 & 15 & 8.865$\times 10^{1}$\\
\hline
H1 & 2.9 & & 0 & - & - & - & - & - & - & - &- & - & - &- &- &- & 35.48076 & -0.31016 & 15 & 8.717$\times 10^{1}$\\
\hline
H2 & 2.9 & & 0 & - & - & - & - & - & - & - &- & - & - &-&- &- & 35.48380 & -0.28637 & 10 & 3.910$\times 10^{1}$\\
\hline
H3 & 2.9 & 3.077$\times 10^{7}$ & 1 & 35.48633 & -0.29391 & 0.291 & H3c3 & 35.48693 & -0.29513 & 5.86  & N &Y &N & N&N & 43.6 & 35.48765 & -0.29326 & 11 & 3.370$\times 10^{1}$ \\
\hline
H4 & 2.9 & & 0 & - & - & - & - & - & - & - &- & - & - &-&- &- & 35.48504 & -0.28344 & 10 & 6.373$\times 10^{1}$\\
\hline
H5 & 2.9 & 1.584$\times 10^{8}$ & 1 & 35.49724 & -0.28803 & 1.499 & H5c3 & 35.49611 & -0.28813 &  6.39 & N &N &Y & N & N & 44.7 & 35.49653 & -0.28683 & 11 & 5.030$\times 10^{1}$ \\
\hline
H6 & 2.9 & 1.864$\times 10^{8}$ & 1 & 35.51886 & -0.27317 & 0.647 & H6c7 & 35.51908 & -0.27330 & 0.363 & N &Y &Y & N&N & 47.5 & 35.52357 & -0.27351 & 12 & 8.366$\times 10^{1}$ \\
& & & 2 & 35.52338 & -0.27343 & 1.764 & H6c8 & 35.52352 & -0.27337 & 2.53 & N &Y &Y & N & Y & 45.2 & & & & \\
\enddata
\end{deluxetable*}
\end{longrotatetable}

\subsection{Strong SiO Sources} \label{sec:strong}

There are 6 clumps (B1, B2, C2, C6, C9, H6) in which the peak of the
SiO integrated intensity is stronger than 10$\sigma$, as well as being clearly
associated with identified continuum sources. They are likely
protostars driving SiO outflows. In this section we present the
outflow morphologies, outflow kinematics and SED fitting results of
the 6 protostar candidates which show strongest SiO outflow
emission. As described in the previous sub-section, the radial velocity of the
protostar is estimated from the dense gas tracers $\rm C^{18}O$(2-1),
DCN(3-2) and DCO$^{+}$(3-2). 
Once the source velocity is defined, the velocity range of the SiO(5-4) emission for the source is determined by looking at the SiO spectra pixel by pixel in the outflow area.

\subsubsection{Morphology and Kinematics} \label{sec:morph}

\begin{figure*}[htbp]
\centering{\includegraphics[width=18cm]{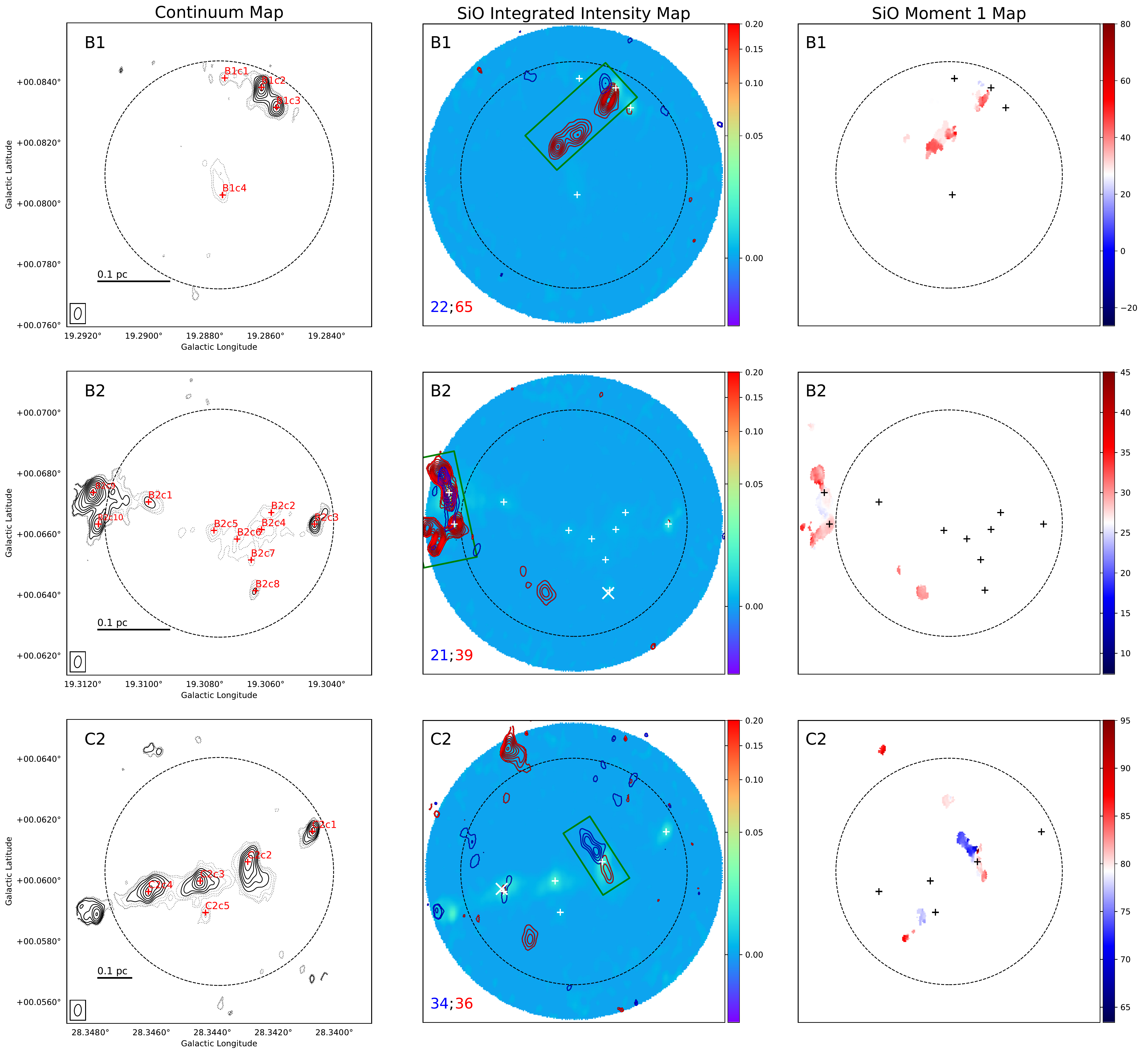}}
\caption{
\textit{Left Column}: 1.3~mm continuum maps. Contour levels are evenly
spaced logarithmically, corresponding to 0.800, 1.15, 1.65, 2.36, 3.39,
4.86, 6.98, 10.0, 14.4, 20.6, 29.6, 42.5, 60.9, 87.4, 125, 180 $\rm
mJy \ beam^{-1}$. The rms noise level in each image is $\sim \ 0.2
\ \rm mJy \ beam^{-1}$, except for C9 that is dynamic range
dominated and has an rms noise level of $\sim \ 0.6 \ \rm mJy
\ beam^{-1}$. The contours below 2 $\rm mJy \ beam^{-1}$ are
dotted. Note, here we only show emission above the 3~$sigma$ noise level.
The small red plus signs denote the peaks of the dendrogram-identified
continuum cores (Liu et al. 2018). The large black plus sign denotes
the center of the observation. The dashed circle shows the FWHM of the
primary beam. A scale bar and beam size are shown in the lower left
corner.
\textit{Middle Column}: Integrated intensity maps of SiO(5-4) emission
(contours) over 1.3~mm continuum emission (color scale in
Jy~beam$^{-1}$).  Contour levels start at 5$\sigma$ in steps of
4$\sigma$ noise level of the integrated intensity. The 1$\sigma$ noise values of the blue- and red-shifted outflows are written in the left corner in unit of mJy beam$^{-1}$ s$^{-1}$. The velocity ranges are listed in Table~\ref{tab:outflow}. The green rectangles denote the apertures for
deriving the averaged spectra and averaged PV diagrams. The small
white plus signs denote the peaks of the dendrogram-identified
continuum cores (Liu et al. 2018). White X signs denote water
masers detected by Wang et al. (2016). The dashed circle shows the
primary beam.
\textit{Right Column}: Intensity weighted velocity (first moment) maps of
SiO(5-4) emission (color scale in km s$^{-1}$). Note we only use
pixels with emission stronger than 3$\sigma$ noise level for H6 and
4$\sigma$ noise level for the other sources.
The small black plus signs denote the peaks of the
dendrogram-identified continuum cores (Liu et al. 2018). The dashed
circle shows the primary beam.} \label{fig:new}
\end{figure*}

\begin{figure*}[!htb]
    \ContinuedFloat
    \centering
    \captionsetup{list=off,format=cont, labelsep=space}
    \includegraphics[width=18cm]{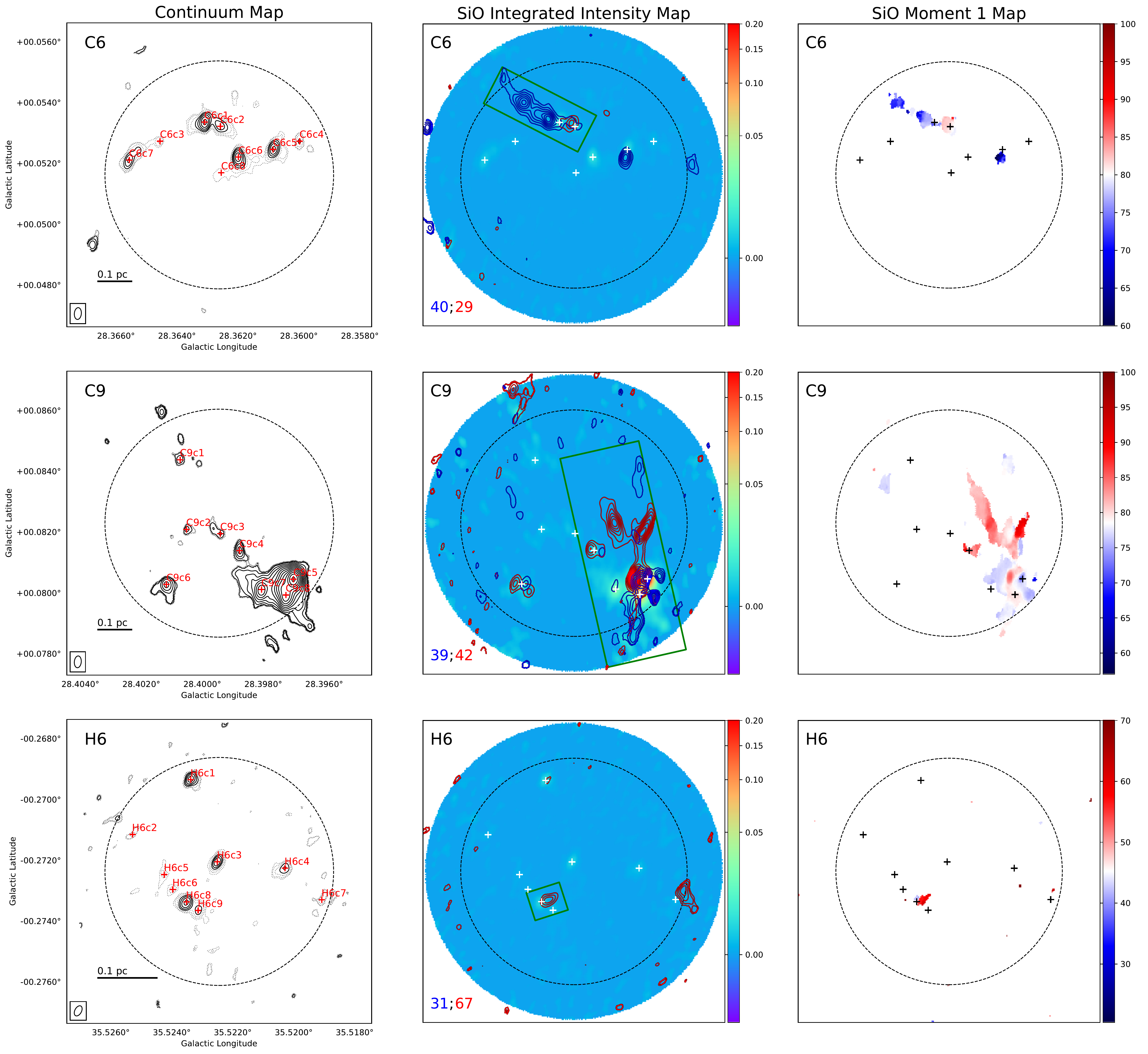}
    \caption{(cont.)}
\end{figure*}

\begin{figure*}[htbp]
\centering{\includegraphics[width=15cm]{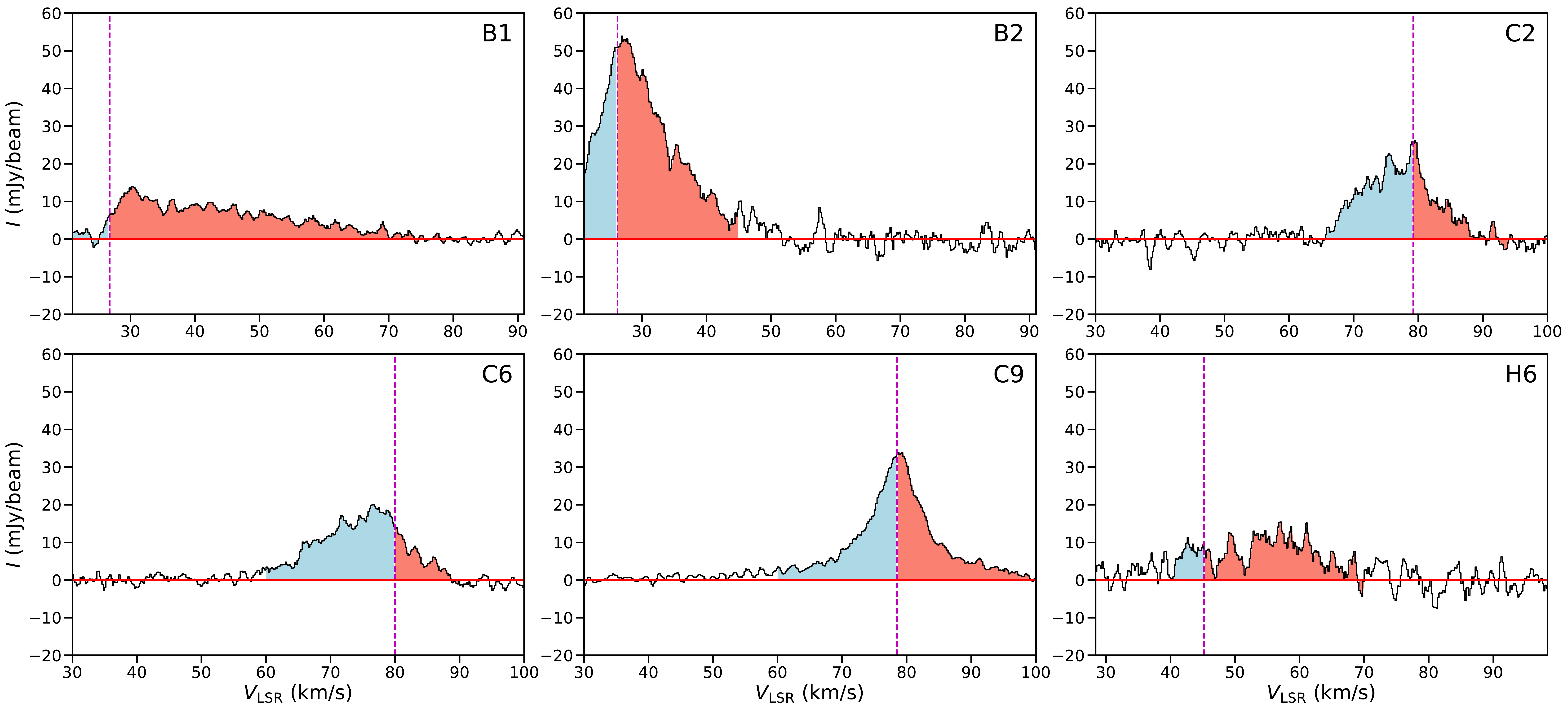}}
\caption{
Averaged SiO(5-4) spectra extracted from the defined rectangular
aperture of each source (see Figure~\ref{fig:new}). The dashed line
denotes the systemic velocity of the 1.3 mm continuum sources. The
blue and red areas denote the velocity range used to derive the blue-
and red-shifted outflows, respectively.}\label{fig:spectra}
\end{figure*}

\begin{figure*}[htbp]
\centering{\includegraphics[width=14cm]{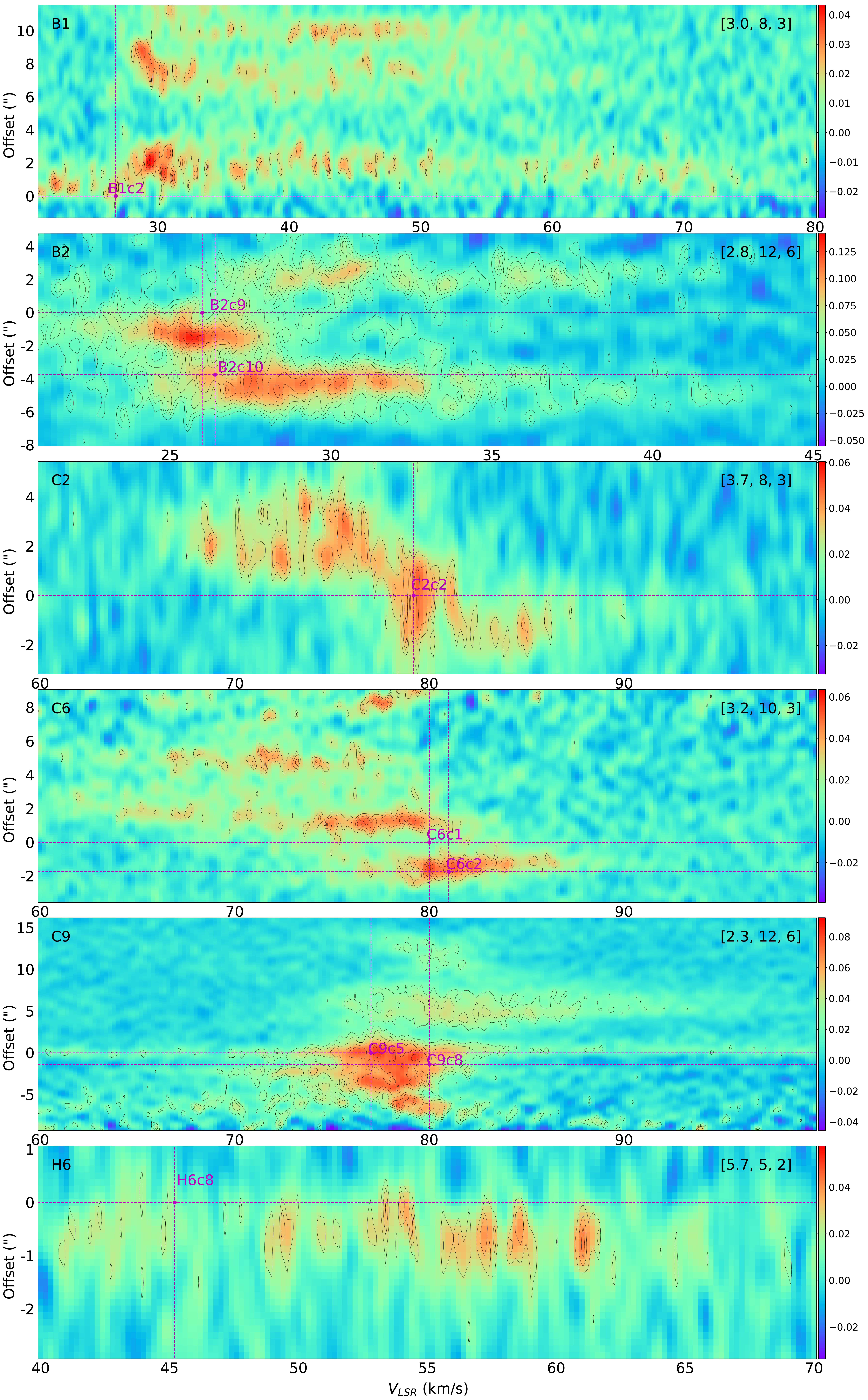}}
\caption{
Averaged position-velocity diagram of SiO(5-4) emission along the long
axis of the rectangles shown in Figure~\ref{fig:new}, i.e., the
outflow axes. The offset from top to bottom corresponds to the offset
from the reference position along the axis from east to west. The
reference position is where the continuum peak is located. Other
continuum peaks in the rectangle are also shown if any. Color scale is
in Jy~beam$^{-1}$. Contour level information is given in upper right:
1 $\sigma$ noise level in mJy beam$^{-1}$; lowest contour level in
number of $\sigma$; then step size between each contour in number of
$\sigma$.} \label{fig:pv}
\end{figure*}

We show the continuum maps, integrated intensity maps of SiO(5-4) and
velocity maps of SiO(5-4) of these sources in
Figure~\ref{fig:new}. The names of the continuum cores are the same
with those in Liu et al. (2018)\footnote{We add two more continuum
  cores in B2 which are outside the primary beam, but associated with
  the strong SiO emission.}. 
By checking the channel maps, we find that the SiO emission with very low velocity ($\rm \la 3 km~\rm s^{-1}$) also follows the shape of the outflow. Thus in our fiducial method we do not consider ambient gas as one would do with CO outflows.
The averaged spectra of SiO(5-4), extracted from the defined rectangular apertures
around each source, are shown in Figure ~\ref{fig:spectra}. The
velocity span of the SiO emission from these sources is
$\sim$30~km~$\rm s^{-1}$.  The averaged position-velocity (PV)
diagrams in the rectangular apertures along the outflow axes are shown
in Figure~\ref{fig:pv}. We can see there is large velocity dispersion
at a certain position in all the sources.

Overall, some outflows appear quite collimated, like C2 and C6, while
others are less ordered, like C9. The morphologies and kinematics of
the six sources are discussed individually below.

{\bf Source B1:} Since the spectral set-up only starts at 21 $\rm km
\ s^{-1}$, we may miss some emission from blue-shifted velocities. The
velocity range of SiO emission is $\sim$ 50 $\rm km \ s^{-1}$.  As
shown in Figure~\ref{fig:new}, the red-shifted component consists of
three peaks. We do not see any more emission beyond the current
displayed area. However, there may be more blue-shifted emission
emerging to the west of the view and in that case the outflow axis
would be oriented NW-SE. Otherwise, if the outflow axis is oriented
N-S, then the origin of the two red-shifted SiO peaks in the east
becomes less clear.

{\bf Source B2:} There is very strong SiO emission located at the
eastern edge of the primary beam. We used the same method in Liu et
al. (2018) to identify continuum cores in the outflow region outside
of the primary beam. Even though the noise is higher than inside the
primary beam, we were still able to identify two cores B2c9 and
B2c10. B2c9 has a mass of 14.4 $M_{\odot}$ and a mass surface density
of 3.14 g cm$^{-2}$ (assuming 20 K dust temperature). Similarly, B2c10
has a mass of 2.31 $M_{\odot}$ and a mass surface density of 1.32 g
cm$^{-2}$. The velocity range of SiO emission is $\sim$24 $\rm km
\ s^{-1}$. Like B1, given the spectral setting we may only see part of
the blue-shifted component. The outflow does not reveal a clear
bipolar structure and the blue lobes and red lobes overlap as shown in
Figure~\ref{fig:new}.
There is also some red-shifted emission within the primary beam which
may be connected to the eastern source(s).

{\bf Source C2:} The 1.3\,mm continuum image reveals five peaks. The
velocity range of the SiO emission is $\sim$30 $\rm km \ s^{-1}$. The
outflow is highly symmetric and highly collimated with a half-opening
angle of about 23$\degr$, as shown in Figure~\ref{fig:new}. Zhang et
al. (2015) also observed this clump (G28.34 P1) via 1.3 mm continuum
and multiple molecular lines including CO(2-1) and SiO(5-4) with a
higher sensitivity, which shows similar results. They determined
sub-fragmentation in the 1.3\,mm continuum cores with 2D Gaussian
fitting. From Figure~\ref{fig:pv} we can see C2 shows a ``Hubble-law"
velocity structure (e.g., Lada \& Fich, 1996) and similar to the
expectation of a single bow shock (Lee et al. 2000), in which the
highest velocity appears at the tip followed by a low-velocity ``wake"
or ``bow wing" and the velocity dispersion decreases significantly in
the post shocked region closer to the protostar. Arc-like PV
structures, similar to those seen in Fig.~\ref{fig:pv}, have been seen
in pulsed jet simulations (e.g., Stone \& Norman 1993; Lee \& Sahai
2004), indicating that this mechanism may be operating here.

{\bf Source C6:} The velocity range of the SiO emission is about 30
$\rm km \ s^{-1}$. As shown in Figure~\ref{fig:new}, the two lobes are
very asymmetric with the blue lobe extending much further and
exhibiting higher velocities (see Figure~\ref{fig:pv}). Similar
asymmetry is also revealed in the CO(2-1) outflow, where there is
strong blue-shifted emission but little red-shifted emission (Kong et
al. 2019). It may be due to an inhomogeneous ambient cloud
environment, which is denser in the south, or due to intrinsically
variable jets. The blue-shifted outflow is highly collimated, consists
of a chain of knots and also has a small wiggle, which resembles the
SiO jets revealed in the HH 212 low-mass protostellar system (Lee et
al. 2015), the V380 Ori NE region (Choi et al. 2017), and the CO
outflow in Serpens South (Plunkett et al. 2015). The knotty feature
suggests an episodic ejection mechanism (e.g., Qiu \& Zhang 2009;
Plunkett et al. 2015; Chen et al. 2016) or alternatively oblique
shocks (Reipurth 1992; Guilloteau et al. 1992). The wiggle may be
caused by jet precession (e.g., Choi et al. 2017) or instability in a
magnetized jet (e.g., Lee et al. 2015). Since we do not see symmetric
features in the red-shifted outflow, the orbiting source jet model of
a protobinary (e.g., Lee et al. 2010) is not favored, though it is
still possible if there is very dense ambient gas located where the
red-shifted outflow should be. The PV diagram (see
Figure~\ref{fig:pv}) is similar to the SiO jet in H212 (Codella et
al. 2007, see their Figure 2 Left). They suggested that the SiO lobes
include a narrower and faster jet-like component distinct from the
swept-up cavity and the high-velocity SiO is probably tracing the base
of the large-scale molecular jet. The velocity structures look like
``Hubble wedges" (Arce \& Goodman 2001). Recent theoretical models of
episodic protostellar outflows (e.g., Federrath et al. 2014; Offner \&
Arce 2014; Offner \& Chaban 2017; Rohde et al. 2019) have been built
to reproduce such features. A separate blue-shifted emission feature
seems to be driven by the continuum core C6c5, which is associated
with a CO(2-1) outflow (Kong et al. 2019). However, the systemic
velocity of C6c5 cannot be determined accurately due to the weak
emission of its dense gas tracers.

\begin{figure*}[htbp]
%\figurenum{}
\gridline{\fig{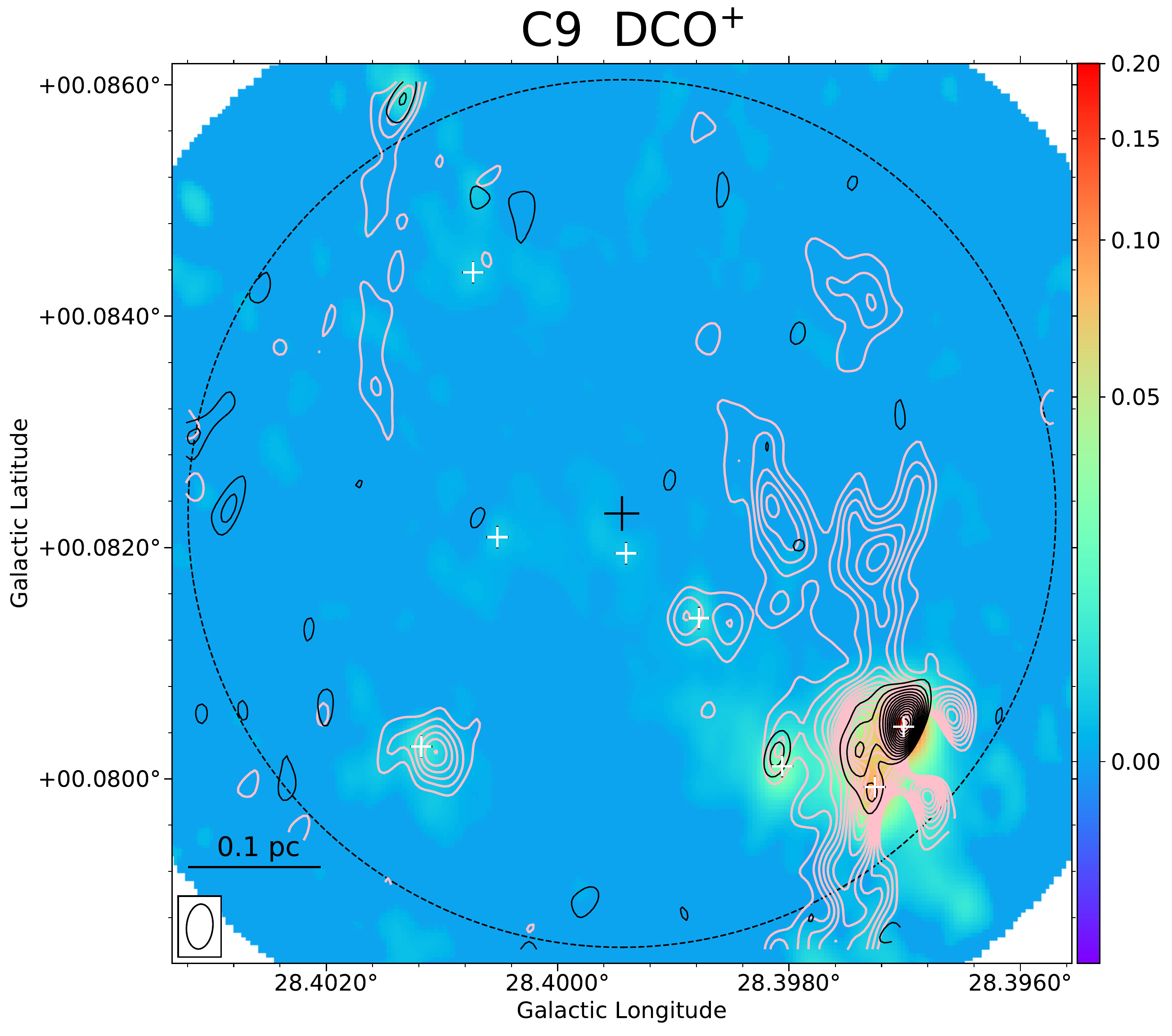}{0.45\textwidth}{(a)}
          \fig{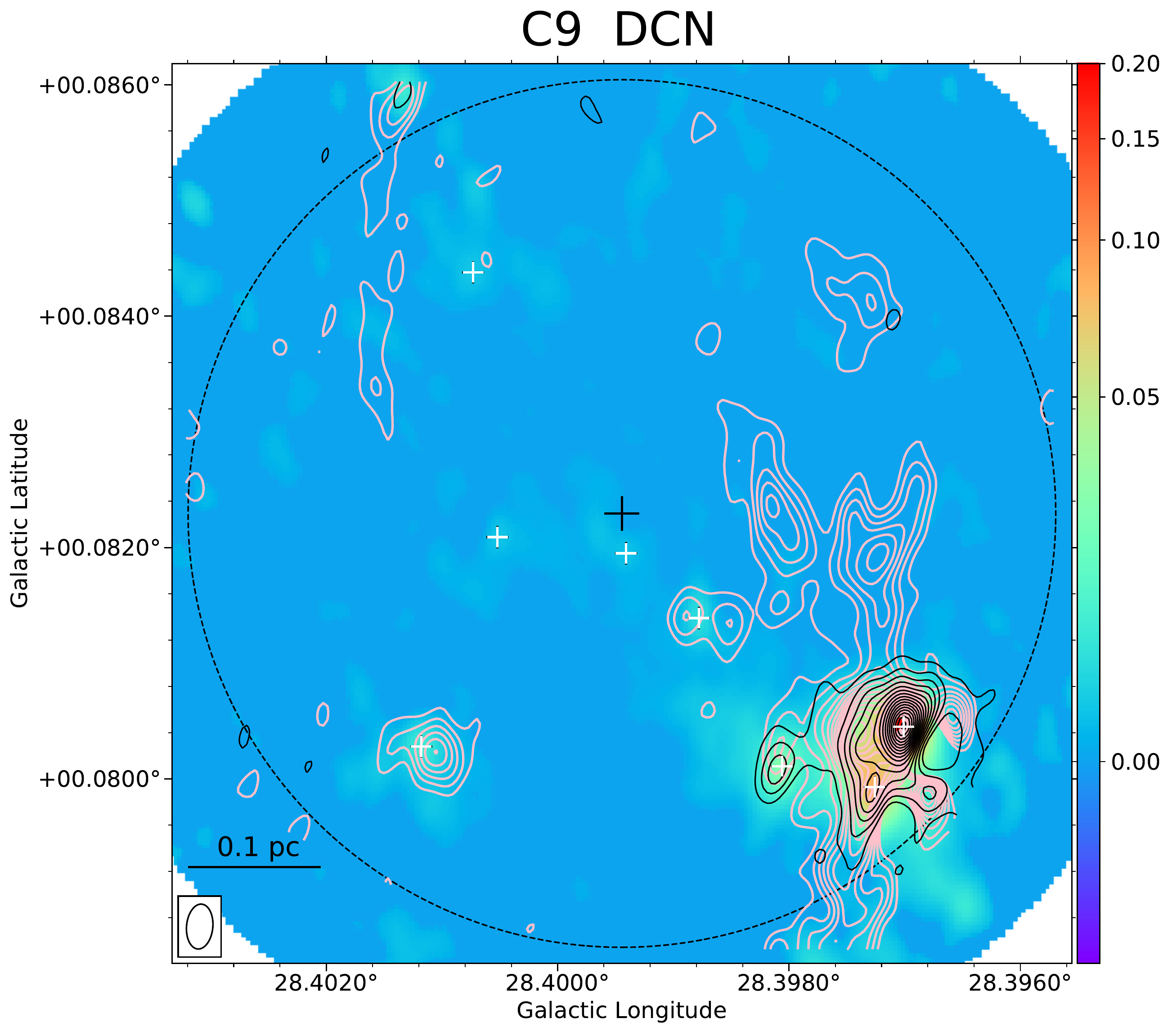}{0.45\textwidth}{(b)}
          }
\gridline{\fig{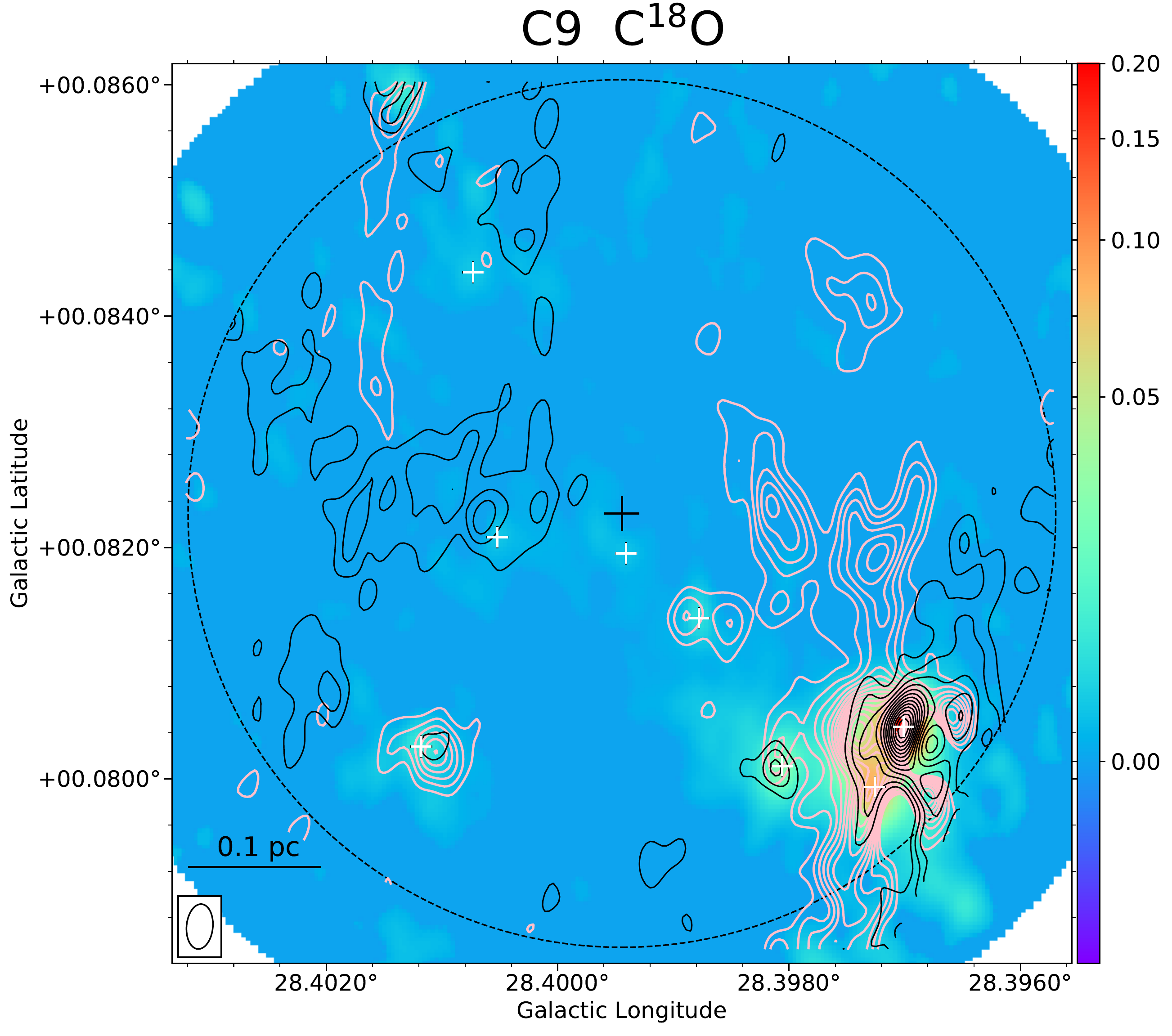}{0.45\textwidth}{(c)}
          \fig{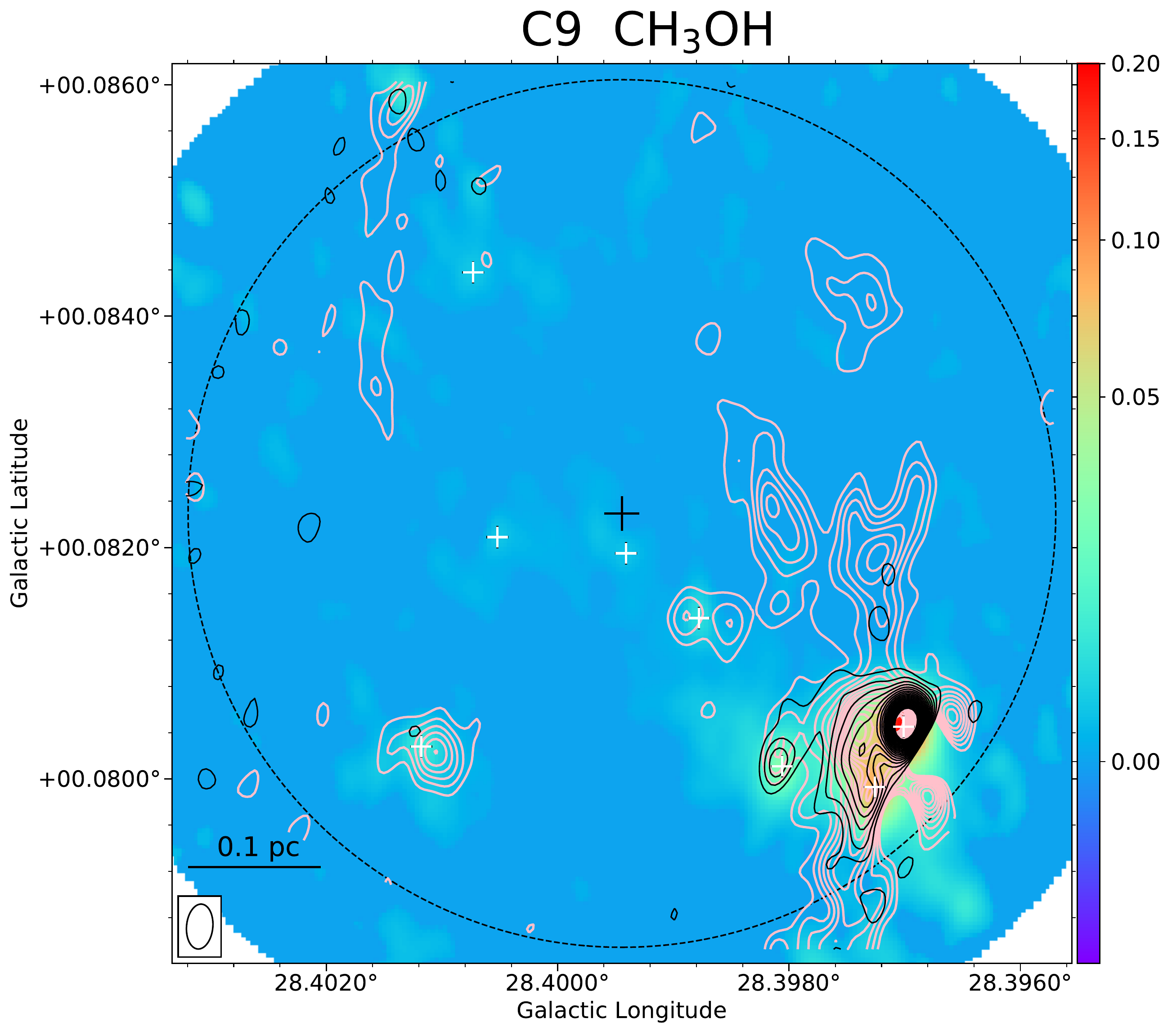}{0.45\textwidth}{(d)}
          }
\caption{
Integrated intensity map in C9 of: (a) $\rm DCO^{+}$(3-2), $\sigma$ =
19.8 mJy ${\rm beam^{-1}}$ km $\rm s^{-1}$; (b) DCN(3-2), $\sigma$ =
19.8 mJy ${\rm beam^{-1}}$ km $\rm s^{-1}$; (c) $\rm C^{18}O$(2-1),
$\sigma$ = 35.3 mJy ${\rm beam^{-1}}$ km $\rm s^{-1}$; (d) $\rm
CH_{3}OH (5_{1,4}-4_{2,2})$, $\sigma$ = 19.8 mJy ${\rm beam^{-1}}$ km $\rm s^{-1}$. All
of these are integrated within $\pm$ 5 km $\rm s^{-1}$ respect to the
cloud velocity. Contour levels start at 5$\sigma$ in steps of
4$\sigma$. The pink contours denote the SiO(5-4) emission of all the
360 velocity channels from 30 km $\rm s^{-1}$ to 102 km $\rm s^{-1}$,
starting at 380 mJy ${\rm beam^{-1}}$ km $\rm s^{-1}$ in steps of 304
mJy ${\rm beam^{-1}}$ km $\rm s^{-1}$. The small white plus signs
denote the positions of the continuum peaks. The large black plus sign
denotes the center of view. The dashed circle shows the primary
beam. A scale bar and beam size are shown in the lower left
corner.} \label{fig:C9_dense}
\end{figure*}

\begin{figure*}
%\figurenum{}
\gridline{\fig{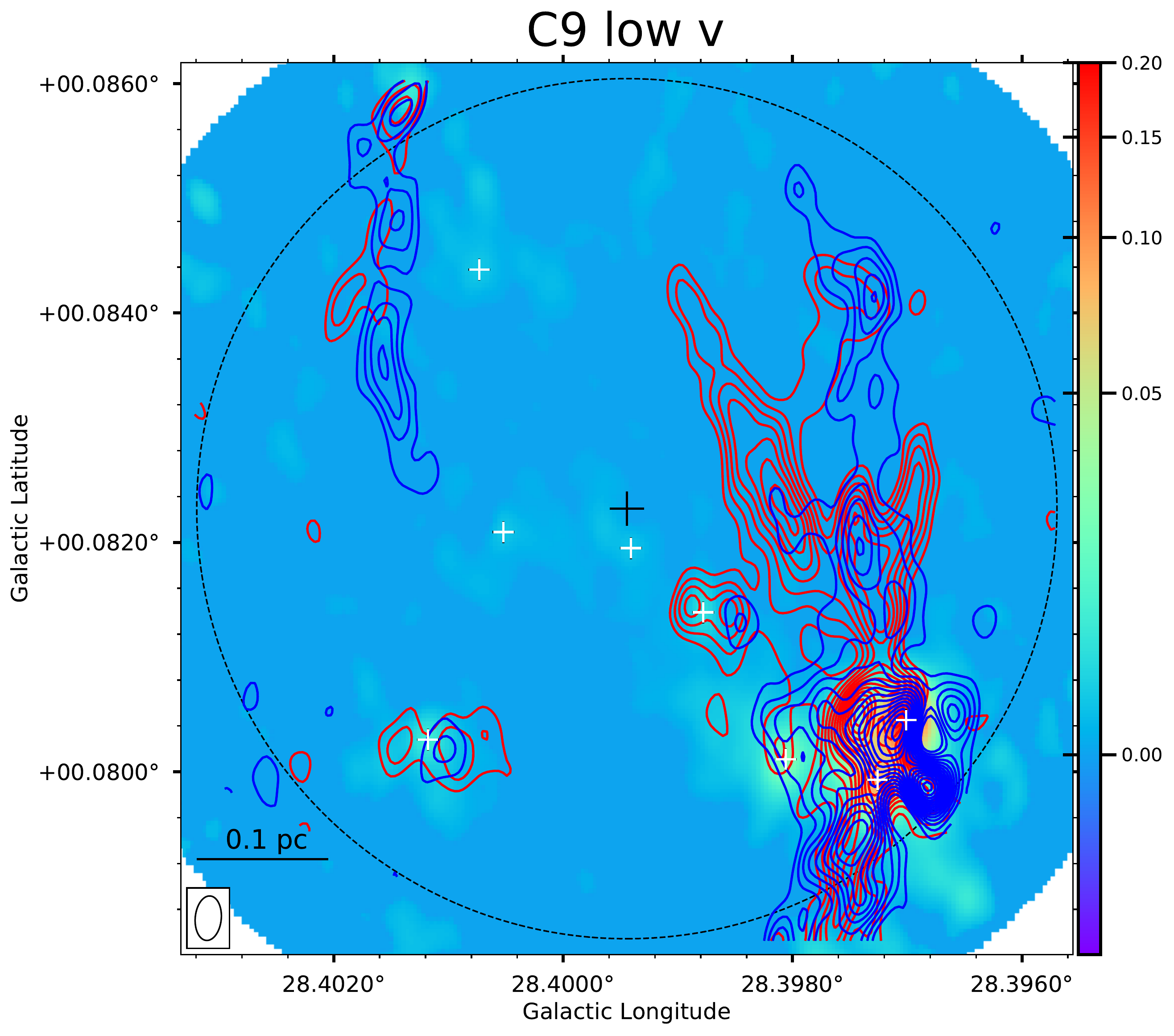}{0.4\textwidth}{(a)}
          \fig{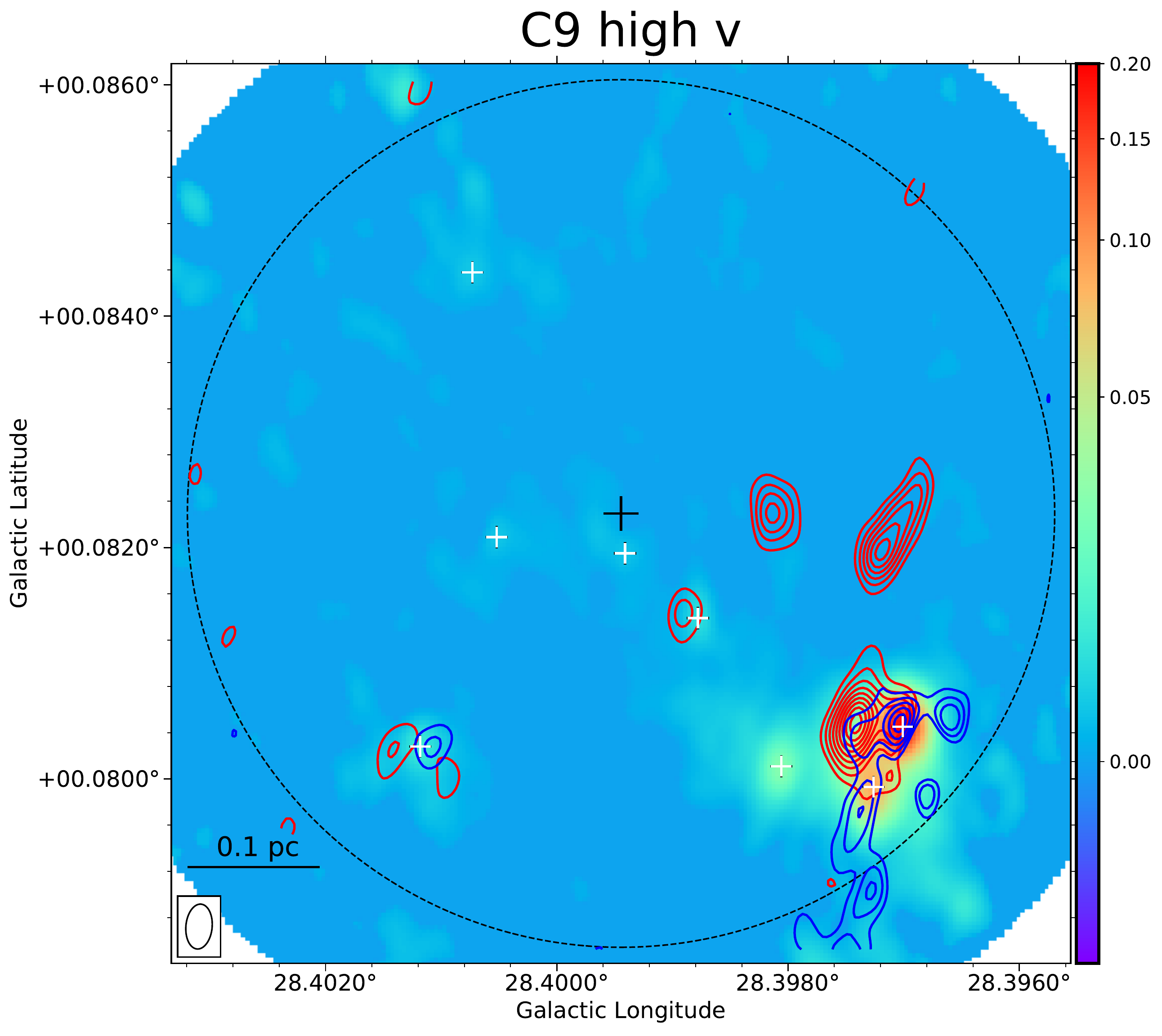}{0.4\textwidth}{(b)}
          }

\caption{
C9 SiO velocity components. (a) Integrated intensity maps of SiO(5-4)
emission with low velocity components over continuum emission. Blue
contours show emission from 67.0 $\rm km \ s^{-1}$ to 78.5 $\rm km
\ s^{-1}$. Red contours show emission from 78.5 $\rm km \ s^{-1}$ to
90.0 $\rm km \ s^{-1}$. Contour levels start at 5$\sigma$ in steps of
4$\sigma$ with with $\sigma$ = 30.5 mJy ${\rm beam^{-1}}$ km $\rm s^{-1}$
for the red-shifted component and $\sigma$ = 30.7 mJy ${\rm beam^{-1}}$ km $\rm s^{-1}$ for the blue-shifted
component. The small white plus signs
denote the positions of the continuum peaks. The large black plus sign
denotes the center of view. The dashed circle shows the primary
beam. A scale bar and beam size are shown in the lower left
corner. (b) The same as (a), but with high velocity components. Blue
contours show emission from 60.0 $\rm km \ s^{-1}$ to 67.0 $\rm km
\ s^{-1}$. Red contours show emission from 90.0 $\rm km \ s^{-1}$ to
100 $\rm km \ s^{-1}$. Contour levels start at 5$\sigma$ in steps of
4$\sigma$ with $\sigma$ = 28.6 mJy ${\rm beam^{-1}}$ km $\rm s^{-1}$
for the red-shifted component and $\sigma$ = 24.0 mJy ${\rm
  beam^{-1}}$ km $\rm s^{-1}$ for the blue-shifted
component.}\label{fig:C9_v}
\end{figure*}

{\bf Source C9:} The 1.3 mm continuum emission reveals 3 cores in the
main outflow area as shown in Figure~\ref{fig:new}. The brightest
core, C9c5, has a peak intensity as high as 197 mJy beam$^{-1}$, while
the second brightest core, C9c8, to its south has a peak intensity of
85.3 mJy beam$^{-1}$. The systemic velocities of C9c5 and C9c8 are
listed in Table~\ref{tab:sum}. At the position of continuum core C9c7,
the peak velocity of DCN is $\sim$80.0 $\rm km \ s^{-1}$ and the peak
velocity of $\rm DCO^{+}$ is $\sim$79.6 $\rm km \ s^{-1}$, while that
of $\rm C^{18}O$ is $\sim$75.9 $\rm km \ s^{-1}$. The systemic
velocity of C9c4 and C9c6 cannot be determined accurately due to their
weak emission in dense gas tracers. The velocity range of the SiO
emission is about 40 $\rm km\ s^{-1}$. Figure~\ref{fig:C9_dense} shows
the emission from the dense gas tracers $\rm DCO^{+}$, DCN, $\rm
C^{18}O$ and the hot gas tracer $\rm CH_{3}OH$. Their peaks
essentially overlap with the continuum peaks. DCN and $\rm CH_{3}OH$
also show extended structures associated with the SiO outflows, while
$\rm C^{18}O$ shows additional emission elsewhere. The disordered and
asymmetric morphology of the SiO outflows is probably due to the
crowded nature of the core region. Several velocity components are
revealed from the PV diagram in Figure~\ref{fig:pv}, with a hint of
``Hubble wedge". The most extended red-shifted emission lies mostly in
the north and the most extended blue-shifted emission in the
south. The morphology could be a result of a combination of the
extended outflows from both C9c5 and C9c8. We further display the
outflows in low-velocity ($<$ 10 $\rm km \ s^{-1}$ with respect to the
systemic velocity) channels and high-velocity ($>$ 10 $\rm km
\ s^{-1}$ with respect to the systemic velocity) channels in
Figure~\ref{fig:C9_v}. Together with further investigation in the
channel map (not shown here), it is more likely that the two farthest
high-velocity red-shifted components revealed in
Figure~\ref{fig:C9_v}(b) come from two distinctive outflows rather
than consisting of one outflow cavity wall. In addition to the most
extended north-south outflows, there seem to be three other smaller
scale outflows from Figure~\ref{fig:C9_v}(b). One has its blue- and
red-shifted components overlapping at C9c5 and is probably driven by
this source. The second one has its blue-shifted component to the west
of C9c5 and red-shifted components to the east of C9c5, and these
features are likely also driven by C9c5. The third one has its
blue-shifted component to the southwest of C9c5. From the spacial
distribution, this outflow could be driven by either C9c5 or C9c8, and
its red-shifted component either lies also to the east of C9c5 or
extends further. It is possible that there are more unresolved
protostars other than C9c5 and C9c8 in the region that drive the
multiple outflows, which would indicate a protobinary or multiple
system within our resolution limit of 5000 AU and that they each drive
an outflow of a different direction. Another possibility is that the
outflow orientation may change over time as reported in other
protostars (e.g., Cunningham et al. 2009; Plambeck et al. 2009;
Principe et al. 2018; Goddi et al. 2018; Brogan et
al. 2018). Nevertheless, the outflows associated with C9c6 and C9c4
are quite clear and relatively separate.

{\bf Source H6:} The blue-shifted emission is quite weak in this
source as shown in Figure~\ref{fig:new}. This may be partly due to
missing blue-shifted channels in our spectral setup. The velocity
range of the SiO outflow is about 30 $\rm km \ s^{-1}$.

\subsubsection{Outflow Mass and Energetics}

Following the method of Goldsmith \& Langer (1999) with an assumption
of optically thin thermal SiO(5-4) emission in local thermodynamic
equilibrium (LTE), we calculate the mass of the SiO(5-4) outflows
using
\begin{equation}
N_{u} = \frac{4\pi}{hcA_{\rm ul}}\int (S_{\nu}/\Omega) dv,
\end{equation}
\begin{equation}
N_{\rm tot} = N_{u} \frac{U(T_{\rm ex})}{g_{u}}e^{\frac{E_{u}}{kT_{\rm ex}}},
\end{equation}
\begin{equation}
M_{\rm out} = N_{\rm tot} \left[ \frac{\rm H_{2}}{\rm SiO} \right] \mu_{g}m_{\rm H_{2}}D^{2}, 
\end{equation}
where $S_{\nu}/\Omega$ is the SiO intensity at frequency $\nu$, $D$ is
the source distance, $\mu_{g} \ = \ 1.36$ is the mean atomic weight and $m_{\rm H_{2}}$ is
the mass of hydrogen molecule. We adopt an excitation temperature of
18~K and a ratio [H$_2$]/[SiO] of $10^{9}$, which are the typical values
of IRDC protostellar sources in the survey of Sanhueza et
al. (2012). The momentum and energy of the outflow are then derived
following
\begin{equation}
P_{\rm out} = \sum M_{\rm out}(\Delta v) \Delta v
\end{equation}
and  
\begin{equation}
E_{\rm out} = \frac{1}{2} \sum M_{\rm out}(\Delta v) \Delta v^{2},
\end{equation}
where $\Delta v$ denotes the outflow velocity relative to $v_{\rm cloud}$.

\begin{longrotatetable} % Landscape page
\centering
\setlength{\tabcolsep}{0.5pt}
\renewcommand{\arraystretch}{0.8}
\begin{table*}[htbp]
\centering
\setlength{\tabcolsep}{2pt}
\begin{center}
 \caption{Estimated Physical Parameters for SiO Outflows}\label{tab:outflow}
\small
\begin{tabular}{cccccccccccccccccc}
  \hline\noalign{\smallskip}
  \hline\noalign{\smallskip}
  Source & $v_{\rm sys}$ & $v_{\rm blue}$ & $M_{\rm out}^{\rm blue}$ & $P_{\rm out}^{\rm blue}$  & $L_{\rm flow}^{\rm blue}$ & $t_{\rm dyn}^{\rm blue}$ & $v_{\rm red}$ & $M_{\rm out}^{\rm red}$ & $P_{\rm out}^{\rm red}$ & $L_{\rm flow}^{\rm red}$ & $t_{\rm dyn}^{\rm red}$  & $M_{\rm out}$ & $P_{\rm out}$ & $E_{\rm out}$ & $\dot{M}_{\rm out}$ & $\dot{P}_{\rm out}$ \\
  & $(\rm km \ s^{-1})$ & $(\rm km \ s^{-1})$ & $(M_{\odot})$ & $(M_{\odot} $  & (pc) & $(10^{3} $ & $(\rm km \ s^{-1})$ & $(M_{\odot})$ & $(M_{\odot} $  & (pc) & $(10^{3} $ &   $(M_{\odot})$ & $(M_{\odot} $ & $(10^{43}$ & $(10^{-4} \ M_{\odot} $ & $(10^{-4} \ M_{\odot} $\\
 & &  &  & $ \rm km \ s^{-1})$ &  & $ \rm yr)$ &  &  & $ \rm km \ s^{-1})$ &  & $\rm yr)$  &   & $ \rm km \ s^{-1})$ & $\rm erg)$ & $\rm yr^{-1})$ & $\rm km \ s^{-1} yr^{-1})$\\
 \hline\noalign{\smallskip}
B1 & 26.8 & 21.0 $-$ 26.8 & 0.011 & 0.031 & 0.033 & 11.75 (8.10) & 26.8 $-$ 80.0 & 0.192 & 2.703 & 0.131 & 9.10 (5.13) & 0.204 & 2.735 & 33.738 & 0.221 & 5.310 \\
& 26.8 & 21.0 $-$ 26.8 & 0.011 & 0.031 & 0.033 & 5.59 & 26.8 $-$ 80.0 & 0.192 & 2.703 & 0.131 & 2.40 & 0.204 & 2.735 & 33.738 & 0.822 & 11.312 \\
& 26.8 & 21.0 $-$ 23.8 & 0.006 & 0.026 & 0.033 & 7.54 (7.26) & 29.8 $-$ 80.0 & 0.174 & 2.292 & 0.131 & 9.70 (5.50) & 0.180 & 2.318 & 26.672 & 0.187 & 4.202 \\
& 26.8 & 21.0 $-$ 23.8 & 0.006 & 0.026 & 0.033 & 5.59 & 29.8 $-$ 80.0 & 0.174 & 2.292 & 0.131 & 2.40 & 0.180 & 2.318 & 26.672 & 0.735 & 9.589 \\
\hline\noalign{\smallskip}
B2 & 26.2 & 21.0 $-$ 26.2 & 0.138 & 0.294 & 0.079 & 35.98 (24.06) & 26.2 $-$ 45.0 & 0.384 & 1.531 & 0.150 & 36.79 (23.00) & 0.522 & 1.825 & 5.354 & 0.143 & 0.788 \\
& 26.2 & 21.0 $-$ 26.2 & 0.138 & 0.294 & 0.079 & 14.77 & 26.2 $-$ 45.0 & 0.384 & 1.531 & 0.150 & 7.81 & 0.522 & 1.825 & 5.354 & 0.585 & 2.160 \\
& 26.2 & 21.0 $-$ 23.2 & 0.044 & 0.177 & 0.079 & 19.29 (18.78) & 29.2 $-$ 45.0 & 0.267 & 0.956 & 0.150 & 41.00 (28.40) & 0.311 & 1.133 & 2.831 & 0.088 & 0.431 \\
& 26.2 & 21.0 $-$ 23.2 & 0.044 & 0.177 & 0.079 & 14.77 & 29.2 $-$ 45.0 & 0.267 & 0.956 & 0.150 & 7.81 & 0.311 & 1.133 & 2.831 & 0.372 & 1.344 \\
\hline\noalign{\smallskip}
C2 & 79.2 & 65.0 $-$ 79.2 & 0.225 & 1.155 & 0.121 & 22.97 (15.74) & 79.2 $-$ 95.0 & 0.145 & 1.431 & 0.081 & 8.06 (7.03) & 0.370 & 2.587 & 12.441 & 0.278 & 2.771 \\
& 79.2 & 65.0 $-$ 79.2 & 0.225 & 1.155 & 0.121 & 8.31 & 79.2 $-$ 95.0 & 0.145 & 1.431 & 0.081 & 5.04 & 0.370 & 2.587 & 12.441 & 0.558 & 4.230 \\
& 79.2 & 65.0 $-$ 76.2 & 0.154 & 1.063 & 0.121 & 17.15 (14.79) & 82.2 $-$ 95.0 & 0.079 & 0.797 & 0.081 & 7.87 (7.11) & 0.233 & 1.860 & 8.705 & 0.190 & 1.839 \\
& 79.2 & 65.0 $-$ 76.2 & 0.154 & 1.063 & 0.121 & 8.31 & 82.2 $-$ 95.0 & 0.079 & 0.797 & 0.081 & 5.04 & 0.233 & 1.860 & 8.705 & 0.342 & 2.860 \\
\hline\noalign{\smallskip}
C6 & 80.0 & 60.0 $-$ 80.0 & 0.571 & 4.354 & 0.212 & 27.20 (18.50) & 80.0 $-$ 90.0 & 0.179 & 2.953 & 0.079 & 4.67 (4.55) & 0.750 & 7.307 & 49.392 & 0.593 & 8.840 \\
& 80.0 & 60.0 $-$ 80.0 & 0.571 & 4.354 & 0.212 & 10.37 & 80.0 $-$ 90.0 & 0.179 & 2.953 & 0.079 & 7.70 & 0.750 & 7.307 & 49.392 & 0.783 & 8.032 \\
& 80.0 & 60.0 $-$ 77.0 & 0.442 & 4.172 & 0.212 & 21.99 (17.85) & 83.0 $-$ 90.0 & 0.092 & 1.586 & 0.079 & 4.45 (4.39) & 0.534 & 5.758 & 38.150 & 0.407 & 5.950 \\
& 80.0 & 60.0 $-$ 77.0 & 0.442 & 4.172 & 0.212 & 10.37 & 83.0 $-$ 90.0 & 0.092 & 1.586 & 0.079 & 7.70 & 0.534 & 5.758 & 38.150 & 0.545 & 6.082 \\
\hline\noalign{\smallskip}
C9 & 78.5 & 60.0 $-$ 78.5 & 1.942 & 10.621 & 0.203 & 36.36 (20.25) & 78.5 $-$ 100.0 & 2.256 & 28.378 & 0.392 & 30.48 (25.87) & 4.197 & 38.999 & 262.490 & 1.274 & 16.214 \\
& 78.5 & 60.0 $-$ 78.5 & 1.942 & 10.621 & 0.203 & 10.75 & 78.5 $-$ 100.0 & 2.256 & 28.378 & 0.392 & 17.84 & 4.197 & 38.999 & 262.490 & 3.071 & 25.790 \\
& 78.5 & 60.0 $-$ 75.5 & 1.123 & 9.489 & 0.203 & 23.53 (18.48) & 81.5 $-$ 100.0 & 1.360 & 16.752 & 0.392 & 31.14 (26.44) & 2.483 & 26.241 & 172.524 & 0.914 & 11.469 \\
& 78.5 & 60.0 $-$ 75.5 & 1.123 & 9.489 & 0.203 & 10.75 & 81.5 $-$ 100.0 & 1.360 & 16.752 & 0.392 & 17.84 & 2.483 & 26.241 & 172.524 & 1.807 & 18.219 \\
\hline\noalign{\smallskip}
H6 & 45.2 & 40.0 $-$ 45.2 & 0.005 & 0.011 & 0.011\tablenotemark{a} & 5.15 (3.77) & 45.2 $-$ 70.0 & 0.025 & 0.176 & 0.039 & 5.34 (3.67) & 0.030 & 0.187 & 0.925 & 0.057 & 0.510 \\
& 45.2 & 40.0 $-$ 45.2 & 0.005 & 0.011 & 0.011 & 2.11 & 45.2 $-$ 70.0 & 0.025 & 0.176 & 0.039 & 1.52 & 0.030 & 0.187 & 0.925 & 0.188 & 1.208 \\
& 45.2 & 40.0 $-$ 42.2 & 0.001 & 0.006 & 0.011 & 2.90 (2.86) & 48.2 $-$ 70.0 & 0.023 & 0.122 & 0.039 & 7.25 (4.56) & 0.025 & 0.128 & 0.518 & 0.037 & 0.288 \\
& 45.2 & 40.0 $-$ 42.2 & 0.001 & 0.006 & 0.011 & 2.11 & 48.2 $-$ 70.0 & 0.023 & 0.122 & 0.039 & 1.52 & 0.025 & 0.128 & 0.518 & 0.161 & 0.829 \\
\hline\noalign{\smallskip}
\end{tabular}
\end{center}
\tablenotetext{a}{We adopt 0.8" for the length of the blue-shifted flow of H6 as an upper limit, which is the minor axis of the beam size.}
\tablecomments{
For each source, the values in the first row are derived using our
fiducial method that uses the full velocity range for the integration
of SiO emission and with $t_{\rm dyn}$ evaluated with an average
outflow velocity. Values of $t_{\rm dyn}$ outside and inside
parentheses use mass and momentum weighted velocities,
respectively. The results in the second row for each source are
derived in the same way as the first row, except that $t_{\rm dyn}$ is
evaluated by using the maximum observed outflow velocity. The values
in the third row are as as the first row, except that SiO emission
within $\pm$ 3 km s$^{-1}$ of the systemic velocity is excluded. The
values in the fourth row are derived in the same way as the third row,
except that $t_{\rm dyn}$ is evaluated by using the maximum observed
outflow velocity.}
\end{table*}
\end{longrotatetable}

 We investigate different methods to calculate the outflow
  properties by varying the velocity range of outflows and the assumed
  dynamical timescale. Our fiducial method involves integrating the
  outflow emission all the way to the systemic velocity of the source
  to calculate the outflow mass. As an alternative, we also try a case
  where velocities within $v_{\rm sys} \pm 3 \rm km \ s^{-1}$ are
  assumed to be ambient material and so are not counted as being part
  of the outflow.
  For the dynamical timescale, our fiducial method assumes $t_{\rm
    dyn} = L_{\rm flow}/v_{\rm avg}$, where $L_{\rm flow}$ is the
  length of the flow extension and $v_{\rm avg}$ is the mass or
  momentum weighted average flow velocity relative to $v_{\rm sys}$.
  As an alternative way, we also consider $t_{\rm dyn} = L_{\rm
    flow}/v_{\rm max}$, where $v_{\rm max}$ is the maximum observed
  flow velocity relative to $v_{\rm sys}$. The combination of these
  two choices, leads to four methods and for each of these we estimate
  the total mass flow rate as $\dot{M}_{\rm out} = M_{\rm out}^{\rm
    blue}/t_{\rm dyn}^{\rm blue} + M_{\rm out}^{\rm red}/t_{\rm
    dyn}^{\rm red}$ and the total momentum flow rate as $\dot{P}_{\rm
    out} = P_{\rm out}^{\rm blue}/t_{\rm dyn}^{\rm blue} + P_{\rm
    out}^{\rm red}/t_{\rm dyn}^{\rm red}$. Note, if $t_{\rm dyn}$ is
  derived with an average outflow velocity, $\dot{M}_{\rm out}$ is
  derived using the mass weighted average velocity and $\dot{P}_{\rm
    out}$ is derived using the momentum weighted average
  velocity. Note also we only count those pixels with integrated
  intensity higher than the 3\,$\sigma$ noise level. No correction for
  inclination to the line of sight, which is uncertain, is applied
  here. These results of the outflow properties derived using these
  various methods are listed in Table~\ref{tab:outflow}.

The derived total outflow masses range from about 0.03 to 4 $M_{\odot}$. The choice of whether to integrate all the way in to the systemic velocity can make a difference of up to about a factor of 1.7 in the mass estimation. The outflow dynamical timescales range from about 1,500~yr up to about 40,000~yr. The choice of whether to use average or maximum velocities to define this timescale can make a difference of up to a factor of about 5, which we thus see is one of the most important factors leading to systematic uncertainties in determination of outflow properties. However, in general we prefer to adopt values based on averaged velocities and consider those based on maximum velocities to be an extreme limiting case. Total mass outflow rates derived using our fiducial method range from (0.06 to 1.3) $\times 10^{-4} \: M_{\odot}
\: \rm yr^{-1}$, while the fiducial method total momentum flow rates range from (0.5 to 16) $\times 10^{-4}\: M_{\odot}\: \rm km \: s^{-1} \: yr^{-1}$.

For comparison, low-mass protostars typically have molecular
outflow mass fluxes as high as $\sim 10^{-6} \ M_{\odot} \ \rm yr^{-1}$ and momentum flow rates
of $\sim 10^{-5} \ M_{\odot} \ \rm km \ s^{-1} \ yr^{-1}$,
while mid- to early-B type protostars have mass outflow rates
$10^{-5}$ to a few $\times \ 10^{-3} \ M_{\odot} \ \rm yr^{-1}$ and
momentum flow rates $10^{-4}$ to $10^{-2} \ M_{\odot} \ \rm km
\ s^{-1} \ yr^{-1}$ (e.g., Arce et al. 2007). Our sources are at the
typical lower limit of mid- to early-B type protostar outflows. Based
on the 1.3 mm continuum core mass, the C2 and C9 cores are likely to
be high-mass protostellar objects. Compared with the massive molecular
outflows traced by CO(2-1) in Beuther et al. (2002) and the massive
outflows traced by SiO(5-4) in Gibb et al. (2007) and
S{\'a}nchez-Monge et al. (2013b), the outflow mass $(M_{\rm out})$, mass
outflow rate ($\dot{M}_{\rm out}$) and the mechanical force
($\dot{P}_{\rm out}$) of C9 are comparable to those in their sample
with a similar outflow length.
There are several possibilities causing the relatively low outflow
parameters of the high-/intermediate-mass protostars in our
sample. First, the high-mass protostars may be still at an early stage
when the outflows have not formed completely. However, there are
protostars in S\'{a}nchez-Monge et al. (2013b) and Csengeri et
al. (2016) that are also very young (indicated by $L/M$) but having
strong SiO outflow emission (see further discussion in
\S\ref{sec:vs}). Second, SiO, as observed here, may not be tracing the
full extent of the outflows. We will return to this point in \S5.1, where a comparison of SiO and CO morphologies is made for a subset of the sources. This may also explain the low SiO-derived outflow
masses compared with CO outflows in Beuther et al. (2002).

\subsubsection{SED Modeling} \label{sec:SED}

\begin{figure*}[htbp]
\centering{\includegraphics[width=15cm]{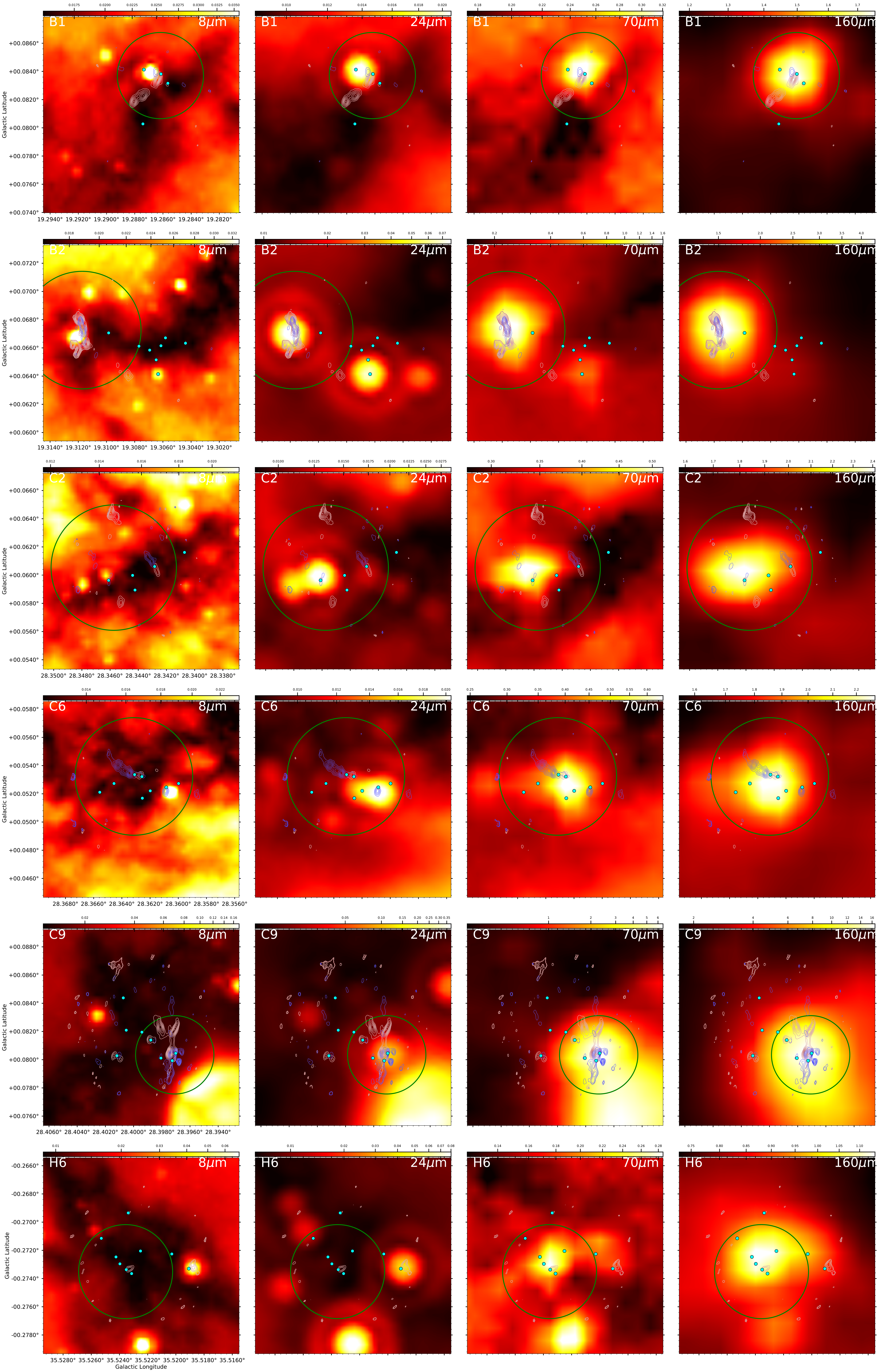}}
\caption{
8\,$\mu$m, 24\,$\mu$m, 70\,$\mu$m and 160\,$\mu$m emission of the
detected protostars. The color scale is mJy pixel$^{-1}$. The blue and
red contours denote the blue lobe and the red lobe of SiO outflows,
contour levels the same as those in Figure~\ref{fig:new}. The green circle
denotes the aperture size used for building SEDs. The aperture radius
is 11\arcsec for B1, 15\arcsec \ for B2, 16\arcsec \ for C2, 15\arcsec
\ for C6, 10\arcsec \ for C9 and 12\arcsec \ for H6. The cyan dots
denotes the positions of the 1.3\,mm cores.}\label{fig:M0_ir}
\end{figure*}

\begin{figure*}[htbp]
\epsscale{1.1}
\plotone{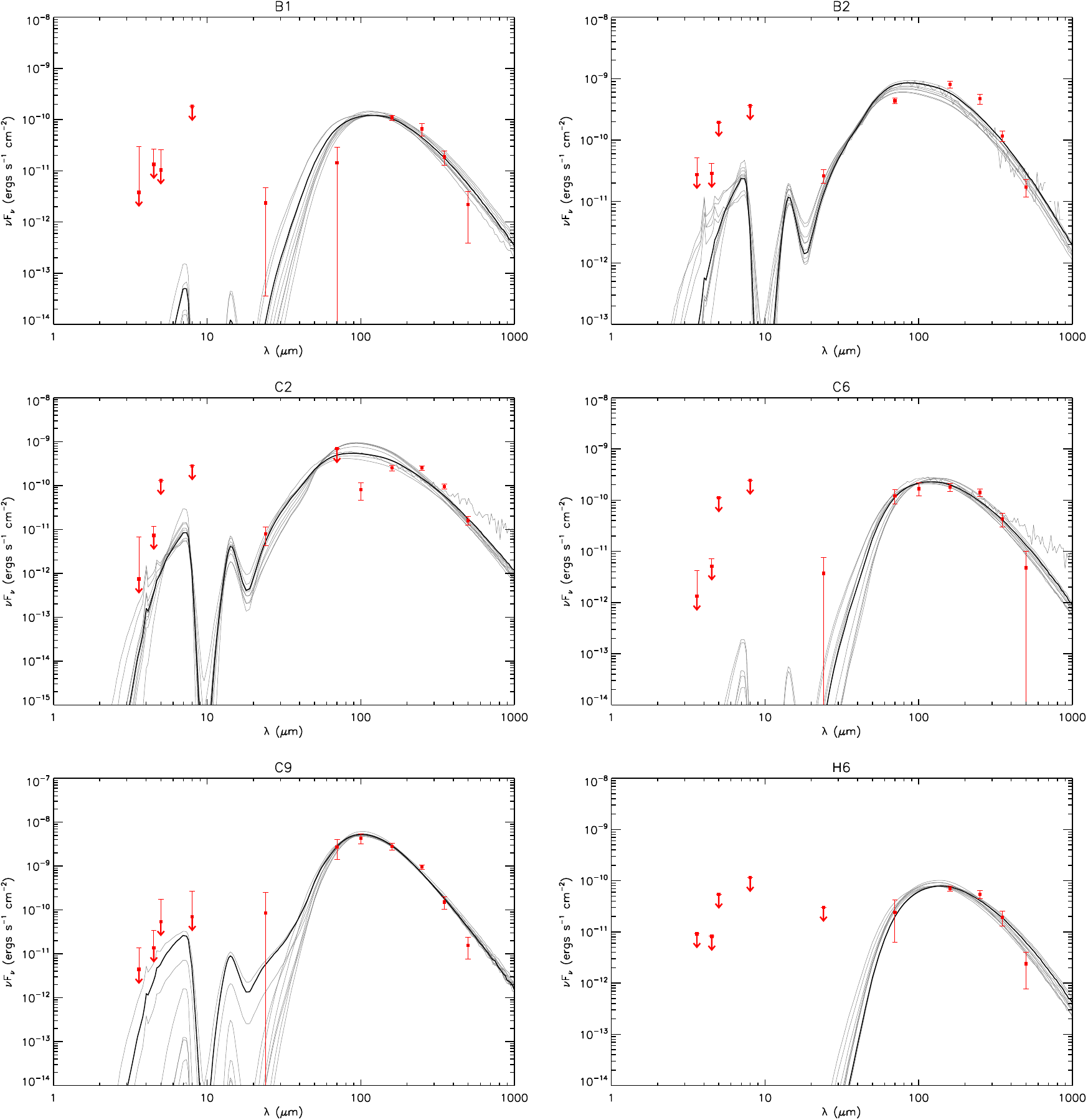}
\caption{
Protostar model fitting to the fixed aperture, background-subtracted
SED data using the ZT model grid. For each source, the best fit model
is shown with a solid black line and the next nine best models are
shown with solid gray lines. Note that the data at $\lesssim8\:{\rm
  \mu m}$ are treated as upper limits. If a background subtracted flux
density is negative, the flux density without background subtraction
is used as upper limits with a negligible error bar (see text). The
model parameter results are listed in
Table~\ref{tab:SEDfit}.}  \label{fig:SEDfit}
\end{figure*}

We investigate the IR emission of the sources that drive the six
strongest SiO outflows with \textit{Spitzer} and \textit{Herschel}
archival data, as shown in Figure~\ref{fig:M0_ir}. This
  allows us to better understand the nature of the protostars, i.e., by
  understanding the IR morphologies and comparing to SED models. Note the dynamic range of
Figure~\ref{fig:M0_ir} is set in such a way that faint sources can
still show strong contrast. However, the SNR can be very low as in B1,
C2, C6 at all wavelengths, B2 at 8~$\mu$m, and H6 at 70~$\mu$m and
160~$\mu$m. The C2, C6, C9 and H6 cores appear dark against the
Galactic background at 8 $\mu$m, which indicates they are at an early
evolutionary stage. We build spectral energy distributions (SEDs) of
the six sources from 3.6~$\rm \mu m$ up to 500~$\rm \mu m$. Given the
relatively large beam size of \textit{Herschel}, we cannot resolve the
individual cores revealed with ALMA and the SEDs represent emission
from a larger scale region. The circular apertures are determined to
include most of the source flux based on their 70 $\mu$m and 160
$\mu$m emission. For B1, C6, C9 and H6, we try to make the apertures
centered at the protostar driving the main outflow. The typical
aperture size is comparable to the primary beam size of our ALMA 1.3
mm observations.

We fit the fixed aperture, background-subtracted SEDs with the
radiative transfer (RT) model developed by Zhang \& Tan (2018), which
describes the evolution of massive (and intermediate-mass) protostars
based on the Turbulent Core model (MT03). The model is described by
five physical parameters: the initial core mass ($M_{c}$), the mean
mass surface density of the surrounding clump ($\Sigma_{\rm cl}$), the
current protostellar mass ($m_{*}$), the foreground extinction
($A_{V}$), and the inclination angle of the outflow axis to the line
of sight ($\theta_{\rm view}$). The models describe collapsing cores
with bolometric luminosities ranging from 10 $L_{\odot}$ to $10^{7}
\ L_{\odot}$ and envelope temperatures from 10 K to 100 K. The
evolutionary timescales range from $10^{3}$ yr to $10^{6}$ yr. In the
grid of models, $M_{c}$ is sampled at 10, 20, 30, 40, 50, 60, 80,
100, 120, 160, 200, 240, 320, 400, 480 $M_{\odot}$ and $\Sigma_{\rm
  cl}$ is sampled at 0.10, 0.32, 1, 3.2 $\rm g \ cm^{-2}$, for a total
of 60 evolutionary tracks. Then along each track, $m_{*}$ is sampled
at 0.5, 1, 2, 4, 8, 12, 16, 24, 32, 48, 64, 96, 128, 160 $M_{\odot}$
(but on each track, the sampling is limited by the final achieved
stellar mass, with star formation efficiencies from the core typically
being $\sim$ 0.5). There are then, in total, 432 physical models
defined by different sets of ($M_{c}$, $\Sigma_{\rm cl}$, $m_{*}$).

The method of deriving SEDs is described in Appendix~\ref{sec:more
  SED}. Here we set the data points at wavelengths $\lesssim8\:{\rm
  \mu m}$ as upper limits given that PAH emission and transiently
heated small grain emission are not well treated in the RT models. In
addition, given the high background that can be present in IRDCs at
certain wavelengths, sometimes the background subtracted flux density
can be negative. In this situation, we use the flux density without
background subtraction as upper limits too. These upper limits can be
distinguished by a negligible error bar. We note that different
aperture sizes can make a significant difference in the flux derived
especially for faint sources whose boundaries are not clear. Since we
do not know the real distribution of the measurement error, the
absolute value of the $\chi^{2}$ is currently dominated by the size of
measurement error and does not indicate the goodness of the model
well. However, for the same source under the theoretical model we can
tell which set of parameters describes the status of the protostar
better by comparing their relative values of $\chi^{2}$.  For
convenience we show the 10 best models. Amongst the best 10 models
there can be a significant variation in model parameters, even though
the shape of the model SED does not change much, which illustrates
degeneracies that exist in trying to constrain protostellar properties
from only their MIR to FIR SEDs (see also De Buizer et al. 2017;
Rosero et al. 2019a; Liu et al. 2019, 2020). Based on experience, when
the best model returns a $\chi^{2}$ smaller than 1 it indicates there
are too few valid data points constraining the fitting, so we would
consider all the models with $\chi^{2}<2$ among the 10 best models as
valid. When the best model returns a $\chi^{2}$ higher than 1, we
would consider all the models with $\chi^{2}$ smaller than twice the
$\chi^{2}$ of the best model among the 10 best models as valid. For
more detailed discussion of the sensitivity of the model to choices in
SED construction for faint sources in IRDCs, see Moser et
al. (2020). We show the 10 best models for each source in
Figure~\ref{fig:SEDfit}. The physical parameters derived are listed in
Table~\ref{tab:SEDfit}. Note that these are distinct physical models
with differing values of $M_{c}$, $\Sigma_{\rm cl}$ and $m_{*}$, i.e.,
we do not display simple variations of $\theta_{\rm view}$ or $A_{V}$
for each of these different physical models. The bottom line of each source shows the average results of the valid models. We take the geometric mean for all quantities of the accepted models except $A_{V}$ , $\theta_{\rm view}$, and $\theta_{w,\rm esc}$, where the arithmetic mean is used as in Moser et al. (2020).

By fitting the SEDs with the models, we assume there is one source
dominating the infrared luminosity in an aperture. Overall the fitting
is reasonable except that the SEDs of B1 and C2 are not clearly
characterized due to their bright infrared background. The peaks of
the model SEDs seem to locate at a shorter wavelength than the
observed SED, which may result in a more evolved stage. In general,
considering these six sources and all their 10 best fit models,
we find protostellar masses $m_{*} \sim 0.5-8\: M_{\odot}$ accreting
at rates of $\sim 10^{-5} - 5 \times 10^{-4} \: M_{\odot} \: \rm
yr^{-1}$ inside cores of initial masses $M_{c} \sim 10 - 500 \:
M_{\odot}$ embedded in clumps with mass surface densities $\Sigma_{\rm
  cl} \sim 0.1-3 \: \rm g \: cm^{-2}$ (the full range of $M_{c}$ and
$\Sigma_{\rm cl}$ covered by the model grid, though individual
  sources are more constrained). The disk accretion rates are close
to the SiO outflow mass loss rates. The isotropic bolometric
luminosity $L_{\rm bol,iso}$ of C9, the most luminous source, is no
larger than $10^{4} \: L_{\odot}$. For the other sources $L_{\rm
  bol,iso}\ \sim 10^{2} - 10^{3} \: L_{\odot}$. The half opening angle
returned by the best ten models is comparable to the measured half
opening angle of the SiO outflow from the ALMA observations for
B1. For the other sources generally the half opening angles returned
by the models are smaller than what may be inferred from SiO
morphologies, if these are to be explained with a single protostellar
source. The implications of these SED fitting results are discussed in more detail in \S\ref{sec:characteristics}.

\begin{deluxetable*}{ccccccccccccc}
\tabletypesize{\scriptsize}
\tablecaption{Parameters of the Ten Best Fitted Models \label{tab:SEDfit}} 
\tablewidth{16pt}
\tablehead{
\colhead{Source} &\colhead{$\chi^{2}$/N} & \colhead{$M_{\rm c}$} & \colhead{$\Sigma_{\rm cl}$} & \colhead{$R_{core}$}  &\colhead{$m_{*}$} & \colhead{$\theta_{\rm view}$} &\colhead{$A_{V}$} & \colhead{$M_{\rm env}$} &\colhead{$\theta_{w,\rm esc}$} & \colhead{$\dot {M}_{\rm disk}$} & \colhead{$L_{\rm bol, iso}$}  & \colhead{$L_{\rm bol}$} \\
\colhead{} & \colhead{} & \colhead{($M_\odot$)} & \colhead{(g $\rm cm^{-2}$)} & \colhead{(pc) ($\arcsec$)} & \colhead{($M_{\odot}$)} & \colhead{(deg)} & \colhead{(mag)} & \colhead{($M_{\odot}$)} & \colhead{(deg)} &\colhead{($M_{\odot}$/yr)} & \colhead{($L_{\odot}$)} & \colhead{($L_{\odot}$)}  \\
\vspace{-0.4cm}
}
\startdata
B1
& 1.07 & 30 & 0.1 & 0.13 ( 11 ) & 0.5 & 13 & 657.7 & 29 & 10 & 1.1$\times 10^{-5}$ & 2.9$\times 10^{2}$ & 9.0$\times 10^{1}$ \\
$d$ = 2.4 kpc
& 1.09 & 20 & 0.1 & 0.10 ( 9 ) & 1.0 & 22 & 711.7 & 17 & 20 & 1.3$\times 10^{-5}$ & 2.7$\times 10^{2}$ & 1.5$\times 10^{2}$ \\
$R_{ap}$ = 11 \arcsec
& 1.10 & 40 & 0.1 & 0.15 ( 13 ) & 0.5 & 13 & 717.7 & 39 & 8 & 1.1$\times 10^{-5}$ & 2.2$\times 10^{2}$ & 8.8$\times 10^{1}$ \\
= 0.13 pc
& 1.20 & 50 & 0.1 & 0.16 ( 14 ) & 0.5 & 13 & 755.8 & 49 & 7 & 1.2$\times 10^{-5}$ & 1.9$\times 10^{2}$ & 8.7$\times 10^{1}$ \\
& 1.31 & 20 & 0.1 & 0.10 ( 9 ) & 4.0 & 51 & 999.0 & 10 & 43 & 2.1$\times 10^{-5}$ & 2.9$\times 10^{2}$ & 6.8$\times 10^{2}$ \\
& 1.32 & 10 & 0.3 & 0.04 ( 4 ) & 1.0 & 29 & 922.9 & 8 & 28 & 2.5$\times 10^{-5}$ & 5.3$\times 10^{2}$ & 2.6$\times 10^{2}$ \\
& 1.34 & 60 & 0.1 & 0.18 ( 15 ) & 0.5 & 13 & 798.8 & 59 & 6 & 1.3$\times 10^{-5}$ & 1.7$\times 10^{2}$ & 8.7$\times 10^{1}$ \\
& 1.35 & 20 & 0.1 & 0.10 ( 9 ) & 0.5 & 22 & 468.5 & 19 & 13 & 9.6$\times 10^{-6}$ & 8.6$\times 10^{1}$ & 9.0$\times 10^{1}$ \\
& 1.37 & 20 & 0.1 & 0.10 ( 9 ) & 2.0 & 13 & 617.6 & 15 & 30 & 1.7$\times 10^{-5}$ & 8.3$\times 10^{2}$ & 1.9$\times 10^{2}$ \\
& 1.40 & 30 & 0.1 & 0.13 ( 11 ) & 2.0 & 89 & 1000.0 & 25 & 23 & 2.0$\times 10^{-5}$ & 1.4$\times 10^{2}$ & 2.4$\times 10^{2}$ \\
Averages & 1.25 & 27 & 0.1 & 0.11 ( 10 ) & 0.9 & 28 & 765.0 & 22 & 19 & 1.5$\times 10^{-5}$ & 2.5$\times 10^{2}$ & 1.5$\times 10^{2}$ \\
\hline\noalign{\smallskip}
B2
& 6.26 & 200 & 0.1 & 0.33 ( 28 ) & 2.0 & 13 & 233.2 & 194 & 7 & 3.5$\times 10^{-5}$ & 7.3$\times 10^{2}$ & 3.5$\times 10^{2}$ \\
$d$ = 2.4 kpc
& 6.64 & 240 & 0.1 & 0.36 ( 31 ) & 1.0 & 13 & 126.1 & 240 & 4 & 2.6$\times 10^{-5}$ & 3.2$\times 10^{2}$ & 2.4$\times 10^{2}$ \\
$R_{ap}$ = 15 \arcsec
& 6.65 & 100 & 0.1 & 0.23 ( 20 ) & 2.0 & 13 & 279.3 & 97 & 11 & 2.9$\times 10^{-5}$ & 1.2$\times 10^{3}$ & 3.8$\times 10^{2}$ \\
= 0.17 pc
& 6.72 & 120 & 0.1 & 0.25 ( 22 ) & 2.0 & 13 & 287.3 & 117 & 9 & 3.0$\times 10^{-5}$ & 1.2$\times 10^{3}$ & 4.3$\times 10^{2}$ \\
& 6.80 & 320 & 0.1 & 0.42 ( 36 ) & 1.0 & 13 & 81.1 & 315 & 3 & 2.8$\times 10^{-5}$ & 2.5$\times 10^{2}$ & 2.0$\times 10^{2}$ \\
& 7.04 & 160 & 0.1 & 0.29 ( 25 ) & 2.0 & 13 & 273.3 & 156 & 8 & 3.3$\times 10^{-5}$ & 1.0$\times 10^{3}$ & 4.3$\times 10^{2}$ \\
& 7.68 & 80 & 0.1 & 0.21 ( 18 ) & 2.0 & 13 & 277.3 & 75 & 12 & 2.7$\times 10^{-5}$ & 1.3$\times 10^{3}$ & 3.5$\times 10^{2}$ \\
& 7.92 & 320 & 0.3 & 0.23 ( 20 ) & 96.0 & 89 & 378.4 & 7 & 85 & 6.6$\times 10^{-5}$ & 2.2$\times 10^{3}$ & 1.2$\times 10^{6}$ \\
& 7.99 & 200 & 0.1 & 0.33 ( 28 ) & 1.0 & 13 & 101.1 & 197 & 4 & 2.5$\times 10^{-5}$ & 2.7$\times 10^{2}$ & 1.8$\times 10^{2}$ \\
& 7.99 & 160 & 0.1 & 0.29 ( 25 ) & 1.0 & 13 & 128.1 & 156 & 5 & 2.3$\times 10^{-5}$ & 3.3$\times 10^{2}$ & 2.0$\times 10^{2}$ \\
Averages & 7.14 & 173 & 0.1 & 0.29 ( 25 ) & 2.2 & 20 & 216.5 & 114 & 15 & 3.1$\times 10^{-5}$ & 6.9$\times 10^{2}$ & 6.7$\times 10^{2}$ \\
\hline\noalign{\smallskip}
C2
& 5.69 & 480 & 0.1 & 0.51 ( 21 ) & 4.0 & 13 & 253.3 & 474 & 6 & 6.1$\times 10^{-5}$ & 1.8$\times 10^{3}$ & 1.0$\times 10^{3}$ \\
$d$ = 5.0 kpc
& 6.13 & 400 & 0.1 & 0.47 ( 19 ) & 4.0 & 13 & 266.3 & 390 & 7 & 5.8$\times 10^{-5}$ & 2.0$\times 10^{3}$ & 1.0$\times 10^{3}$ \\
$R_{ap}$ = 16 \arcsec
& 6.50 & 320 & 0.3 & 0.23 ( 10 ) & 96.0 & 62 & 520.5 & 7 & 85 & 6.6$\times 10^{-5}$ & 1.3$\times 10^{6}$ & 1.2$\times 10^{6}$ \\
= 0.39 pc
& 8.54 & 480 & 0.1 & 0.51 ( 21 ) & 2.0 & 13 & 133.1 & 477 & 4 & 4.3$\times 10^{-5}$ & 7.4$\times 10^{2}$ & 5.7$\times 10^{2}$ \\
& 8.66 & 240 & 0.1 & 0.36 ( 15 ) & 4.0 & 13 & 289.3 & 229 & 9 & 5.1$\times 10^{-5}$ & 2.7$\times 10^{3}$ & 1.0$\times 10^{3}$ \\
& 9.19 & 320 & 0.3 & 0.23 ( 10 ) & 2.0 & 13 & 169.2 & 314 & 4 & 9.2$\times 10^{-5}$ & 1.4$\times 10^{3}$ & 1.1$\times 10^{3}$ \\
& 9.35 & 200 & 0.3 & 0.19 ( 8 ) & 4.0 & 13 & 328.3 & 193 & 9 & 1.1$\times 10^{-4}$ & 4.3$\times 10^{3}$ & 1.4$\times 10^{3}$ \\
& 9.53 & 200 & 0.3 & 0.19 ( 8 ) & 2.0 & 13 & 232.2 & 196 & 6 & 8.2$\times 10^{-5}$ & 2.0$\times 10^{3}$ & 1.2$\times 10^{3}$ \\
& 9.53 & 240 & 0.3 & 0.20 ( 8 ) & 2.0 & 13 & 226.2 & 235 & 5 & 8.6$\times 10^{-5}$ & 1.8$\times 10^{3}$ & 1.2$\times 10^{3}$ \\
& 9.55 & 160 & 0.3 & 0.17 ( 7 ) & 4.0 & 13 & 314.3 & 153 & 11 & 1.1$\times 10^{-4}$ & 4.3$\times 10^{3}$ & 1.2$\times 10^{3}$ \\
Averages & 8.12 & 285 & 0.2 & 0.28 ( 11 ) & 4.2 & 18 & 273.3 & 188 & 15 & 7.3$\times 10^{-5}$ & 3.9$\times 10^{3}$ & 2.1$\times 10^{3}$ \\
\hline\noalign{\smallskip}
C6
& 0.88 & 480 & 0.1 & 0.51 ( 21 ) & 2.0 & 13 & 642.6 & 477 & 4 & 4.3$\times 10^{-5}$ & 7.4$\times 10^{2}$ & 5.7$\times 10^{2}$ \\
$d$ = 5.0 kpc
& 0.96 & 240 & 0.1 & 0.36 ( 15 ) & 4.0 & 13 & 935.9 & 229 & 9 & 5.1$\times 10^{-5}$ & 2.7$\times 10^{3}$ & 1.0$\times 10^{3}$ \\
$R_{ap}$ = 15 \arcsec
& 0.97 & 400 & 0.1 & 0.47 ( 19 ) & 2.0 & 13 & 591.6 & 391 & 4 & 4.1$\times 10^{-5}$ & 7.7$\times 10^{2}$ & 5.5$\times 10^{2}$ \\
= 0.36 pc
& 1.10 & 400 & 0.1 & 0.47 ( 19 ) & 4.0 & 89 & 961.0 & 390 & 7 & 5.8$\times 10^{-5}$ & 9.4$\times 10^{2}$ & 1.0$\times 10^{3}$ \\
& 1.13 & 320 & 0.1 & 0.42 ( 17 ) & 4.0 & 13 & 536.5 & 308 & 7 & 5.5$\times 10^{-5}$ & 1.2$\times 10^{3}$ & 5.4$\times 10^{2}$ \\
& 1.20 & 480 & 0.1 & 0.51 ( 21 ) & 4.0 & 89 & 974.0 & 474 & 6 & 6.1$\times 10^{-5}$ & 9.2$\times 10^{2}$ & 1.0$\times 10^{3}$ \\
& 1.25 & 160 & 0.1 & 0.29 ( 12 ) & 4.0 & 13 & 757.8 & 151 & 12 & 4.5$\times 10^{-5}$ & 3.1$\times 10^{3}$ & 9.1$\times 10^{2}$ \\
& 1.27 & 200 & 0.3 & 0.19 ( 8 ) & 1.0 & 13 & 725.7 & 196 & 4 & 5.8$\times 10^{-5}$ & 6.5$\times 10^{2}$ & 5.2$\times 10^{2}$ \\
& 1.34 & 200 & 0.1 & 0.33 ( 14 ) & 4.0 & 13 & 605.6 & 194 & 10 & 4.8$\times 10^{-5}$ & 1.9$\times 10^{3}$ & 6.7$\times 10^{2}$ \\
& 1.42 & 320 & 0.3 & 0.23 ( 10 ) & 96.0 & 68 & 1000.0 & 7 & 85 & 6.6$\times 10^{-5}$ & 1.1$\times 10^{6}$ & 1.2$\times 10^{6}$ \\
Averages & 1.14 & 299 & 0.1 & 0.36 ( 15 ) & 4.2 & 33 & 773.1 & 198 & 15 & 5.2$\times 10^{-5}$ & 2.4$\times 10^{3}$ & 1.5$\times 10^{3}$ \\
\hline\noalign{\smallskip}
C9
& 1.14 & 160 & 3.2 & 0.05 ( 2 ) & 2.0 & 13 & 214.2 & 158 & 5 & 4.4$\times 10^{-4}$ & 8.5$\times 10^{3}$ & 5.2$\times 10^{3}$ \\
$d$ = 5.0 kpc
& 1.20 & 120 & 3.2 & 0.05 ( 2 ) & 4.0 & 13 & 646.6 & 113 & 11 & 5.7$\times 10^{-4}$ & 4.0$\times 10^{4}$ & 1.1$\times 10^{4}$ \\
$R_{ap}$ = 10 \arcsec
& 1.25 & 120 & 3.2 & 0.05 ( 2 ) & 8.0 & 13 & 951.0 & 106 & 15 & 7.9$\times 10^{-4}$ & 1.2$\times 10^{5}$ & 1.9$\times 10^{4}$ \\
= 0.24 pc
& 1.33 & 240 & 1.0 & 0.11 ( 5 ) & 8.0 & 13 & 854.9 & 223 & 11 & 4.0$\times 10^{-4}$ & 4.4$\times 10^{4}$ & 1.4$\times 10^{4}$ \\
& 1.35 & 400 & 1.0 & 0.15 ( 6 ) & 4.0 & 13 & 158.2 & 394 & 5 & 3.3$\times 10^{-4}$ & 6.7$\times 10^{3}$ & 4.7$\times 10^{3}$ \\
& 1.42 & 240 & 0.3 & 0.20 ( 8 ) & 12.0 & 86 & 1000.0 & 216 & 15 & 2.0$\times 10^{-4}$ & 3.1$\times 10^{4}$ & 4.1$\times 10^{4}$ \\
& 1.43 & 480 & 1.0 & 0.16 ( 7 ) & 4.0 & 13 & 233.2 & 478 & 5 & 3.4$\times 10^{-4}$ & 6.6$\times 10^{3}$ & 5.0$\times 10^{3}$ \\
& 1.43 & 400 & 0.1 & 0.47 ( 19 ) & 16.0 & 89 & 767.8 & 364 & 17 & 1.1$\times 10^{-4}$ & 2.7$\times 10^{4}$ & 3.8$\times 10^{4}$ \\
& 1.43 & 320 & 0.1 & 0.42 ( 17 ) & 24.0 & 89 & 927.9 & 256 & 27 & 1.2$\times 10^{-4}$ & 4.4$\times 10^{4}$ & 8.4$\times 10^{4}$ \\
& 1.44 & 100 & 3.2 & 0.04 ( 2 ) & 12.0 & 77 & 1000.0 & 77 & 20 & 9.4$\times 10^{-4}$ & 2.5$\times 10^{4}$ & 5.2$\times 10^{4}$ \\
Averages & 1.34 & 225 & 0.9 & 0.12 ( 5 ) & 7.3 & 42 & 675.4 & 204 & 13 & 3.4$\times 10^{-4}$ & 2.4$\times 10^{4}$ & 1.7$\times 10^{4}$ \\
\hline\noalign{\smallskip}
H6
& 0.56 & 80 & 0.1 & 0.21 ( 15 ) & 0.5 & 89 & 1000.0 & 79 & 5 & 1.4$\times 10^{-5}$ & 8.6$\times 10^{1}$ & 9.2$\times 10^{1}$ \\
$d$ = 2.9 kpc
& 0.57 & 60 & 0.1 & 0.18 ( 13 ) & 0.5 & 34 & 888.9 & 59 & 6 & 1.3$\times 10^{-5}$ & 8.6$\times 10^{1}$ & 8.7$\times 10^{1}$ \\
$R_{ap}$ = 12 \arcsec
& 0.61 & 50 & 0.1 & 0.16 ( 12 ) & 0.5 & 22 & 841.8 & 49 & 7 & 1.2$\times 10^{-5}$ & 8.7$\times 10^{1}$ & 8.7$\times 10^{1}$ \\
= 0.17 pc
& 0.65 & 100 & 0.1 & 0.23 ( 17 ) & 0.5 & 89 & 1000.0 & 99 & 4 & 1.5$\times 10^{-5}$ & 8.7$\times 10^{1}$ & 9.1$\times 10^{1}$ \\
& 0.68 & 30 & 0.1 & 0.13 ( 9 ) & 1.0 & 86 & 1000.0 & 27 & 15 & 1.5$\times 10^{-5}$ & 1.3$\times 10^{2}$ & 1.7$\times 10^{2}$ \\
& 0.71 & 120 & 0.1 & 0.25 ( 18 ) & 0.5 & 86 & 1000.0 & 118 & 4 & 1.5$\times 10^{-5}$ & 8.4$\times 10^{1}$ & 8.8$\times 10^{1}$ \\
& 0.72 & 40 & 0.1 & 0.15 ( 10 ) & 0.5 & 13 & 806.8 & 39 & 8 & 1.1$\times 10^{-5}$ & 2.2$\times 10^{2}$ & 8.8$\times 10^{1}$ \\
& 0.73 & 30 & 0.1 & 0.13 ( 9 ) & 2.0 & 89 & 1000.0 & 25 & 23 & 2.0$\times 10^{-5}$ & 1.4$\times 10^{2}$ & 2.4$\times 10^{2}$ \\
& 0.98 & 30 & 0.1 & 0.13 ( 9 ) & 0.5 & 22 & 703.7 & 29 & 10 & 1.1$\times 10^{-5}$ & 8.8$\times 10^{1}$ & 9.0$\times 10^{1}$ \\
& 1.01 & 40 & 0.1 & 0.15 ( 10 ) & 1.0 & 89 & 1000.0 & 38 & 12 & 1.6$\times 10^{-5}$ & 1.3$\times 10^{2}$ & 1.7$\times 10^{2}$ \\
Averages & 0.71 & 51 & 0.1 & 0.17 ( 12 ) & 0.7 & 62 & 924.1 & 49 & 10 & 1.4$\times 10^{-5}$ & 1.1$\times 10^{2}$ & 1.1$\times 10^{2}$ \\
\enddata
\tablenotetext{a}{Average results of the valid models. See text in \S~\ref{sec:SED} for more detail. }
\end{deluxetable*}

\subsection{Weak SiO Sources}

As shown in Figure~\ref{fig:trunk}, while in some clumps there is a
good indication of the location of the driving source of the SiO
emission, e.g., A1, D9, and E2, where a continuum source is located in
between two patches of SiO emission, in other clumps the situation is
more uncertain. The shape of the SiO emission also varies -- in A1, A2
and C4 the main SiO emission is relatively elongated, in D8 and H5
it appears more rounded, and in other cases SiO is hardly
resolved. We investigate the kinematics of the SiO(5-4) emission via
the channel maps of each source and try to locate the driving source
assuming the SiO emission comes from an outflow.

For the sources that show relatively strong SiO emission (see
Figure~\ref{fig:spec30}), like A1, A2, C4, D8, H5, from their channel
maps the velocity range of the SiO emission above 3$\sigma$ noise
level exceeds 6~km~$\rm s^{-1}$. Duarte-Cabral et al. (2014) suggested
that narrow line SiO emission ($\sigma_{v} < $ 1.5 km s$^{-1}$, i.e.,
line width $<$ 3.5 km s$^{-1}$) appears unrelated to outflows, but
rather traces large scale collapse of material onto massive dense
cores (see also Cossentino et al. 2018, 2020). Thus, in this context,
in these five sources it is more likely that SiO traces shocks from
outflows. In A1 there are two components revealed in
Figure~\ref{fig:trunk}, while the northern component appears to be
red-shifted and the southern component appears to be blue-shifted (see
also \S\ref{sec:vla}). In A2 the large extended emission in the center
of the field (see Figure~\ref{fig:trunk}) shows a hint of bipolar
structure with the blue-shifted emission to the east of the field
center and the red-shifted emission to the west. In C4, D8 and H5, no
clear bipolar structure is seen.

In other sources the signal to noise ratio of SiO emission is low and
most of the time the emission above the 3$\sigma$ noise level in the
channel maps appears as an unresolved peak and the velocity range does
not exceed 4~km~$\rm s^{-1}$. In D9 the northern component appears
blue-shifted and the southern component red-shifted. In E2 there is a
hint that the eastern component is blue-shifted relative to the source
velocity and the western component red-shifted, though the emission is
not stronger than the 2$\sigma$ noise level.  There is hardly any
bipolar outflow structure revealed in the other sources. Thus for
these sources we are not sure whether the SiO emission comes from
protostellar outflows.

Overall, in A1, A2, D9, and E2 there is a hint of bipolar structures
revealed in the channel maps. For A1, D9, and E2, the continuum source
located in the middle of the two patches of SiO emission is likely
responsible for driving the elongated SiO outflows. In sources like
A2, the driving source of SiO is ambiguous. The emission of the dense
gas tracers at the continuum peaks and the SiO peaks are all very
weak. D8 is another example of an ambiguous SiO driving source. The
velocity range of the central SiO emission is as wide as 30 km $\rm
s^{-1}$, thus the SiO is not likely from large-scale cloud collision
(Duarte-Cabral et al. 2014). The integrated intensity of the central
SiO is even higher than H6. In D1, D3 and D8 the 1.3~mm continuum
cores are not associated with any detectable dense gas tracers.

The SED fitting results for these sources are presented in
Appendix~\ref{sec:more SED}. Most of the SEDs are not well defined,
likely because most sources are very young and even appear dark
against the background at 70\,$\mu$m. The stellar masses of the valid
models range from 0.5 $M_{\odot}$ to 4 $M_{\odot}$. The isotropic
luminosities ranges from $10^{2} \: L_{\odot}$ to $10^{3} \:
L_{\odot}$.

\section{6\,cm Radio Emission}\label{sec:vla}

\begin{figure*}[htbp]
%\figurenum{}
\gridline{\fig{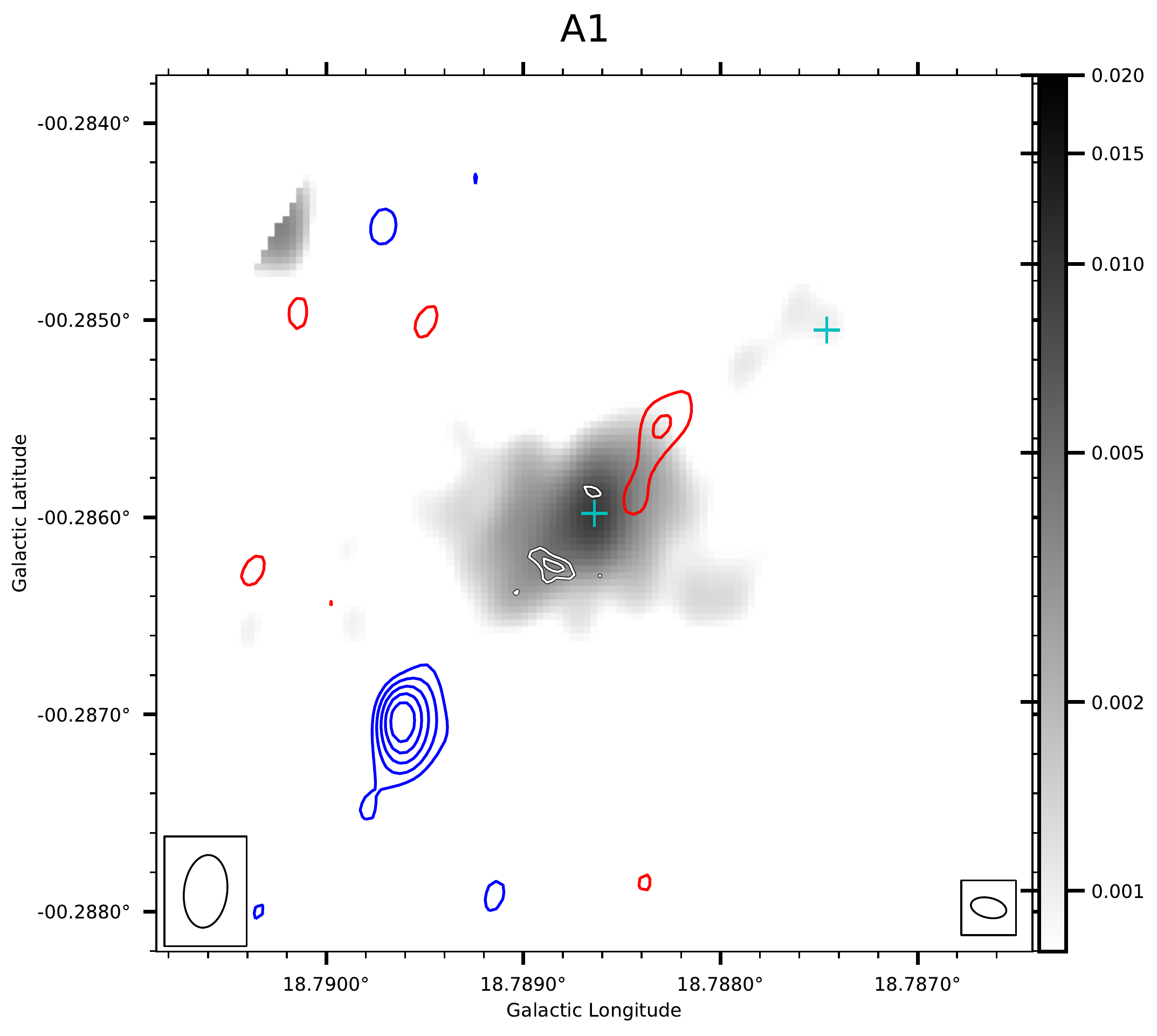}{0.5\textwidth}{White contours: starting level: 3$\sigma$, step: 1$\sigma$, $\sigma$ value: 3.3\,$\mu$Jy beam$^{-1}$. \\Blue contours: starting level: 4$\sigma$, step: 2$\sigma$, $\sigma$ value: 42\,mJy beam$^{-1}$ km s$^{-1}$, velocity range: 45-66.5 km s$^{-1}$. \\Red contours: starting level: 4$\sigma$, step: 2$\sigma$, $\sigma$ value: 39\,mJy beam$^{-1}$ km s$^{-1}$, velocity range: 66.5-85 km s$^{-1}$. }
          \fig{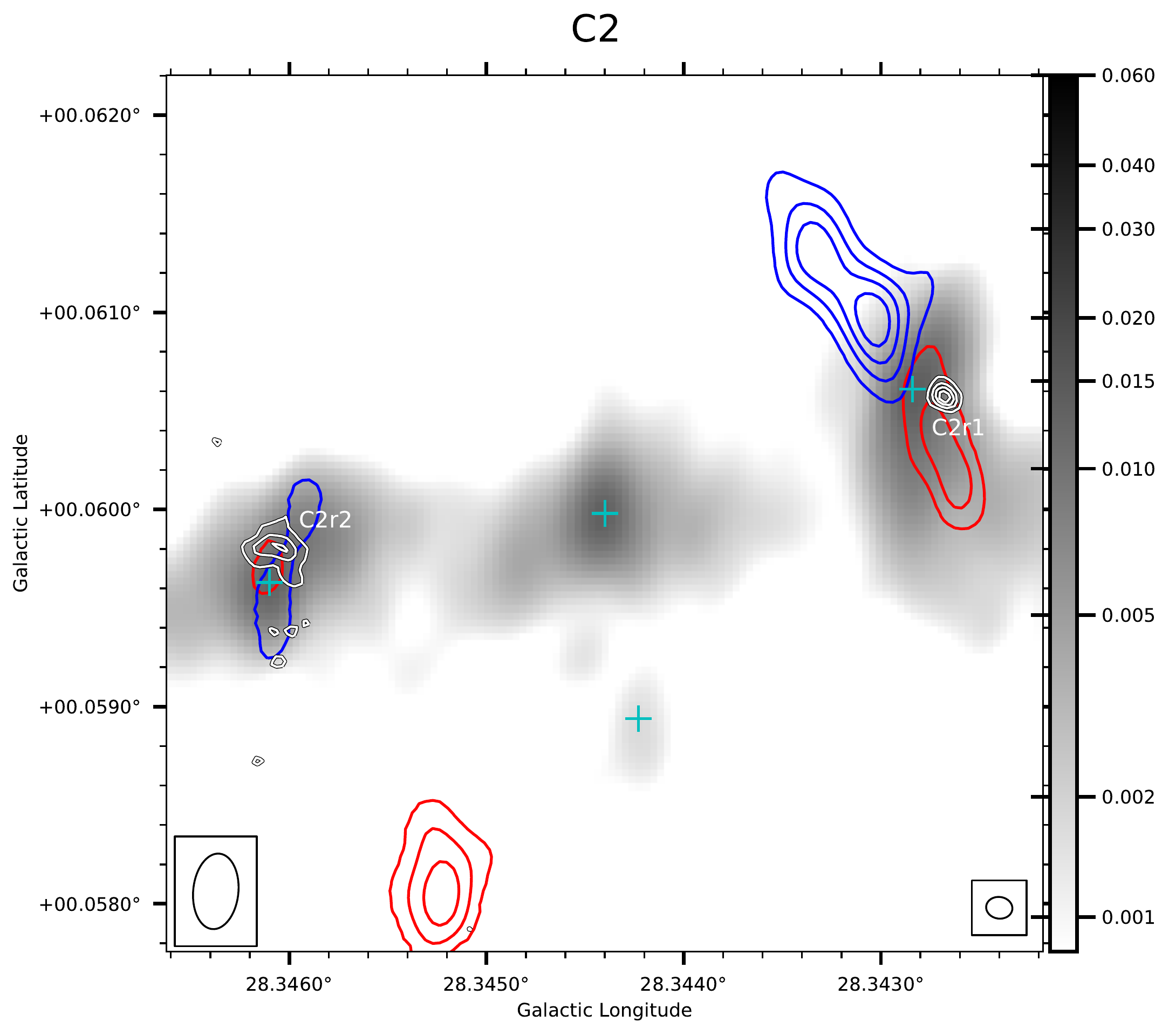}{0.5\textwidth}{White contours: starting level: 3$\sigma$, step: 2$\sigma$, $\sigma$ value: 3.0\,$\mu$Jy beam$^{-1}$. \\Blue contours: starting level: 5$\sigma$, step: 4$\sigma$, $\sigma$ value: 34\,mJy beam$^{-1}$ km s$^{-1}$, velocity range: 65-79.2 km s$^{-1}$. \\Red contours: starting level: 5$\sigma$, step: 4$\sigma$, $\sigma$ value: 36\,mJy beam$^{-1}$ km s$^{-1}$, velocity range: 79.2-95 km s$^{-1}$. }
          }
\gridline{\fig{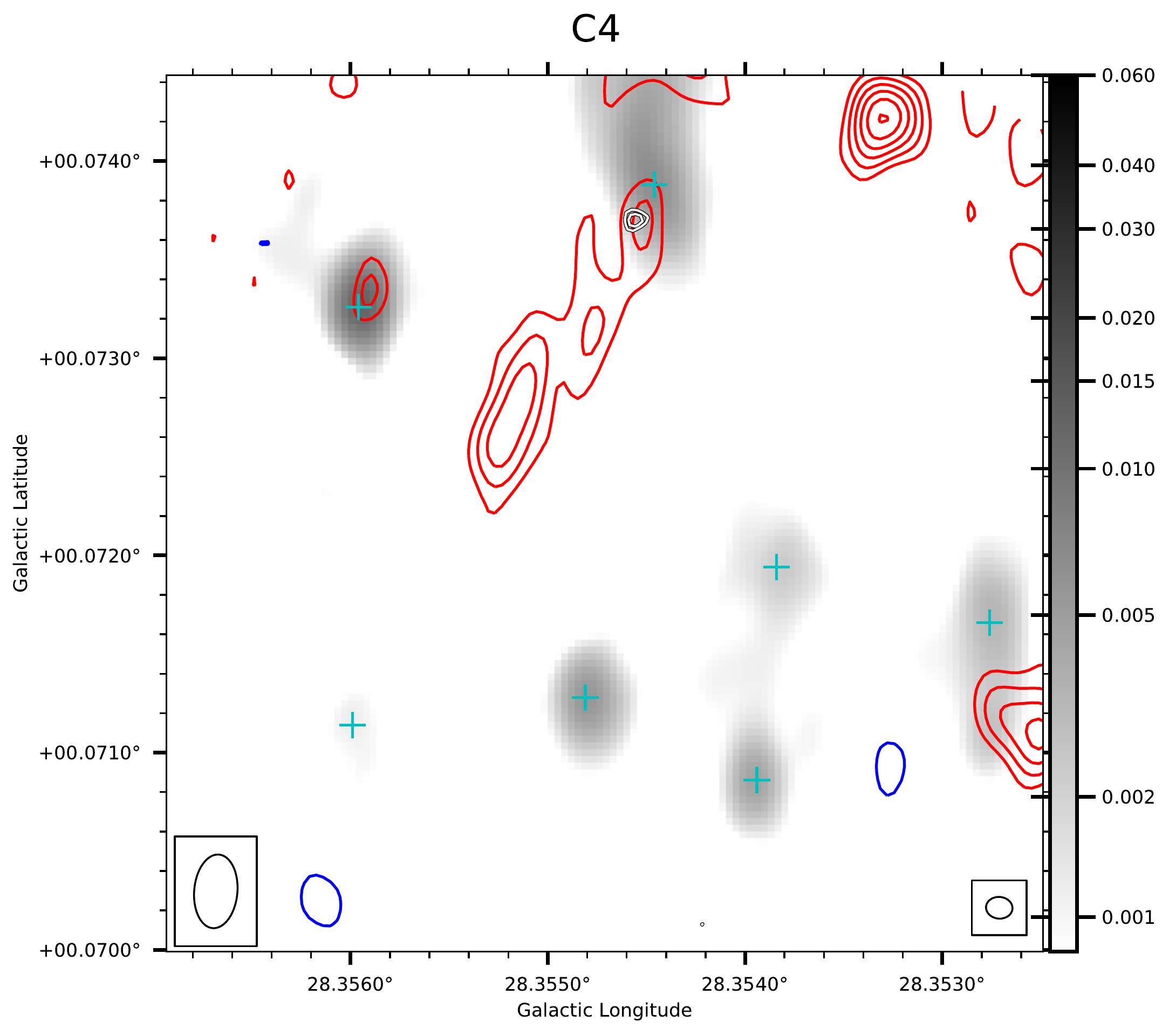}{0.5\textwidth}{White contours: starting level: 3$\sigma$, step: 1$\sigma$, $\sigma$ value: 2.7\,$\mu$Jy beam$^{-1}$. \\Blue contours: starting level: 5$\sigma$, step: 3$\sigma$, $\sigma$ value: 39\,mJy beam$^{-1}$ km s$^{-1}$, velocity range: 60-79 km s$^{-1}$. \\Red contours: starting level: 5$\sigma$, step: 3$\sigma$, $\sigma$ value: 41\,mJy beam$^{-1}$ km s$^{-1}$, velocity range: 79-100 km s$^{-1}$. }
          \fig{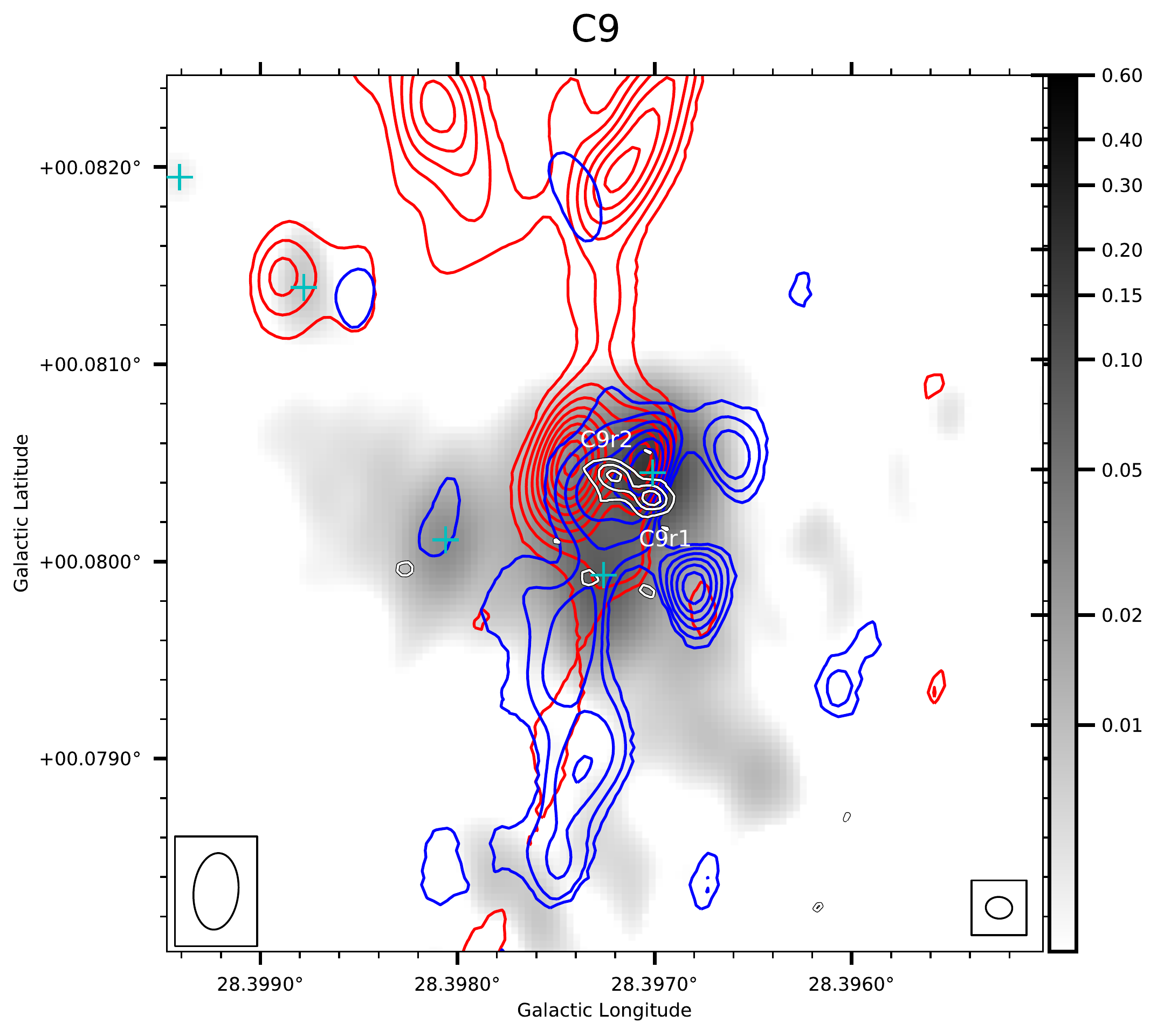}{0.5\textwidth}{White contours: starting level: 3$\sigma$, step: 2$\sigma$, $\sigma$ value: 4.6\,$\mu$Jy beam$^{-1}$. \\Blue contours: starting level: 5$\sigma$, step: 5$\sigma$, $\sigma$ value: 38\,mJy beam$^{-1}$ km s$^{-1}$, velocity range: 60-78.5 km s$^{-1}$. \\Red contours: starting level: 5$\sigma$, step: 5$\sigma$, $\sigma$ value: 42\,mJy beam$^{-1}$ km s$^{-1}$, velocity range: 78.5-100 km s$^{-1}$. }
          }
\caption{6\,cm radio emission and SiO outflow emission over 1.3\,mm continuum. White contours are the 6\,cm radio continuum. Blue and red contours are the blue- and red-shifted SiO integrated intensity. Grayscale is the 1.3mm continuum. Plus signs denote the 1.3mm continuum peaks. The beam size of the ALMA 1.3\,mm continuum observations is shown in the lower left corner. The beam size of the VLA 6\,cm continuum observations is shown in the lower right corner. The images have a field of view of 16\arcsec $\times$ 16\arcsec.} \label{fig:radio}
\end{figure*}

We searched for 6\,cm radio emission in the ALMA field of view ($\sim$
27\arcsec in diameter) of each clump. Out of the 15 clumps that exhibit strong SiO outflows observed with VLA, we only detected
radio emission above the 3\,$\sigma$ noise level in four sources A1,
C2, C4, C9. The peak positions and flux measurements of the 6\,cm
radio continuum are reported in Table~\ref{tab:radioflux}. The radio
continuum emission is shown in Figure~\ref{fig:radio}. The emission in
A1, C2 and C4 is hardly resolved, while the emission in C9 is resolved
into two peaks. The negative artifacts are not significant with no
stronger than -2 $\sigma$ level in C2 and C9 and no stronger than -3
$\sigma$ level in A1 and C4. We note that the low detection rate of
6\,cm radio emission may be partly due to limited sensitivities. The
sensitivities of the 6\,cm images vary by a factor of almost 3, 
  and the detections in A1, C2, and C4 come from images that achieves
  a higher sensitivity than others in the same cloud, while in the B
and D cloud, where no 6\,cm emission is detected, the sensitivity is
much worse.

In Sanna et al. (2018), the detection rate of 22 GHz continuum
emission towards 25 H$_2$O maser sites is 100\% and they suggested
H$_2$O masers are preferred signposts of bright radio thermal jets
($\gg$ 1mJy). Here we see coincidence of 6 GHz radio continuum
emission and H$_2$O masers at C2c4 and C4c1.

We attempt to derive the in-band spectra index, $\alpha$, of the two
sources in C9, i.e., assuming $F_{\nu} \propto \nu ^{\alpha}$, by
dividing the continuum data into two centered at 5.03\,GHz and
6.98\,GHz respectively. In A1, C2 and C4 the sources detected in the
combined continuum data do not have enough signal to noise for such an
estimate. We derive an $\alpha$ of -0.52 for C9r1 and -2.36 for
C9r2. However, as discussed in Rosero et al. (2016), the in-band
spectral index derived from only two data points can be highly
uncertain and more measurements at other wavelengths are required for
confirmation of these results.

There are offsets between the radio continuum peak and the 1.3mm
continuum peak in all the sources, typically about 500 mas, i.e.,
about 1 VLA synthesized beam width and corresponding to $\sim$ 2500
AU. Such offsets likely indicate that the radio emission comes from
jet lobes and/or that the offset is due to a gradient in the optical
depth. Another possibility is that the offset is due to astrometric
uncertainties in the VLA data, but this would require an uncertainty
several times larger than we have estimated ($\lesssim$ 170 mas). This
error takes into account the accuracy of the phase calibrator and VLA
antenna positions, the transfer of solutions from the phase calibrator
to the target, and the statistical error in measuring the source's
peak position. The unresolved emission in A1, C2 and C4 basically
follows the direction and shape of the VLA beam and it is hard to
compare with the direction of the outflows. The extension of the
emission in C9 is not along the direction of the large scale
north-south outflow. However, it could be related to the small scale
outflows in the region.

\begin{table}[htbp]
\centering
\setlength{\tabcolsep}{2pt}
\begin{center}
 \caption{Parameters of 6 cm Radio Continuum} \label{tab:radioflux}
\small
\begin{tabular}{cccccccccccc}
  \hline\noalign{\smallskip}
  \hline\noalign{\smallskip}
  Source & R.A. & Decl. & $I_{\rm peak}$ & $S_{\rm 6GHz}$  \\
   & (J2000) & (J2000) &  ($\mu$Jy beam$^{-1}$) & ($\mu$Jy)  \\
  \hline\noalign{\smallskip}
A1 & 18:26:15.442 & -12:41:37.505 & 14.44 & 24.14   \\
C2r1 & 18:42:50.228 & -4:03:21.022 & 31.37 & 24.80  \\
C2r2 & 18:42:50.762 & -4:03:11.534 & 21.79 & 68.65  \\ 
C4 & 18:42:48.724 & -4:02:21.433 & 13.16 & 5.85 \\
C9r1 & 18:42:51.979 & -3:59:54.534 & 40.05 & 47.70 \\
C9r2 & 18:42:51.979 & -3:59:53.734 & 35.87 & 46.97 \\
\noalign{\smallskip}\hline
\end{tabular}
\end{center}
\end{table}

\section{Discussion}

\subsection{SiO Detection Rate}

We have detected SiO(5-4) emission in 20 out of 32 IRDC clumps, a
detection rate of 62\%. Our sample has a distance range of 2.4-5.7
kpc. The cloud with the lowest detection rate, Cloud F, is located at
a moderate distance of 3.7 kpc. The cloud with the highest detection
rate, cloud D, happens to be located at the largest distance. Thus the
detection rate does not appear to have a strong dependence on
distance.
L\'{o}pez-Sepulcre et al. (2011) derived an SiO detection rate of 88\%
towards 20 high-mass ($M_{\rm clump} \gtrsim \ 100 \: M_{\odot}$)
IR-dark clumps (not detected at 8 $\mu$m with the \textit{Midcourse
  Space eXperient} (MSX)) with SiO(2-1) and SiO(3-2). Csengeri et
al. (2016) derived an SiO detection rate of 61\% towards 217 IR-quiet
clumps (a threshold of 289 Jy at 22 $\mu$m at 1 kpc) with SiO(2-1). In
particular, the SiO detection rate is 94\% for a subsample of clumps
with $M_{\rm clump} > 650 \: M_{\odot}$ and 1 kpc $<d<$ 7 kpc. With a
more evolved sample, Harju et al. (1998) derived an SiO detection rate
of 38\% for protostars above $10^{3} \: L_{\odot}$ with SiO(2-1) and
SiO(3-2). Gibb et al. (2007) detected SiO emission in five out of 12
(42\%) massive protostars with SiO(5-4). Li et al. (2019a) detected
SiO(5-4) emission in 25 out of 44 IRDCs with a detection rate of 57\%,
and 32 out of 86 protostars in massive clumps (a subsample defined to
be at the intermediate evolutionary stage between IRDCs and HII
regions) with a detection rate of 37\%.

Our result is similar to that of the full sample of Csengeri et
al. (2016) and the IRDC sample of Li et al. (2019a). The clumps in our
sample are not as massive as those of L\'{o}pez-Sepulcre et
al. (2011). In Liu et al. (2018), seven of the 1.3~mm
  continuum cores (C2c2, C2c3, C2c4, C9c5, C9c7, C9c8, D9c2) are
  high-mass protostellar candidates, i.e., with masses
  $>25\:M_{\odot}$, assuming a core-to-star efficiency of about
  30\%. Twelve others (A1c2, A3c3, C2c1, C4c1, C4c2, C6c1, C6c5, C6c6,
  C9c6, D5c6, D6c1, D6c4) are intermediate-mass protostellar
  candidates, i.e., with masses in the range 9 to 24~$M_{\odot}$. The
non-detection of SiO tends to be in clumps with weak or non-detected
1.3 mm continuum emission ($M_{\rm core} < 5 \: M_{\odot}$, Liu et
al. 2018), and dark against background up to 100\,$\mu$m (except
  for source H2, see Figure~\ref{fig:more SEDfit} in
  Appendix~\ref{sec:more SED}). This indicates in IRDCs,
representative of the earliest evolutionary phases, shocked gas is
more common in the higher mass regime. The non-detections may reflect
the more diffuse clumps without star formation as suggested in
Csengeri et al. (2016). However, as discussed in \S\ref{sec:sio
  detection}, there is possible large scale SiO emission in those
regions, which is resolved out with our observations. On the other
hand, we do see SiO emission in clumps with only low-mass 1.3\,mm
cores detected and low luminosity. Overall, compared with more evolved
IR bright protostars in Harju et al. (1998) and Gibb et al. (2007),
the detection rate of SiO in our sample, most of which have a
luminosity $< \: 10^{3} \: L_{\odot}$ and still appear dark against
background at 70\,$\mu$m, and other early stage IR-quiet protostars
(L\'{o}pez-Sepulcre et al. 2011, Csengeri et al. 2016, Li et
al. 2019a), is higher.

\begin{figure*}[htbp]
\epsscale{1.2}
\plotone{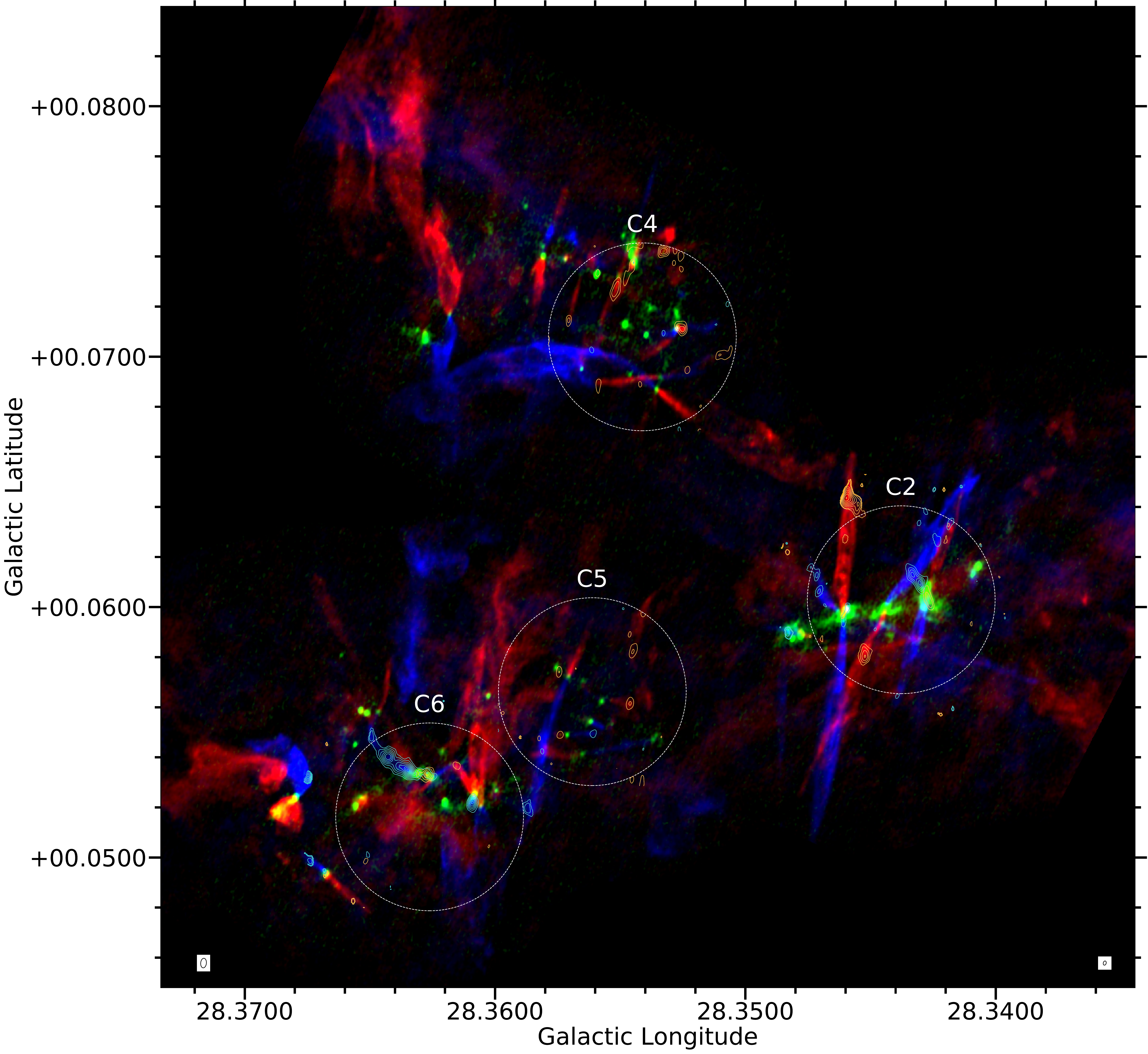}
\caption{Map of molecular outflows in part of Cloud C. The RGB colorscale mosaic is the same with Figure 1 in Kong et al. (2019), where green represents 1.3~mm continuum emission observed with 0.5\arcsec resolution, blue represents blue-shifted CO(2-1) emission, integrated from 11.7 to 75.4 km s$^{-1}$, and red represents red-shifted CO(2-1) emission, integrated from 83.2 to 146.9 km s$^{-1}$. The cyan and orange contours represent blue- and red-shifted SiO outflows observed in this paper. Contour levels start at 5$\sigma$ in steps of 4$\sigma$ noise level of the integrated intensity (see Section~\ref{sec:morph}. The dashed circles denote the primary beams of the observations presented in this paper. The beam size of the SiO outflows is shown in the lower left corner. The beam size of Kong et al. (2019) observations is shown in the lower right corner.}\label{fig:CO}
\end{figure*}

For sources in the C2, C4, C5 and C6 regions of Cloud C we are able to
make a direct comparison of the SiO emission with previously reported
CO(2-1) outflow data from Kong et al. (2019). As shown in
Figure~\ref{fig:CO}, we see that CO outflow emission is generally more
common and more widespread than that from SiO, which is more
``sparse'', i.e., localized to parts of a given CO outflow. The CO
outflows are highly collimated. Because of its sparse appearance, SiO typically appears to have a less collimated morphology
than the CO outflows. One reason why we do not see clear collimated outflows with SiO(5-4) as with CO(2-1) is the high critical density of the SiO(5-4) transition ($\sim 2.5 \times 10^6 \rm \: cm ^{-3}$), implying that only high density shocked regions can be detected with this transition. Such high-density gas may also be confined in regions narrower than the synthesized beam, resulting in a diluted signal that is more difficult to detect than the more extended CO(2-1). Comparing to the results of Kong et al. (2019) and Liu et
al. (2018), we find 3 overlapping identified 1.3~mm continuum sources
(those with $<$ 1\arcsec difference in core coordinates) in C2 (C2c2,
C2c3, C2c4), 4 in C4 (C4c1, C4c2, C4c6, C4c8), 4 in C5 (C5c2, C5c4,
C5c5, C5c6), and 4 in C6 (C6c1, C6c2, C6c5, C6c6). In C2, the main SiO
outflow driven by C2c2 does not align with the main CO outflow, but
there seems to be a secondary weak CO in that direction especially
revealed in the blue-shifted side. Other than that we see several
patches of SiO emission overlapping with CO outflows with the same
direction relative to the plane of sky. In C6 the main blue-shifted
SiO outflow driven by C6c1 aligns very well with the blue-shifted CO
outflow. The red-shifted SiO outflow in the opposite direction is
relatively weak, and we do not see strong red-shifted CO outflow
coming from the west of C6c1 either. The SiO emission associated with
C6c5 aligns well with the bipolar CO outflow. Outside of the primary
beam of C6 we also see several overlaps of the SiO emission with the
CO outflows. In C4 and C5 the SiO emission is relatively weak, but,
still, where we detect SiO emission, most of the time there are also
CO outflows with the same direction relative to the plane of sky. Our
results partly agree with the results in Li et al. (2019b), who detect
SiO(5-4) emission only in 6 out of 17 IRDC protostellar sources with
detected CO(2-1) outflows. They find that the SiO emission is much
fainter and has a narrower velocity range ($10 - 15 \rm km \ s^{-1}$)
than that of the CO emission. While the strong SiO emission in their
sample still overlaps with CO outflows, the weak SiO emission features
are generally not associated with CO (see their Figure 2).

\subsection{Characteristics of the Protostellar Sources} \label{sec:characteristics}

\begin{figure*}[htbp]
\epsscale{1.1}
\plotone{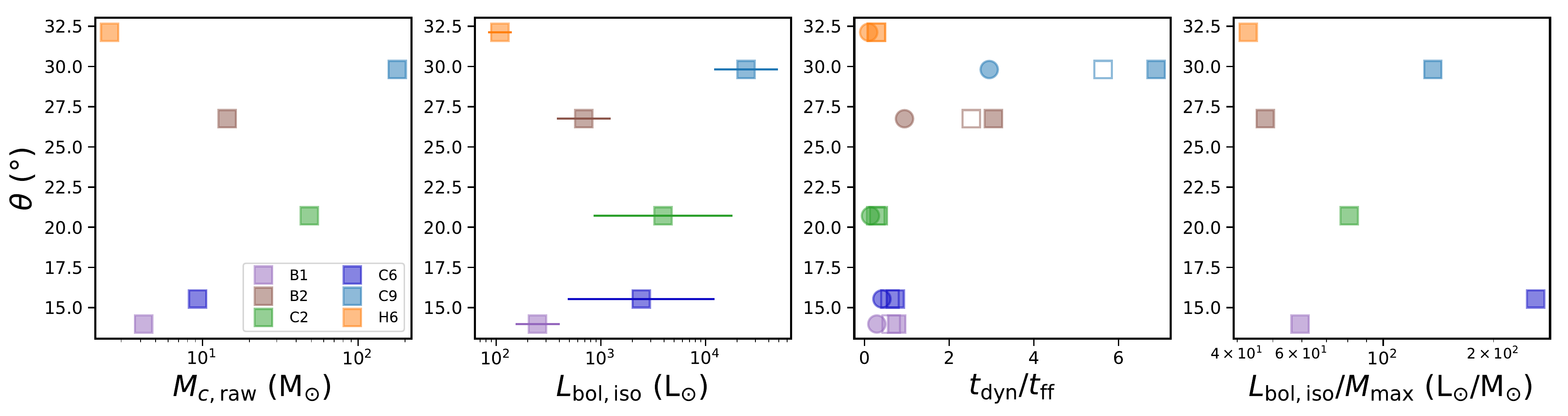}
\caption{Outflow half opening angle $\theta$ versus the mm core mass $M_{c, \rm raw}$, isotropic bolometric luminosity $L_{\rm bol,iso}$, the ratio of outflow dynamical time scale over free-fall time scale $t_{\rm dyn}/t_{\rm ff}$, and luminosity to mass ratio $L_{\rm bol,iso}/M_{\rm max}$. The filled squares represent outflow parameters derived with the fiducial method, i.e., using the full velocity range and $t_{\rm dyn}$ derived with an average outflow velocity. The filled circles represent outflow parameters derived with the full velocity range and $t_{\rm dyn}$ derived with the maximum outflow velocity. The empty squares represent outflow parameters derived excluding the emission within $\pm$ 3 km s$^{-1}$ to the systemic velocity, which we consider coming from ambient gas, and $t_{\rm dyn}$ derived with an average outflow velocity. The error bar represents 95\% geometric confidence interval of $L_{\rm bol,iso}$ among the valid models.}\label{fig:corr6_hoa}
\end{figure*}

We investigate the potential correlations between outflow collimation
and core mass, luminosity, time scale, luminosity-to-mass ratio, etc.,
for the strong SiO sources as shown in
Figure~\ref{fig:corr6_hoa}. The half opening angles are directly
  measured from the SiO outflows associated with individual
  protostellar sources observed by {\it ALMA}. $M_{c, \rm raw}$ and
  $M_{\rm max}$ are derived from the 1.3~mm continuum emission. Note
the outflow half opening angle and the dynamical time adopt the values
of the representative flow of each source, i.e., red flow for B1, B2,
H6 and blue flow for C2, C6 and C9. The free-fall time is derived from
$t_{\rm ff} = \sqrt{\frac{3\pi}{32G\rho}}$. $\rho$ is derived from the
1.3mm dust continuum core mass from Liu et al. (2018). Here for
$L_{\rm bol,iso}/M_{\rm max}$ in each source, we use the mass of the
one main core driving the outflows (B1c2, B2c9, C2c2, C6c1, C9c5,
H6c8) from Liu et al. (2018), which happens to be the most massive
core in each field, and assumes the bolometric luminosity derived from
the SED fitting mainly comes from this core. The collimation of the
outflow lobes that are not very extended, like H6, may be influenced
if they were to be corrected for inclination. Overall, if assuming the
same inclination for every source, i.e., with no relative effects of
inclination, we do not see apparent correlations between outflow
opening angles and evolutionary stage. Note that if we exclude B2
  and H6, which have a poorly defined outflow opening angle, we do see
  that the outflow opening angle increases as core mass and luminosity
  increase in the other four sources.

\begin{figure*}[htbp]
\epsscale{1.1}
\plotone{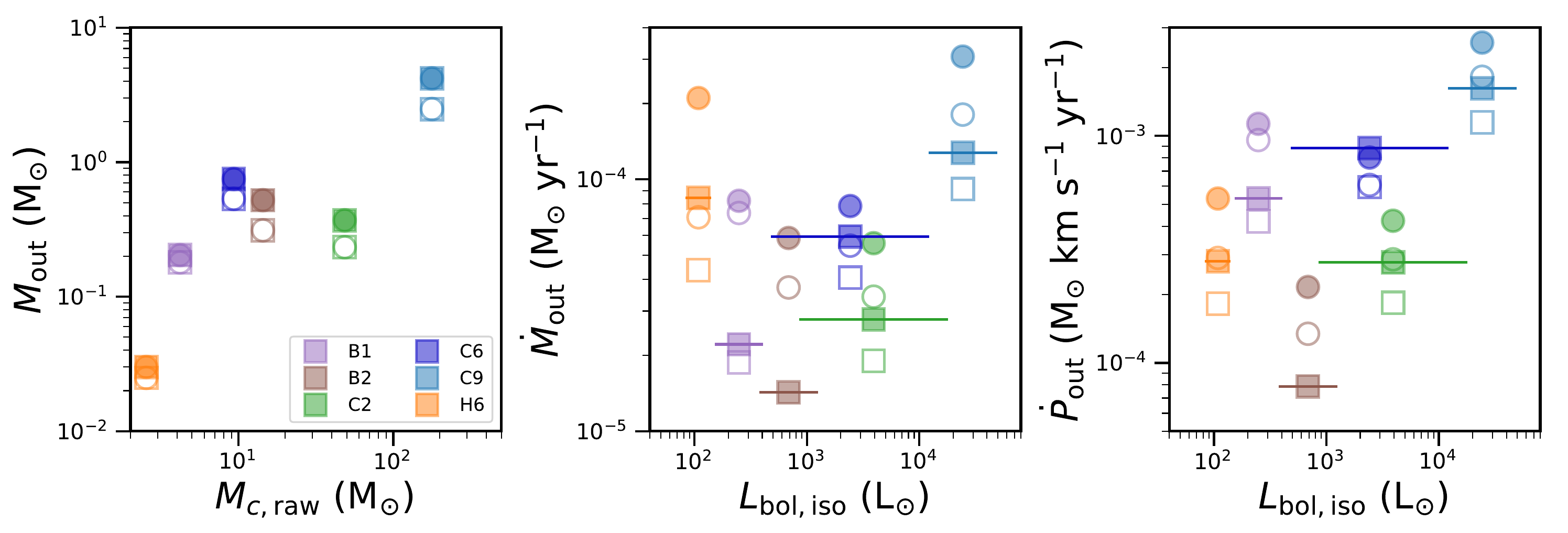}
\caption{From left to right: outflow mass $M_{\rm out}$ versus the mm core mass $M_{c, \rm raw}$, outflow mass rate $\dot{M}_{\rm out}$ versus isotropic bolometric luminosity $L_{\rm bol,iso}$, outflow momentum rate $\dot{P}_{\rm out}$ versus isotropic bolometric luminosity $L_{\rm bol,iso}$. The symbols are the same with Figure~\ref{fig:corr6_hoa}, with the additional empty circles representing outflow parameters derived excluding the emission within $\pm$ 3 km s$^{-1}$ to the systemic velocity, which we consider coming from ambient gas, and $t_{\rm dyn}$ derived with the maximum outflow velocity.}\label{fig:corr6_2}
\end{figure*}

We also investigate the mass entrainment of the strong SiO outflows as
shown in Figure~\ref{fig:corr6_2}. The outflow mass seems to increase
with the core mass, which is consistent with the results of Beuther et
al. (2002). Previous outflow studies have also found correlation
between the bolometric luminosity and the mechanical force and
momentum, which holds over six orders of magnitude of $L_{\rm bol}$
(e.g., Figure 4 in Beuther et al. 2002, Figure 7 in Maud et al. 2015),
and is interpreted as evidence for a single outflow mechanism that
scales with the stellar luminosity, and the outflows motion being
momentum driven. From the six sources presented here, the potential
correlation of the mechanical force and mass entrainment rate with
luminosity has significant scatter.

Since we can only resolve the protostellar cores down to a typical
scale of 5000 AU, we are unable to further distinguish the outflow
launching mechanism (e.g., the X-wind model (Shu et al. 2000) and the
disk-wind model (K\"onigl \& Pudritz 2000)).

H$_2$O maser emission traces shocked gas propagating in dense regions
($n_{H_2} >10^{6} \rm \: cm^{-3}$) at velocities between 10 and 200
km\,s$^{-1}$ (e.g., Hollenbach et al. 2013) and is considered to be a
signpost of protostellar outflows within a few 1000 AU of the driving
source. Surveys have found H$_2$O masers associated with young stellar
objects with luminosities of $1-10^{5} \: L_{\odot}$ (Wouterloot \&
Walmsley 1986; Churchwell et al. 1990; Palla et al. 1993; Claussen et
al. 1996). Wang et al. (2006) observed $\rm H_{2}O$ maser emission
toward a sample of 140 compact, cold IRDC cores with the VLA. All the
32 regions except for C9 are covered by their observation. Only at
B2c8 (0.66 $M_{\odot}$), C2c4 (33 $M_{\odot}$), C4c1 (17 $M_{\odot}$),
D5c1 (3.6 $M_{\odot}$) are water masers detected with emission higher
than the detection limit of 1\,Jy (see Figure~\ref{fig:new}). They have a relatively low detection rate of 12\% and suggest that most of the IRDC cores have not yet formed high-mass protostars or are forming only low- or intermediate-mass protostars. While more of our sources are at an early evolutionary stage, as indicated by the SED fitting, we do have some massive cores in B2, C2 and D9 (C9 has the most massive cores and appears most evolved but it is not covered in the Wang et al. (2006) observations). We think sensitivity is likely to be a major limiting factor here. Our VLA observations are sensitive to CH$_3$OH masers, which trace high-mass
star formation. The CH$_3$OH maser results will be presented in a
later paper (Rosero et al. in prep.). From an initial inspection, it
appears that there is only detection in the B2 and C9 clump
(associated with the C9r1 core).

Compared with the SED fitting results of more evolved massive YSOs (De
Buizer et al. 2017; Liu et al. 2019, 2020), the accretion rates $\dot {M}_{\rm disk}$
derived in our IRDC sample are about one order of magnitude lower even
for the high-/intermediate-mass sources, though the photometry scale
may be smaller by a factor of 2. Further comparison of protostellar
properties of the sources studied here and those of the SOMA survey
sample has been presented by Liu et al. (2020). The low luminosity, low accretion rates and low current stellar mass returned by SED fitting
suggest the sources in our sample are at an early stage.

\subsection{Strength of SiO Emission}\label{sec:vs}

\begin{figure}[htbp]
\epsscale{1.0}
\plotone{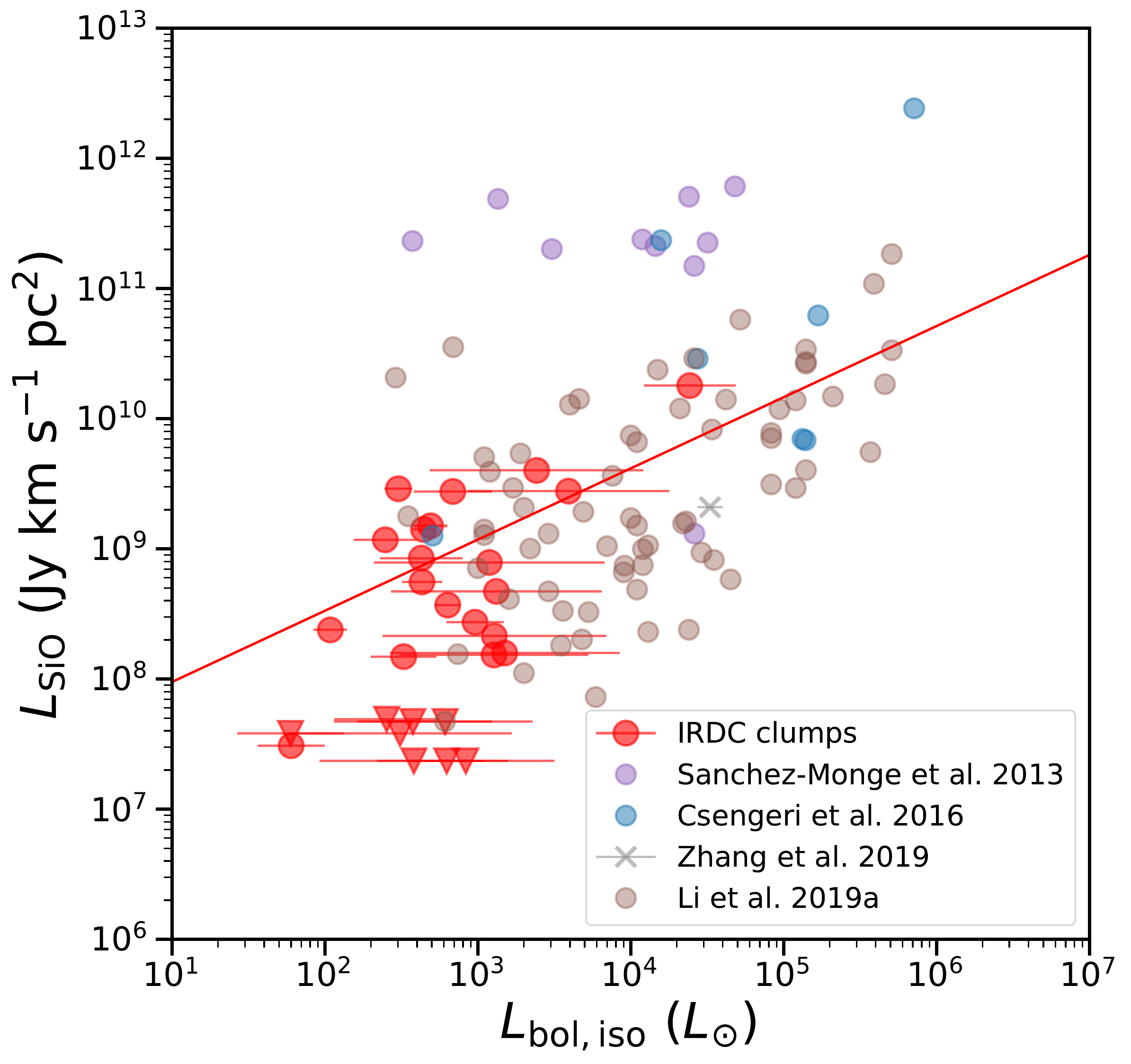}
\caption{
SiO(5-4) line luminosity $L_{\rm SiO}$ as a function of protostar
bolometric luminosity $L_{\rm bol,iso}$. The red dots denote our
SiO(5-4) data, while the red triangles indicate upper limits for
SiO line luminosity. The error bar represents 95\% geometric confidence interval of $L_{\rm bol,iso}$ among the valid models.
The purple dots denote SiO(5-4) data from S{\'a}nchez-Monge et
al. (2013b). The blue dots denote SiO(5-4) data from Csengeri et
al. (2016). The brown dots denote SiO(5-4) data from Li et
al. (2019). The gray cross denotes SiO(5-4) data from Zhang et
al. (2019). The red line shows a linear fit {\bf $f(x) = 0.55x +
7.43$}.}\label{fig:vs1}
\end{figure}

\begin{figure}[htbp]
\epsscale{1.0}
\plotone{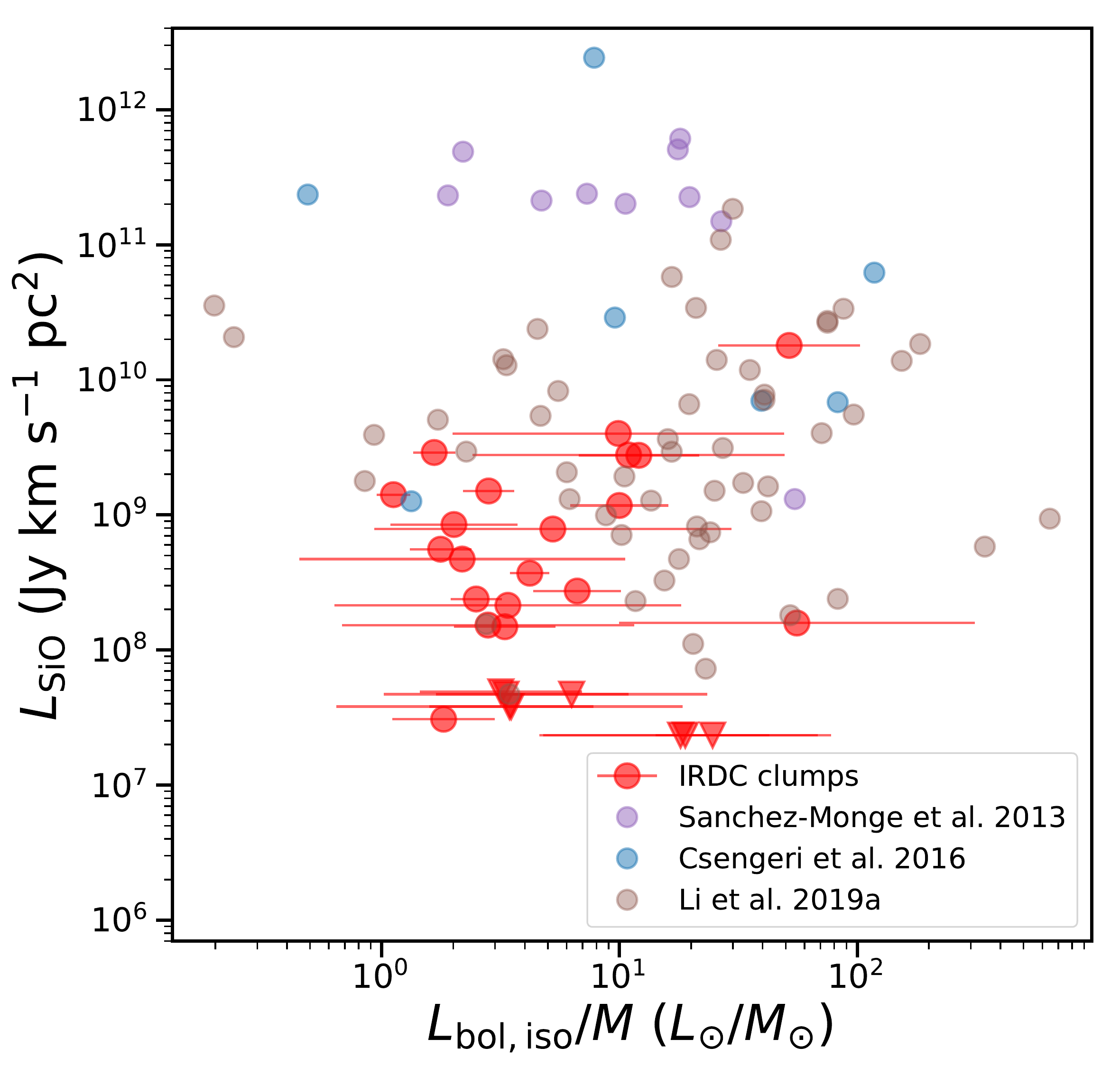}
\caption{
SiO(5-4) line luminosity $L_{\rm SiO}$ as a function of the bolometric
luminosity-mass ratio $L_{\rm bol,iso}/M$. The markers are the same
as in Figure~\ref{fig:vs1}. The error bar represents 95\% geometric confidence interval of $L_{\rm bol,iso}$ among the valid models.
}\label{fig:vs2}
\end{figure}

We measured the bolometric luminosity of the sources defined by their
MIR emission (geometric mean $L_{\rm bol,iso}$ of the valid models in
the ten best models of each source) and the SiO line luminosity inside
the aperture to explore how SiO acts as an outflow tracer across the
luminosity regime. The flux inside the contours denoting
\textit{trunk} structures in Figure~\ref{fig:trunk} is used to
calculate the SiO line luminosity. For the 6 sources for which we have accurate velocity range estimates, we remade their integrated intensity maps using the updated velocity range and derived new \textit{trunk} structures with astrodendro. For the other sources, we adopted a universal
velocity range of $\pm$15\,km s$^{-1}$ relative to the systemic
velocity for all the sources, given that the SNRs for more than half of
the 32 sources are too low to determine a distinct velocity range.
For most sources this velocity range represents the SiO emission
well. We sum up flux from all the \textit{trunks} inside the ALMA field of view for
each source. Note that in B2 and C9, we only include part of the SiO
which is inside the apertures shown in Figure~\ref{fig:M0_ir}. For all
the other sources almost all the detected SiO emission falls inside
the aperture used for photometry.

The results are shown in Figure~\ref{fig:vs1}. Where there is SiO, we
can always find 1.3\,mm continuum emission and some infrared emission
nearby, but not vice versa. Those clumps with no detectable SiO
emission are denoted with 3\,$\sigma$ SiO detection thresholds in
Figure~\ref{fig:vs1}. In C3 and D2 there is neither SiO or 1.3\,mm
continuum cores detected, so we think there are no protostars in these
two clumps and do not measure the luminosities there. We also include
measurement of SiO(5-4) from the literature. We notice there may be
biases to combine the data due to differences in instrument and
method. Nevertheless, we see a slight increasing trend of SiO line
luminosity with the bolometric luminosity. We obtain a linear fit of 
$f(x) = (0.55 \pm 0.09)x + (7.43 \pm 0.37)$ and the Pearson correlation coefficient is
$\sim$ 0.50. Our result is consistent with the results of Codella et
al. (1999). They studied SiO emission towards both low- and high-mass
YSOs and found a trend of brighter SiO emission from higher luminosity
sources, suggesting more powerful shocks in the vicinity of more
massive YSOs.

On the other hand, Motte et al. (2007) found that the SiO integrated
intensities of the infrared-quiet cores are higher than those of the
luminous infrared sources. Sakai et al. (2010) also found the SiO
integrated intensities of some MSX sources are much lower than those
of the MSX dark sources. They suggested that the SiO emission from the
MSX sources traces relatively older shocks, whereas it mainly traces
newly-formed shocks in the MSX dark sources, which results in the
observed decrease in the SiO abundance and SiO line width in the late
stage MSX sources. S\'{a}nchez-Monge et al. (2013b) argued that SiO is
largely enhanced in the first evolutionary stages, probably owing to
strong shocks produced by the protostellar jet. They suggested that as
the object evolves, the power of the jet decreases and so does
the SiO abundance.
 
To help break the degeneracy between mass and evolution, we also plot
the SiO luminosity versus $L_{\rm bol,iso}/M$, as shown in
Figure~\ref{fig:vs2}, together with data from literature. The $L_{\rm
  bol}/M$ ratio is commonly used as a tracer of evolutionary stage
(e.g., Molinari et al. 2008; Ma et al. 2013; Urquhart et al. 2014;
Leurini et al. 2014; Csengeri et al. 2016) with a higher value
corresponding to a later stage.
Here, $M$ is derived from the flux measurement
of the archival 1.1mm Bolocam Galactic Plane Survey (BGPS) data (Ginsburg et al. 2013) in the same aperture used for
constructing SEDs according to
\begin{equation}
M_{\rm dust} = \frac{F_{\nu}D^{2}}{B_{\nu}(T_{\rm dust})\kappa_{\nu}},
\end{equation}
where $M_{\rm dust}$ is the dust mass, $F_{\nu}$ is the continuum flux
at frequency $\nu$, $D$ is the source distance, and $B_{\nu}(T_{\rm
  dust})$ is the Planck function at dust temperature $T_{\rm dust}$ =
20 K. A common choice of $\kappa_{\nu}$ is predicted by the moderately
coagulated thin ice mantle dust model of Ossenkopf \& Henning (1994),
with opacity per unit dust mass of $\kappa_{\rm 1.2mm, \ d} = 0.899
\rm \ cm^{2} g^{-1}$. A gas-to-refractory-component-dust-mass ratio of
141 is estimated by Draine (2011) so $\kappa_{\rm 1.2mm} = 6.376
\times 10^{-3} \rm \ cm^{2} g^{-1}$. Note in S\'{a}nchez-Monge et
al. (2013b) and Csengeri et al. (2016), $M$ also represent the same
scale with their $L_{\rm bol}$. Overall, we do not see clear relation
between the SiO luminosity and the evolutionary stage.

A number of studies have found a decrease in SiO abundance with
increasing $L/M$ in massive star-forming regions (e.g.,
S\'{a}nchez-Monge et al. 2013b; Leurini et al. 2014; Csengeri et
al. 2016). However, in S\'{a}nchez-Monge et al. (2013b), SiO(2-1) and
SiO(5-4) outflow energetics seem to remain constant with time (i.e.,
an increasing $L/M$). In Csengeri et al. (2016), SiO column density
estimated from the LTE assumption and the (2-1) transition also seems
to remain constant with time. In L\'{o}pez-Sepulcre et al. (2011) there seems
to be no apparent correlation between the SiO(2-1) line luminosity and
$L/M$, though they claimed a dearth of points at low $L/M$ and low SiO
luminosity. Li et al. (2019a) also find the SiO luminosities and the SiO abundance do not show apparent differences among various evolutionary stages in their sample from IRDCs to young H\,II regions.

The protostars in our sample mostly occupy a luminosity range of
$10^{2}-10^{3} \ L_{\rm bol}$. Given the fact that most of them have
SiO detection, it seems that as long as a protostar approaches a
luminosity of $\sim \ 10^{2} \ L_{\rm bol}$, the shocks in the outflow
are strong enough to form SiO emission.

\subsection{Nature of the Radio Sources} 

\begin{figure}[htbp]
\centering{\includegraphics[width=0.48\textwidth]{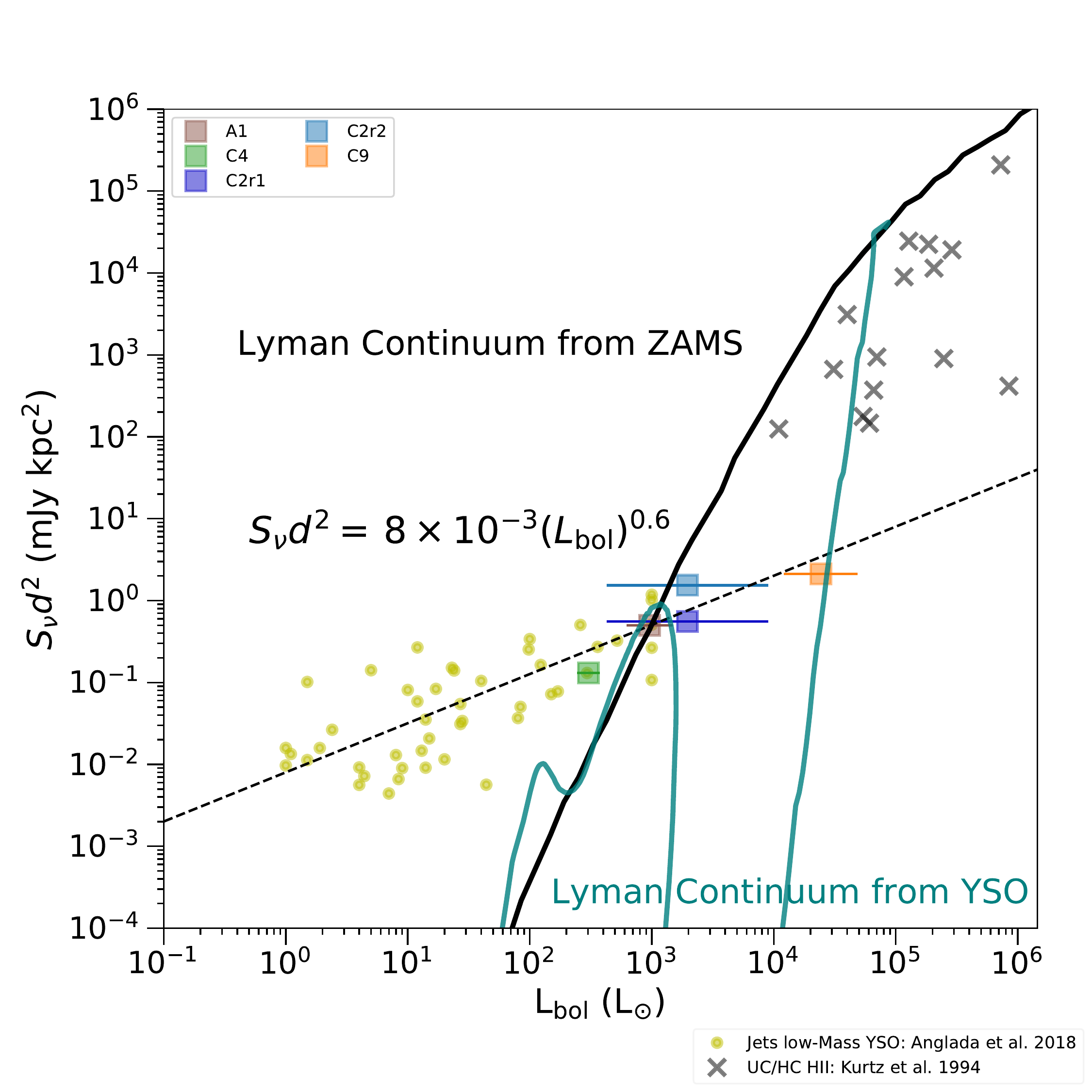}}
\caption{
Radio luminosity scaled to 5\,GHz (assuming a spectral index
$\alpha=0.6$) versus the bolometric luminosity. Note for the two radio
sources in C2 we adopt a luminosity value of $L_{\rm bol}$ of C2
divided by 2. The bolometric luminosity is given by the geometric mean
value of the isotropic luminosity returned by the valid models for
each source. The yellow circles represent ionized jets toward low-mass
stars from Anglada (2018). The dashed line shows a power law relation
for these sources, given by Anglada et al. (2015): $\left(\frac{S_{\nu}d^2}{\rm mJy\:kpc^2}\right) = 8 \times 10^{-3} \left(\frac{L_{\rm bol}}{L_\odot}\right)^{0.6}$.
The $\times$ symbols are UC and HC HII regions from Kurtz et
al. (1994). The black and the cyan continuous lines are the radio emission expected from optically thin H II regions powered by a single zero-age main-sequence (ZAMS) star at a
given luminosity (S\'{a}nchez-Monge et
al. 2013a) and from a YSO with $M_{\rm c} = 60~M_{\odot}$ and $\Sigma_{\rm cl} = 1 ~ \rm g~cm^{-2}$ (Tanaka et al. 2016), respectively.} \label{fig:lradio_vs_lbol}
\end{figure}

\begin{figure}[htbp]
\centering{\includegraphics[width=0.48\textwidth]{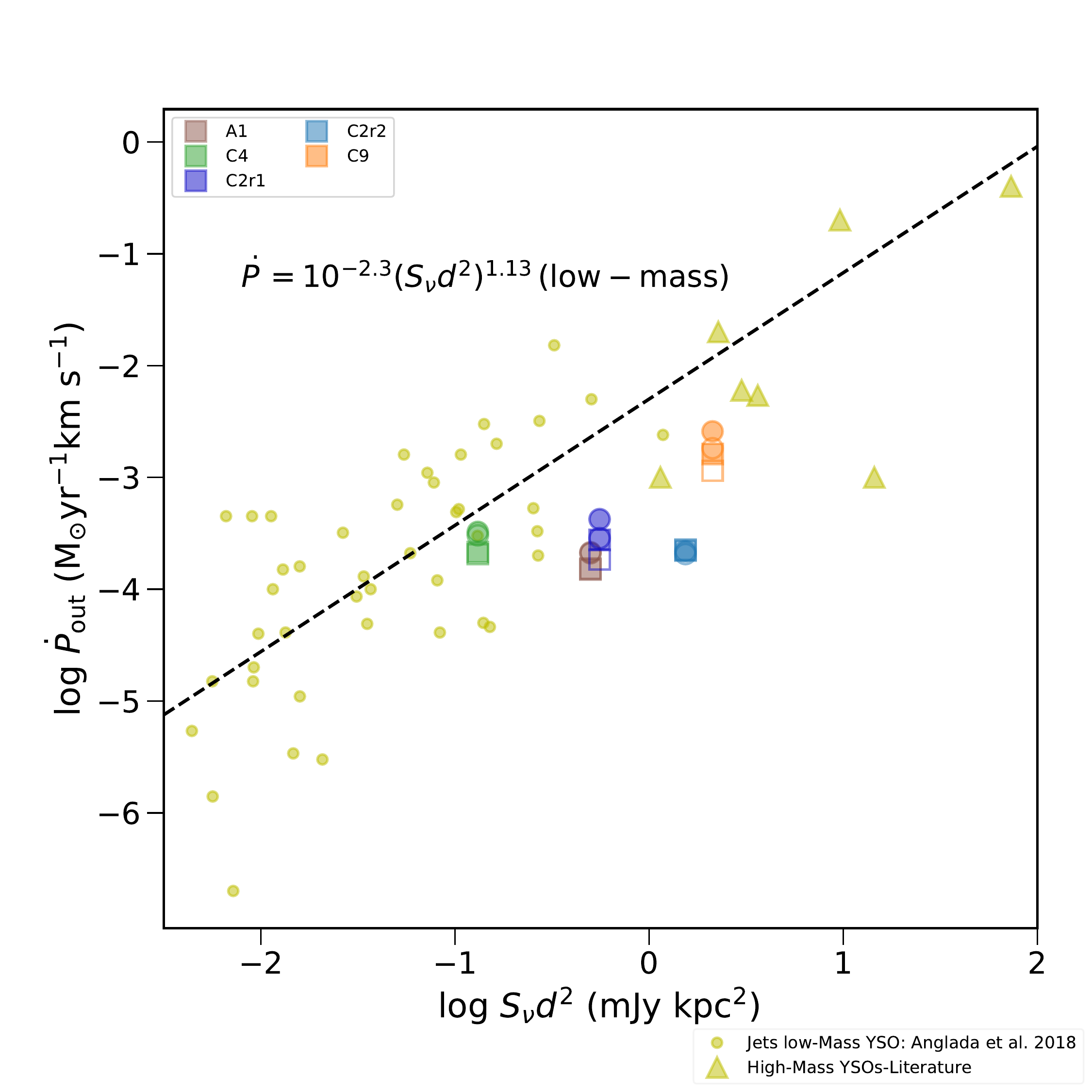}}
\caption{Momentum rate of the molecular outflow as a function of the radio luminosity at 5 GHz. The momentum rate values of the molecular outflow for our sources are measured from our SiO data; for the rest of the data the values are collected from the literature. The symbols for our sources are the same with Figure~\ref{fig:corr6_2}. The yellow circles represent ionized jets associated with low-mass protostars from Anglada et al. (2018). The yellow triangles represent ionized jets associated with high-mass stars (Rodr{\'\i}guez et al. 2008; Moscadelli et al. 2016). The dashed line relation shows the correlation found by Anglada (1995) derived for jets from low-mass stars.} \label{fig:pdot_vs_lradio}
\end{figure}

As photoionization cannot account for the observed radio continuum
emission of low luminosity objects, shock ionization has been proposed
as a viable alternative mechanism (Curiel et al. 1987; Gonz{\'a}lez \&
Cant{\'o} 2002) and the correlation of the bolometric and radio
luminosities is interpreted as a consequence of the accretion and
outflow relationship (Anglada et al. 2018). Additionally, synchrotron emission could also arise from fast shocks in disks and jets (e.g., Carrasco-Gonz{\'a}lez et al. 2010). Based on the radio/bolometric luminosity
comparison as shown in Figure~\ref{fig:lradio_vs_lbol}, it seems the radio emission in A1, C2,
C4 and C9 is more likely to be from shock ionized radio jets
than a HC HII region. Note that for the two radio sources in C2 we adopt a luminosity
value of $L_{\rm bol}$ of C2 divided by 2. More information is needed
to confirm their nature, such as a robust spectral index derived from
multi bands and resolved morphology. If they are shock ionized jets,
they are likely to be detected because the cores are massive enough to
drive strong shocks. We did not see clear relation between the radio emission detection
and $L/M$ and $t_{\rm dyn}/t_{\rm ff}$ (not shown here), which indicate the evolutionary stage. It is likely that our sample is overall at an early stage and
photoionization has not become significant enough yet to form a HC HII
region, even for the high-mass cores C2 and C9, which would eventually
evolve to HII regions based on their current core mass. A similar early-stage population prior to the onset of a HC H II region is also seen in Lim \& De Buizer (2019), where half of the 41 MYSO candidates revealed by MIR in W51A have no detected radio continuum emission at centimeter wavelengths. However, we
are not sure why other intermediate-mass cores with strong SiO
outflows, e.g., the cores in C6, do not show detected shock ionized radio jets.

Also, it has been found previously that the outflow momentum rate is
correlated with the radio luminosity, which can be explained by the
shock ionization mechanism working in radio jets (Anglada et
al. 2018). In particular, the shocked-induced ionization model implies
$\left(\frac{S_{\nu}d^{2}}{\rm mJy\,kpc^{2}}\right) = 10^{3.5} \eta
\left(\frac{\dot{P}}{M_{\odot}\,\rm yr^{-1}\,km\,s^{-1}}\right)$ at
$\nu=5$ GHz where $\eta$ is the ionization fraction. In
Figure~\ref{fig:pdot_vs_lradio} we show the momentum rate ($\dot{P}$)
of our SiO molecular outflows that are associated with centimeter
emission as a function of the radio luminosity ($S_{\nu}d^{2}$) of the
ionized jet estimated from our flux values at 5 GHz. The yellow
triangles represent ionized jets associated with high-mass stars as
collected from the literature (Rodr{\'\i}guez et al. (2008);
Moscadelli et al. (2016)) and the yellow circles represent ionized
jets associated with low-mass protostars from Anglada et
al. (2018). The molecular outflow data from the literature are from
observations using different spectral lines and telescopes, thus the
scatter in the data. The outflow momentum rate estimates typically have considerable uncertainties of more than one order of magnitude (Anglada et al. 2018). The dashed line relation in
Figure~\ref{fig:pdot_vs_lradio} shows the observational correlation
found by Anglada (1995) derived for jets associated with low-mass
stars with an ionization fraction of $\sim 10\%$. Despite the scatter
in the data in Figure~\ref{fig:pdot_vs_lradio}, our observations
appear to suggest that the ionization fraction or the fraction of
material that gets ionized by shocks may be higher than $\sim 10\%$
for high-mass protostars than for the low-mass counterpart (see Rosero
et al. 2019b for a further discussion). However, a larger and
homogeneous sample is required for drawing any conclusions.

If we assume the outflows are materials entrained by jets, there is a higher chance that the radio flux of the jet can be detected in sources that show brighter SiO outflows (see Figure ~\ref{fig:lradio_vs_lbol} and \ref{fig:pdot_vs_lradio}). However, the mm cores associated with detected radio emission are not always
the cores that drive the strongest SiO outflows or that have the highest bolometric luminosity (e.g., A1 and C4). Nevertheless, they are among the most massive ones in all the
32 clumps in our ALMA survey. A larger number of detections is required to derive any robust correlation between the radio flux and the core mass, the outflow strength. As discussed in \S~\ref{sec:vla}, the detection in our sample is partly limited by sensitivity.

We have a radio detection rate of $\sim$ 27\% out of the 15 IRDC
clumps with both SiO outflows and 1.3\,mm continuum emission. In
Rosero et al. (2016) detection rates of radio sources associated with
the millimeter dust clumps within IRDCs with and without IR sources (CMC--IRs and CMCs, respectively), and hot molecular cores (HMCs) are 53\%, 6\%,
and 100\%, respectively. The majority of our 15 sources should belong
to their CMC category. The low detection rate in our sample is
consistent with increasing high-mass star formation activity from CMCs
to HMCs. The offsets between the 1.3\,mm continuum peak and the 6\,cm
continuum peak are typically $\sim$ 2500 AU, smaller than that of 4000
au and 10000 au for CMC--IRs and HMCs, respectively, in Rosero et
al. (2016).

\section{Conclusions}

In this work, we have presented the results of ALMA observations of
SiO(5-4) and VLA 6\,cm radio observations made towards 32 IRDC clumps
potentially harboring prestellar/protostellar sources. Our goal is to
characterize a large number of protostars from low-mass to high-mass
at the earliest phases with their outflow emission, and investigate
the onset of SiO emission as shock tracers and the onset of
ionization, to understand massive star formation. In summary, our main
results and conclusions are as follows.

1. We have detected SiO(5-4) emission in 20 out of 32 IRDC clumps with
a detection rate of 62\%. From the non-detection in our sample and
comparison with the SiO detection rate in other IR-dark clumps, it
seems that at early evolutionary stages, shocked gas is more common in
the higher mass regime. Compared with more evolved IR-bright
protostars, the SiO detection rate is overall higher in early-stage
protostars. Compared with CO outflows, which are generally highly
  collimated, the SiO outflows appear to be somewhat more disordered
  and fragmented, though most are still consistent with being
  collimated structures or parts of the collimated outflows.

2. In the 20 sources with detected SiO, 11 sources with relatively
strong SiO emission seem to host SiO outflows based on their wide line
widths. Most SiO outflows show bipolar structures though they can be
highly asymmetric. Some SiO outflows are collimated while others are
less ordered. There is evidence for successive ejection events as well
as multiple outflows originating from $\la$ 0.1pc, which can be due to
outflow changing orientation over time. For the six protostellar
sources with strongest SiO outflow emission, we do not see clear
  dependence of the collimation of the outflows on evolutionary
  stage.

3. For the six protostars with strongest SiO emission, we locate the
protostellar sources driving the outflows, which appear as nearby mm
continuum peaks, in position-velocity space utilizing dense gas
tracers DCN(3-2), DCO$^+$(3-2) and C$^{18}$O(2-1). They have
relatively low outflow masses, mass outflow rates and momentum flow
rates, in spite of the fact that some of them are high-mass
protostellar candidates based on their mm core masses. The outflow
masses appear to increase with the core mass. We do not see a strong
dependence of mechanical force and mass entrainment rate on bolometric
luminosity.

4. Where there is SiO(5-4) emission seen, we can always find 1.3 mm
continuum emission and some infrared emission nearby, but not vice
versa. The low luminosity, low accretion rates and low current
  stellar mass returned by SED fitting suggest the sources in our
  sample are at an early evolutionary stage. With the entire sample
and data from literature, we see a slight increasing trend of SiO line
luminosity with the bolometric luminosity, which suggests more
powerful shocks in the vicinity of more massive YSOs. We do not see
clear relation between the SiO luminosity and the evolutionary stage
indicated by $L/M$. Given the fact that most of our sample have SiO
detection, it seems that as long as a protostar approaches a
luminosity of $\sim \ 10^{2} \ L_{\odot}$, the shocks in the outflow
are strong enough to form SiO emission.

5. We detect 6\,cm radio continuum emission in 4 out of the 15
  sources with bright SiO emission. The radio emission is likely due
to shock-ionized jets associated with most massive protostellar
cores. It is likely that our sample is overall at an early stage and
photoionization has not become significant enough yet to form HC HII
regions, even for the high-mass cores. Our VLA detections are
  limited by sensitivity and higher-sensitivity observations are
  needed to find a larger number of sources to be able to determine if
  there is any robust correlation between the radio flux and the core
  mass, outflow strength and evolutionary stage.

6. As an overall summary of implications for massive star
  formation, we conclude that SiO outflows and radio continuum
  emission are important diagnostic tracers of sources that have
  relatively special shock conditions that often have properties
  consistent with being massive protostars in an early stage of
  evolution. While the SiO emission is generally more disordered
  and/or sparse than that seen in CO, the emission features are
  generally consistent with being parts of highly collimated
  outflows. This indicates that these candidate early-stage massive
  protostars are launching such outflows, which is an
  expectation of core accretion models. Among our sample of six strong
  SiO outflow sources, we have one example, C9, that exhibits a much
  more disordered morphology, which may indicate a different type of
  outflow or the presence of multiple protostars in a small projected
  region of the sky that are each launching collimated SiO
  outflows. Higher resolution follow-up observations of this source
  are needed to distinguish these possibilities.

\acknowledgments We thank thank the anonymous referee, whose comments improved this manuscript. M.L. acknowledges funding from the Jefferson Scholars
Foundation. M.L. and J.C.T. acknowledge funding from
NASA/USRA/SOFIA. J.C.T. acknowledges support from NSF grant
AST1411527, VR grant 2017-04522 and ERC project 788829 -
MSTAR. This paper makes use of the following ALMA data: ADS/JAO.ALMA\#2013.1.00806.S. ALMA is a partnership of ESO (representing its member states), NSF (USA) and NINS (Japan), together with NRC (Canada), NSC and ASIAA (Taiwan), and KASI (Republic of Korea), in cooperation with the Republic of Chile. The Joint ALMA Observatory is operated by ESO, AUI/NRAO, and NAOJ. The National Radio Astronomy Observatory is a facility of the National Science Foundation operated under cooperative agreement by Associated Universities, Inc.

\software{CASA (McMullin et al. 2007), astrodendro (http://www.dendrograms.org), Photutils (https://doi.org/10.5281/zenodo.596036).}

\appendix

\section{MIR Photometry and SED fitting} \label{sec:more SED}

We used PHOTUTILS (https://doi.org/10.5281/zenodo.596036), a PYTHON package to measure the flux density of the
entire sample. Note that clumps with neither 1.3 mm continuum nor SiO
emission are not taken into consideration. The circular aperture is
determined based on the 70 $\mu$m and 160 $\mu$m 2D profile to cover
most of the clumpy infrared emission. Sometimes when the boundary of
the core and the background is unclear, we adopt an aperture radius of
15\arcsec, so that the aperture diameter is close to the beam size of
{\it Herschel} 500 $\mu$m images. The aperture radii for most sources
are around 15\arcsec. After measuring the flux inside the aperture, we
carry out background subtraction using the median flux density in an
annular region extending from 1 to 2 aperture radii to remove general
background and foreground contamination and the effect of a cooler,
more massive clump surrounding the core at long wavelengths (see also
De Buizer et al. 2017; Liu et al. 2019, 2020; Moser et al. 2020). Note that if the
annular region overlaps with the apertures of other sources, we
exclude the overlapping part for background estimation. The error of
the flux is estimated to be the quadratic sum of the 10\% of the
background subtracted flux as calibration uncertainty and the
background fluctuation. The background fluctuation is derived in the
following way: first we divide the annular region extending from 1 to
2 aperture radii evenly into six sectors and measure the standard
deviation of the mean values in the six sectors; then we rotate every
sector by 10\arcdeg and measure the standard deviation in the same way
again; we repeat the rotation for another four times and obtain six
measured standard deviations in total; finally we derive the mean
value of the six standard deviation as the estimation of the
background fluctuation.

We use data at wavelengths $< \ 8 \ \mu$m as upper limits due to PAH
emission and thermal emission from very small grains. Additionally, if
the background is stronger than the flux density inside the aperture,
then the flux density without background subtraction is given and
treated as an upper limit. The fitting procedure involves convolving
model SEDs with the filter response functions for the various
telescope bands. Source distances were adopted from Butler \& Tan
(2012).

\begin{figure*}[htbp]
\epsscale{1.2}
\plotone{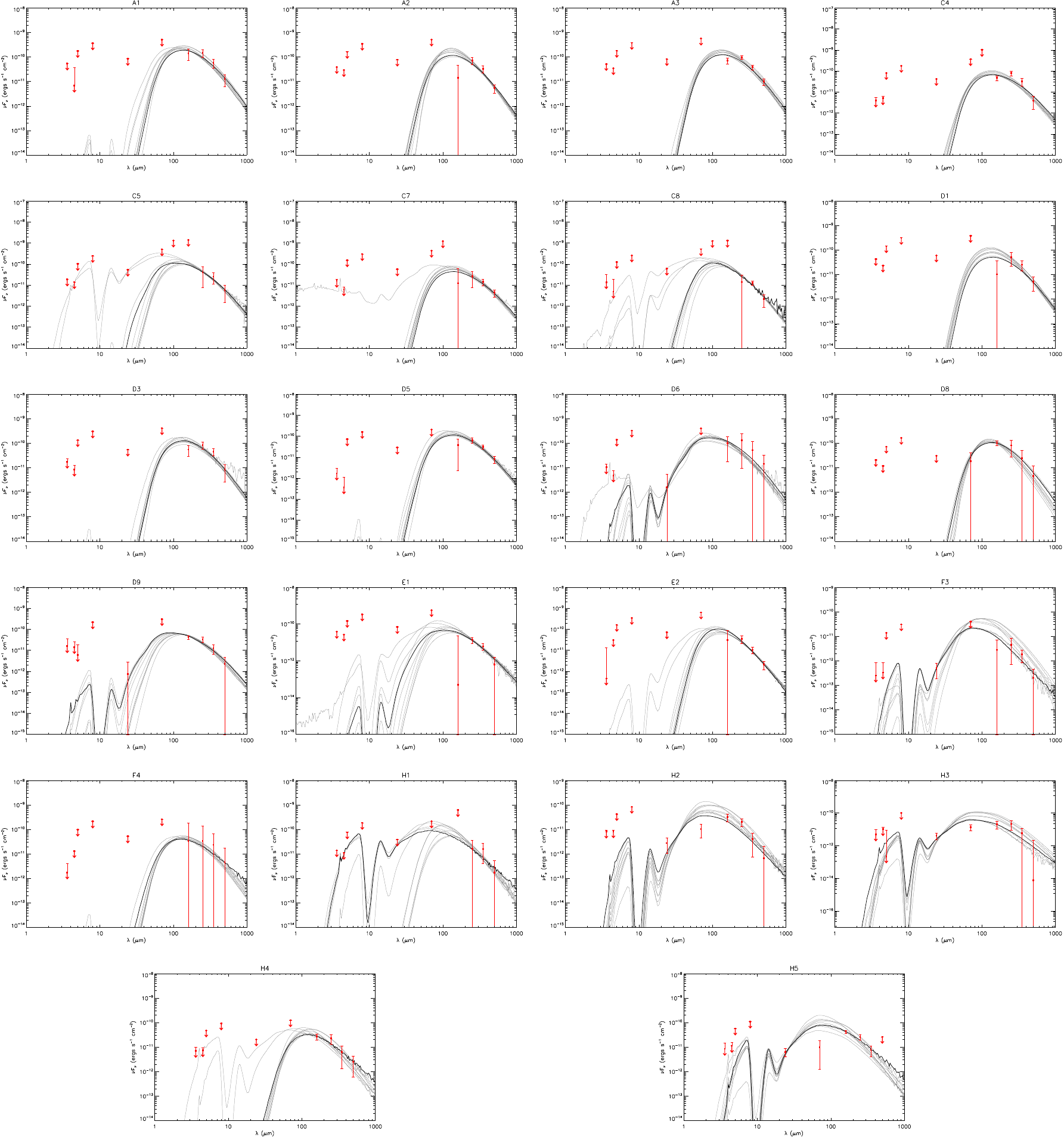}
\caption{The same with Figure~\ref{fig:SEDfit} but for weak SiO sources. The model parameter results are listed in Table~\ref{tab:more SEDfit}.}  \label{fig:more SEDfit}
\end{figure*}

We show the SED fitting of other sources in the sample in
Figure~\ref{fig:more SEDfit}. The model parameters are listed in
Table~\ref{tab:more SEDfit}.

\clearpage
\renewcommand{\arraystretch}{0.8}
\startlongtable
\begin{deluxetable}{ccccccccccccc}
\tabletypesize{\scriptsize}
\tablecaption{Parameters of the Ten Best Fitted Models} \label{tab:more SEDfit}
\tablewidth{16pt}
\tablehead{
\colhead{Source} &\colhead{$\chi^{2}$/N} & \colhead{$M_{\rm c}$} & \colhead{$\Sigma_{\rm cl}$} & \colhead{$R_{core}$}  &\colhead{$m_{*}$} & \colhead{$\theta_{\rm view}$} &\colhead{$A_{V}$} & \colhead{$M_{\rm env}$} &\colhead{$\theta_{w,\rm esc}$} & \colhead{$\dot {M}_{\rm disk}$} & \colhead{$L_{\rm bol, iso}$}  & \colhead{$L_{\rm bol}$} \\
\colhead{} & \colhead{} & \colhead{($M_\odot$)} & \colhead{(g $\rm cm^{-2}$)} & \colhead{(pc) ($\arcsec$)} & \colhead{($M_{\odot}$)} & \colhead{(deg)} & \colhead{(mag)} & \colhead{($M_{\odot}$)} & \colhead{(deg)} &\colhead{($M_{\odot}$/yr)} & \colhead{($L_{\odot}$)} & \colhead{($L_{\odot}$)} \\
\vspace{-0.4cm}
}
\startdata
A1
& 0.16 & 480 & 0.1 & 0.51 ( 22 ) & 2.0 & 48 & 1000.0 & 477 & 4 & 4.3$\times 10^{-5}$ & 5.7$\times 10^{2}$ & 5.7$\times 10^{2}$ \\
$d$ = 4.8 kpc
& 0.19 & 400 & 0.1 & 0.47 ( 20 ) & 2.0 & 13 & 941.9 & 391 & 4 & 4.1$\times 10^{-5}$ & 7.7$\times 10^{2}$ & 5.5$\times 10^{2}$ \\
$R_{ap}$ = 16 \arcsec
& 0.22 & 240 & 0.1 & 0.36 ( 15 ) & 4.0 & 77 & 1000.0 & 229 & 9 & 5.1$\times 10^{-5}$ & 9.1$\times 10^{2}$ & 1.0$\times 10^{3}$ \\
= 0.37 pc
& 0.24 & 320 & 0.1 & 0.42 ( 18 ) & 4.0 & 13 & 656.7 & 308 & 7 & 5.5$\times 10^{-5}$ & 1.2$\times 10^{3}$ & 5.4$\times 10^{2}$ \\
& 0.26 & 200 & 0.3 & 0.19 ( 8 ) & 1.0 & 29 & 1000.0 & 196 & 4 & 5.8$\times 10^{-5}$ & 5.2$\times 10^{2}$ & 5.2$\times 10^{2}$ \\
& 0.29 & 400 & 0.1 & 0.47 ( 20 ) & 4.0 & 89 & 1000.0 & 390 & 7 & 5.8$\times 10^{-5}$ & 9.4$\times 10^{2}$ & 1.0$\times 10^{3}$ \\
& 0.33 & 200 & 0.1 & 0.33 ( 14 ) & 4.0 & 13 & 682.7 & 194 & 10 & 4.8$\times 10^{-5}$ & 1.9$\times 10^{3}$ & 6.7$\times 10^{2}$ \\
& 0.33 & 160 & 0.1 & 0.29 ( 13 ) & 4.0 & 13 & 1000.0 & 151 & 12 & 4.5$\times 10^{-5}$ & 3.1$\times 10^{3}$ & 9.1$\times 10^{2}$ \\
& 0.35 & 480 & 0.1 & 0.51 ( 22 ) & 4.0 & 89 & 1000.0 & 474 & 6 & 6.1$\times 10^{-5}$ & 9.2$\times 10^{2}$ & 1.0$\times 10^{3}$ \\
& 0.36 & 240 & 0.1 & 0.36 ( 15 ) & 2.0 & 29 & 411.4 & 233 & 6 & 3.6$\times 10^{-5}$ & 5.2$\times 10^{2}$ & 5.2$\times 10^{2}$ \\
Averages & 0.26 & 291 & 0.1 & 0.37 ( 16 ) & 2.8 & 41 & 869.3 & 283 & 7 & 4.9$\times 10^{-5}$ & 9.6$\times 10^{2}$ & 7.0$\times 10^{2}$ \\
\hline\noalign{\smallskip}
A2
& 0.53 & 160 & 0.1 & 0.29 ( 13 ) & 2.0 & 65 & 1000.0 & 156 & 8 & 3.3$\times 10^{-5}$ & 4.0$\times 10^{2}$ & 4.3$\times 10^{2}$ \\
$d$ = 4.8 kpc
& 0.54 & 200 & 0.1 & 0.33 ( 14 ) & 2.0 & 74 & 1000.0 & 194 & 7 & 3.5$\times 10^{-5}$ & 3.2$\times 10^{2}$ & 3.5$\times 10^{2}$ \\
$R_{ap}$ = 16 \arcsec
& 0.54 & 100 & 0.1 & 0.23 ( 10 ) & 4.0 & 83 & 1000.0 & 91 & 15 & 4.0$\times 10^{-5}$ & 6.6$\times 10^{2}$ & 8.8$\times 10^{2}$ \\
= 0.37 pc
& 0.56 & 100 & 0.3 & 0.13 ( 6 ) & 1.0 & 86 & 1000.0 & 97 & 6 & 4.9$\times 10^{-5}$ & 4.2$\times 10^{2}$ & 4.4$\times 10^{2}$ \\
& 0.57 & 120 & 0.1 & 0.25 ( 11 ) & 4.0 & 89 & 1000.0 & 111 & 14 & 4.2$\times 10^{-5}$ & 7.3$\times 10^{2}$ & 9.4$\times 10^{2}$ \\
& 0.57 & 80 & 0.3 & 0.12 ( 5 ) & 2.0 & 77 & 1000.0 & 76 & 11 & 6.4$\times 10^{-5}$ & 5.3$\times 10^{2}$ & 6.2$\times 10^{2}$ \\
& 0.57 & 120 & 0.1 & 0.25 ( 11 ) & 2.0 & 22 & 915.9 & 117 & 9 & 3.0$\times 10^{-5}$ & 4.3$\times 10^{2}$ & 4.3$\times 10^{2}$ \\
& 0.58 & 80 & 0.1 & 0.21 ( 9 ) & 4.0 & 22 & 1000.0 & 71 & 18 & 3.7$\times 10^{-5}$ & 9.0$\times 10^{2}$ & 8.5$\times 10^{2}$ \\
& 0.60 & 60 & 1.0 & 0.06 ( 2 ) & 0.5 & 89 & 1000.0 & 59 & 5 & 7.2$\times 10^{-5}$ & 3.9$\times 10^{2}$ & 4.0$\times 10^{2}$ \\
& 0.60 & 50 & 1.0 & 0.05 ( 2 ) & 0.5 & 29 & 1000.0 & 50 & 5 & 6.9$\times 10^{-5}$ & 3.9$\times 10^{2}$ & 4.0$\times 10^{2}$ \\
Averages & 0.57 & 99 & 0.2 & 0.16 ( 7 ) & 1.7 & 64 & 991.6 & 94 & 10 & 4.5$\times 10^{-5}$ & 4.9$\times 10^{2}$ & 5.4$\times 10^{2}$ \\
\hline\noalign{\smallskip}
A3
& 0.94 & 320 & 0.1 & 0.42 ( 18 ) & 2.0 & 13 & 1000.0 & 315 & 5 & 3.9$\times 10^{-5}$ & 5.9$\times 10^{2}$ & 3.8$\times 10^{2}$ \\
$d$ = 4.8 kpc
& 0.99 & 320 & 0.1 & 0.42 ( 18 ) & 4.0 & 89 & 1000.0 & 308 & 7 & 5.5$\times 10^{-5}$ & 4.9$\times 10^{2}$ & 5.4$\times 10^{2}$ \\
$R_{ap}$ = 16 \arcsec
& 1.09 & 240 & 0.1 & 0.36 ( 15 ) & 2.0 & 71 & 1000.0 & 233 & 6 & 3.6$\times 10^{-5}$ & 4.9$\times 10^{2}$ & 5.2$\times 10^{2}$ \\
= 0.37 pc
& 1.21 & 200 & 0.1 & 0.33 ( 14 ) & 4.0 & 89 & 1000.0 & 194 & 10 & 4.8$\times 10^{-5}$ & 5.7$\times 10^{2}$ & 6.7$\times 10^{2}$ \\
& 1.30 & 400 & 0.1 & 0.47 ( 20 ) & 2.0 & 89 & 1000.0 & 391 & 4 & 4.1$\times 10^{-5}$ & 5.3$\times 10^{2}$ & 5.5$\times 10^{2}$ \\
& 1.59 & 160 & 0.1 & 0.29 ( 13 ) & 4.0 & 86 & 1000.0 & 151 & 12 & 4.5$\times 10^{-5}$ & 7.4$\times 10^{2}$ & 9.1$\times 10^{2}$ \\
& 1.60 & 200 & 0.1 & 0.33 ( 14 ) & 2.0 & 13 & 770.8 & 194 & 7 & 3.5$\times 10^{-5}$ & 7.3$\times 10^{2}$ & 3.5$\times 10^{2}$ \\
& 1.62 & 160 & 0.1 & 0.29 ( 13 ) & 2.0 & 13 & 1000.0 & 156 & 8 & 3.3$\times 10^{-5}$ & 1.0$\times 10^{3}$ & 4.3$\times 10^{2}$ \\
& 1.62 & 480 & 0.1 & 0.51 ( 22 ) & 2.0 & 89 & 1000.0 & 477 & 4 & 4.3$\times 10^{-5}$ & 5.5$\times 10^{2}$ & 5.7$\times 10^{2}$ \\
& 1.69 & 120 & 0.1 & 0.25 ( 11 ) & 4.0 & 39 & 1000.0 & 111 & 14 & 4.2$\times 10^{-5}$ & 8.4$\times 10^{2}$ & 9.4$\times 10^{2}$ \\
Averages & 1.33 & 238 & 0.1 & 0.36 ( 15 ) & 2.6 & 59 & 977.1 & 230 & 8 & 4.1$\times 10^{-5}$ & 6.4$\times 10^{2}$ & 5.6$\times 10^{2}$ \\
\hline\noalign{\smallskip}
C4
& 0.97 & 320 & 0.1 & 0.42 ( 17 ) & 1.0 & 44 & 1000.0 & 315 & 3 & 2.8$\times 10^{-5}$ & 2.0$\times 10^{2}$ & 2.0$\times 10^{2}$ \\
$d$ = 5.0 kpc
& 0.98 & 240 & 0.1 & 0.36 ( 15 ) & 1.0 & 48 & 1000.0 & 240 & 4 & 2.6$\times 10^{-5}$ & 2.4$\times 10^{2}$ & 2.4$\times 10^{2}$ \\
$R_{ap}$ = 13 \arcsec
& 1.10 & 200 & 0.1 & 0.33 ( 14 ) & 2.0 & 89 & 1000.0 & 194 & 7 & 3.5$\times 10^{-5}$ & 3.2$\times 10^{2}$ & 3.5$\times 10^{2}$ \\
= 0.32 pc
& 1.19 & 200 & 0.1 & 0.33 ( 14 ) & 1.0 & 13 & 767.8 & 197 & 4 & 2.5$\times 10^{-5}$ & 2.7$\times 10^{2}$ & 1.8$\times 10^{2}$ \\
& 1.19 & 100 & 0.1 & 0.23 ( 10 ) & 2.0 & 44 & 1000.0 & 97 & 11 & 2.9$\times 10^{-5}$ & 3.5$\times 10^{2}$ & 3.8$\times 10^{2}$ \\
& 1.22 & 120 & 0.1 & 0.25 ( 11 ) & 2.0 & 80 & 1000.0 & 117 & 9 & 3.0$\times 10^{-5}$ & 3.7$\times 10^{2}$ & 4.3$\times 10^{2}$ \\
& 1.23 & 160 & 0.1 & 0.29 ( 12 ) & 1.0 & 22 & 828.8 & 156 & 5 & 2.3$\times 10^{-5}$ & 2.1$\times 10^{2}$ & 2.0$\times 10^{2}$ \\
& 1.28 & 160 & 0.1 & 0.29 ( 12 ) & 2.0 & 86 & 1000.0 & 156 & 8 & 3.3$\times 10^{-5}$ & 3.9$\times 10^{2}$ & 4.3$\times 10^{2}$ \\
& 1.32 & 120 & 0.1 & 0.25 ( 11 ) & 1.0 & 13 & 840.8 & 120 & 6 & 2.2$\times 10^{-5}$ & 4.3$\times 10^{2}$ & 2.2$\times 10^{2}$ \\
& 1.34 & 80 & 0.1 & 0.21 ( 9 ) & 2.0 & 22 & 1000.0 & 75 & 12 & 2.7$\times 10^{-5}$ & 3.5$\times 10^{2}$ & 3.5$\times 10^{2}$ \\
Averages & 1.18 & 157 & 0.1 & 0.29 ( 12 ) & 1.4 & 46 & 943.7 & 153 & 7 & 2.7$\times 10^{-5}$ & 3.0$\times 10^{2}$ & 2.8$\times 10^{2}$ \\
\hline\noalign{\smallskip}
C5
& 0.02 & 240 & 0.1 & 0.36 ( 15 ) & 1.0 & 83 & 409.4 & 240 & 4 & 2.6$\times 10^{-5}$ & 2.3$\times 10^{2}$ & 2.4$\times 10^{2}$ \\
$d$ = 5.0 kpc
& 0.02 & 120 & 0.1 & 0.25 ( 11 ) & 2.0 & 22 & 1000.0 & 117 & 9 & 3.0$\times 10^{-5}$ & 4.3$\times 10^{2}$ & 4.3$\times 10^{2}$ \\
$R_{ap}$ = 15 \arcsec
& 0.02 & 320 & 0.1 & 0.42 ( 17 ) & 1.0 & 89 & 241.2 & 315 & 3 & 2.8$\times 10^{-5}$ & 1.9$\times 10^{2}$ & 2.0$\times 10^{2}$ \\
= 0.36 pc
& 0.02 & 200 & 0.1 & 0.33 ( 14 ) & 2.0 & 86 & 1000.0 & 194 & 7 & 3.5$\times 10^{-5}$ & 3.2$\times 10^{2}$ & 3.5$\times 10^{2}$ \\
& 0.02 & 160 & 0.1 & 0.29 ( 12 ) & 2.0 & 89 & 1000.0 & 156 & 8 & 3.3$\times 10^{-5}$ & 3.9$\times 10^{2}$ & 4.3$\times 10^{2}$ \\
& 0.03 & 100 & 0.3 & 0.13 ( 5 ) & 32.0 & 55 & 1000.0 & 4 & 81 & 5.0$\times 10^{-5}$ & 1.7$\times 10^{5}$ & 1.4$\times 10^{5}$ \\
& 0.03 & 100 & 0.3 & 0.13 ( 5 ) & 0.5 & 44 & 651.7 & 99 & 4 & 3.5$\times 10^{-5}$ & 2.4$\times 10^{2}$ & 2.4$\times 10^{2}$ \\
& 0.03 & 60 & 0.1 & 0.18 ( 7 ) & 16.0 & 13 & 207.2 & 7 & 76 & 2.0$\times 10^{-5}$ & 5.3$\times 10^{4}$ & 2.5$\times 10^{4}$ \\
& 0.03 & 80 & 0.1 & 0.21 ( 9 ) & 4.0 & 22 & 1000.0 & 71 & 18 & 3.7$\times 10^{-5}$ & 9.0$\times 10^{2}$ & 8.5$\times 10^{2}$ \\
& 0.03 & 100 & 0.1 & 0.23 ( 10 ) & 2.0 & 13 & 115.1 & 97 & 11 & 2.9$\times 10^{-5}$ & 1.2$\times 10^{3}$ & 3.8$\times 10^{2}$ \\
Averages & 0.02 & 130 & 0.1 & 0.24 ( 10 ) & 2.6 & 51 & 662.5 & 75 & 22 & 3.1$\times 10^{-5}$ & 1.2$\times 10^{3}$ & 9.9$\times 10^{2}$ \\
\hline\noalign{\smallskip}
C7
& 0.08 & 200 & 0.1 & 0.33 ( 14 ) & 0.5 & 22 & 1000.0 & 200 & 3 & 1.7$\times 10^{-5}$ & 1.3$\times 10^{2}$ & 1.3$\times 10^{2}$ \\
$d$ = 5.0 kpc
& 0.09 & 160 & 0.1 & 0.29 ( 12 ) & 1.0 & 89 & 1000.0 & 156 & 5 & 2.3$\times 10^{-5}$ & 1.9$\times 10^{2}$ & 2.0$\times 10^{2}$ \\
$R_{ap}$ = 15 \arcsec
& 0.10 & 120 & 0.1 & 0.25 ( 11 ) & 1.0 & 80 & 1000.0 & 120 & 6 & 2.2$\times 10^{-5}$ & 2.0$\times 10^{2}$ & 2.2$\times 10^{2}$ \\
= 0.36 pc
& 0.10 & 200 & 0.1 & 0.33 ( 14 ) & 1.0 & 86 & 1000.0 & 197 & 4 & 2.5$\times 10^{-5}$ & 1.7$\times 10^{2}$ & 1.8$\times 10^{2}$ \\
& 0.11 & 100 & 0.1 & 0.23 ( 10 ) & 1.0 & 13 & 1000.0 & 98 & 7 & 2.0$\times 10^{-5}$ & 4.4$\times 10^{2}$ & 2.0$\times 10^{2}$ \\
& 0.13 & 60 & 0.1 & 0.18 ( 7 ) & 16.0 & 51 & 1000.0 & 7 & 76 & 2.0$\times 10^{-5}$ & 3.4$\times 10^{4}$ & 2.5$\times 10^{4}$ \\
& 0.13 & 80 & 0.1 & 0.21 ( 9 ) & 2.0 & 48 & 1000.0 & 75 & 12 & 2.7$\times 10^{-5}$ & 3.1$\times 10^{2}$ & 3.5$\times 10^{2}$ \\
& 0.16 & 100 & 0.1 & 0.23 ( 10 ) & 2.0 & 89 & 1000.0 & 97 & 11 & 2.9$\times 10^{-5}$ & 3.2$\times 10^{2}$ & 3.8$\times 10^{2}$ \\
& 0.16 & 60 & 0.1 & 0.18 ( 7 ) & 2.0 & 22 & 1000.0 & 55 & 15 & 2.5$\times 10^{-5}$ & 3.5$\times 10^{2}$ & 3.5$\times 10^{2}$ \\
& 0.17 & 160 & 0.1 & 0.29 ( 12 ) & 0.5 & 13 & 0.0 & 158 & 3 & 1.6$\times 10^{-5}$ & 1.2$\times 10^{2}$ & 9.8$\times 10^{1}$ \\
Averages & 0.12 & 113 & 0.1 & 0.25 ( 10 ) & 1.4 & 51 & 900.0 & 89 & 14 & 2.2$\times 10^{-5}$ & 3.8$\times 10^{2}$ & 3.5$\times 10^{2}$ \\
\hline\noalign{\smallskip}
C8
& 0.17 & 100 & 0.3 & 0.13 ( 5 ) & 32.0 & 74 & 997.0 & 4 & 81 & 5.0$\times 10^{-5}$ & 5.3$\times 10^{4}$ & 1.4$\times 10^{5}$ \\
$d$ = 5.0 kpc
& 0.18 & 60 & 0.1 & 0.18 ( 7 ) & 2.0 & 86 & 632.6 & 55 & 15 & 2.5$\times 10^{-5}$ & 2.6$\times 10^{2}$ & 3.5$\times 10^{2}$ \\
$R_{ap}$ = 15 \arcsec
& 0.18 & 50 & 0.1 & 0.16 ( 7 ) & 4.0 & 89 & 1000.0 & 41 & 24 & 3.2$\times 10^{-5}$ & 4.7$\times 10^{2}$ & 7.9$\times 10^{2}$ \\
= 0.36 pc
& 0.18 & 50 & 0.3 & 0.09 ( 4 ) & 0.5 & 80 & 538.5 & 48 & 6 & 2.9$\times 10^{-5}$ & 1.8$\times 10^{2}$ & 1.9$\times 10^{2}$ \\
& 0.18 & 80 & 0.1 & 0.21 ( 9 ) & 1.0 & 83 & 0.0 & 77 & 8 & 1.9$\times 10^{-5}$ & 1.7$\times 10^{2}$ & 1.9$\times 10^{2}$ \\
& 0.19 & 60 & 0.3 & 0.10 ( 4 ) & 0.5 & 77 & 1000.0 & 60 & 5 & 3.0$\times 10^{-5}$ & 1.8$\times 10^{2}$ & 1.8$\times 10^{2}$ \\
& 0.19 & 30 & 0.3 & 0.07 ( 3 ) & 1.0 & 34 & 697.7 & 28 & 13 & 3.5$\times 10^{-5}$ & 3.5$\times 10^{2}$ & 4.3$\times 10^{2}$ \\
& 0.19 & 30 & 0.3 & 0.07 ( 3 ) & 2.0 & 13 & 873.9 & 26 & 21 & 4.8$\times 10^{-5}$ & 4.0$\times 10^{3}$ & 6.2$\times 10^{2}$ \\
& 0.19 & 40 & 0.1 & 0.15 ( 6 ) & 4.0 & 44 & 219.2 & 30 & 27 & 3.0$\times 10^{-5}$ & 5.1$\times 10^{2}$ & 7.5$\times 10^{2}$ \\
& 0.19 & 50 & 0.1 & 0.16 ( 7 ) & 2.0 & 22 & 26.0 & 46 & 16 & 2.4$\times 10^{-5}$ & 3.1$\times 10^{2}$ & 3.1$\times 10^{2}$ \\
Averages & 0.18 & 51 & 0.2 & 0.13 ( 5 ) & 2.0 & 60 & 598.5 & 34 & 22 & 3.1$\times 10^{-5}$ & 6.1$\times 10^{2}$ & 6.6$\times 10^{2}$ \\
\hline\noalign{\smallskip}
D1
& 0.29 & 320 & 0.1 & 0.42 ( 15 ) & 1.0 & 22 & 1000.0 & 315 & 3 & 2.8$\times 10^{-5}$ & 2.1$\times 10^{2}$ & 2.0$\times 10^{2}$ \\
$d$ = 5.7 kpc
& 0.30 & 200 & 0.1 & 0.33 ( 12 ) & 2.0 & 34 & 1000.0 & 194 & 7 & 3.5$\times 10^{-5}$ & 3.5$\times 10^{2}$ & 3.5$\times 10^{2}$ \\
$R_{ap}$ = 15 \arcsec
& 0.31 & 240 & 0.1 & 0.36 ( 13 ) & 1.0 & 13 & 1000.0 & 240 & 4 & 2.6$\times 10^{-5}$ & 3.2$\times 10^{2}$ & 2.4$\times 10^{2}$ \\
= 0.41 pc
& 0.32 & 160 & 0.1 & 0.29 ( 11 ) & 2.0 & 48 & 1000.0 & 156 & 8 & 3.3$\times 10^{-5}$ & 4.1$\times 10^{2}$ & 4.3$\times 10^{2}$ \\
& 0.33 & 320 & 0.1 & 0.42 ( 15 ) & 2.0 & 89 & 1000.0 & 315 & 5 & 3.9$\times 10^{-5}$ & 3.6$\times 10^{2}$ & 3.8$\times 10^{2}$ \\
& 0.35 & 120 & 0.1 & 0.25 ( 9 ) & 2.0 & 22 & 1000.0 & 117 & 9 & 3.0$\times 10^{-5}$ & 4.3$\times 10^{2}$ & 4.3$\times 10^{2}$ \\
& 0.37 & 240 & 0.1 & 0.36 ( 13 ) & 2.0 & 89 & 1000.0 & 233 & 6 & 3.6$\times 10^{-5}$ & 4.8$\times 10^{2}$ & 5.2$\times 10^{2}$ \\
& 0.38 & 200 & 0.1 & 0.33 ( 12 ) & 4.0 & 89 & 1000.0 & 194 & 10 & 4.8$\times 10^{-5}$ & 5.7$\times 10^{2}$ & 6.7$\times 10^{2}$ \\
& 0.38 & 100 & 0.1 & 0.23 ( 8 ) & 4.0 & 34 & 1000.0 & 91 & 15 & 4.0$\times 10^{-5}$ & 7.8$\times 10^{2}$ & 8.8$\times 10^{2}$ \\
& 0.39 & 120 & 0.1 & 0.25 ( 9 ) & 4.0 & 89 & 1000.0 & 111 & 14 & 4.2$\times 10^{-5}$ & 7.3$\times 10^{2}$ & 9.4$\times 10^{2}$ \\
Averages & 0.34 & 188 & 0.1 & 0.32 ( 12 ) & 2.1 & 53 & 1000.0 & 181 & 8 & 3.5$\times 10^{-5}$ & 4.3$\times 10^{2}$ & 4.5$\times 10^{2}$ \\
\hline\noalign{\smallskip}
D3
& 0.61 & 400 & 0.1 & 0.47 ( 17 ) & 2.0 & 68 & 1000.0 & 391 & 4 & 4.1$\times 10^{-5}$ & 5.3$\times 10^{2}$ & 5.5$\times 10^{2}$ \\
$d$ = 5.7 kpc
& 0.62 & 480 & 0.1 & 0.51 ( 18 ) & 2.0 & 89 & 1000.0 & 477 & 4 & 4.3$\times 10^{-5}$ & 5.5$\times 10^{2}$ & 5.7$\times 10^{2}$ \\
$R_{ap}$ = 15 \arcsec
& 0.62 & 320 & 0.1 & 0.42 ( 15 ) & 4.0 & 39 & 1000.0 & 308 & 7 & 5.5$\times 10^{-5}$ & 5.3$\times 10^{2}$ & 5.4$\times 10^{2}$ \\
= 0.41 pc
& 0.75 & 320 & 0.1 & 0.42 ( 15 ) & 2.0 & 13 & 1000.0 & 315 & 5 & 3.9$\times 10^{-5}$ & 5.9$\times 10^{2}$ & 3.8$\times 10^{2}$ \\
& 0.76 & 240 & 0.1 & 0.36 ( 13 ) & 2.0 & 29 & 1000.0 & 233 & 6 & 3.6$\times 10^{-5}$ & 5.2$\times 10^{2}$ & 5.2$\times 10^{2}$ \\
& 0.78 & 200 & 0.1 & 0.33 ( 12 ) & 4.0 & 29 & 1000.0 & 194 & 10 & 4.8$\times 10^{-5}$ & 6.5$\times 10^{2}$ & 6.7$\times 10^{2}$ \\
& 0.90 & 320 & 0.3 & 0.23 ( 8 ) & 96.0 & 74 & 1000.0 & 7 & 85 & 6.6$\times 10^{-5}$ & 9.5$\times 10^{5}$ & 1.2$\times 10^{6}$ \\
& 0.91 & 160 & 0.1 & 0.29 ( 11 ) & 4.0 & 29 & 1000.0 & 151 & 12 & 4.5$\times 10^{-5}$ & 8.6$\times 10^{2}$ & 9.1$\times 10^{2}$ \\
& 0.93 & 240 & 0.1 & 0.36 ( 13 ) & 4.0 & 86 & 1000.0 & 229 & 9 & 5.1$\times 10^{-5}$ & 9.0$\times 10^{2}$ & 1.0$\times 10^{3}$ \\
& 0.95 & 200 & 0.3 & 0.19 ( 7 ) & 1.0 & 80 & 1000.0 & 196 & 4 & 5.8$\times 10^{-5}$ & 5.1$\times 10^{2}$ & 5.2$\times 10^{2}$ \\
Averages & 0.77 & 273 & 0.1 & 0.34 ( 12 ) & 3.6 & 54 & 1000.0 & 180 & 15 & 4.8$\times 10^{-5}$ & 1.3$\times 10^{3}$ & 1.3$\times 10^{3}$ \\
\hline\noalign{\smallskip}
D5
& 0.32 & 320 & 0.1 & 0.42 ( 15 ) & 4.0 & 39 & 1000.0 & 308 & 7 & 5.5$\times 10^{-5}$ & 5.3$\times 10^{2}$ & 5.4$\times 10^{2}$ \\
$d$ = 5.7 kpc
& 0.33 & 400 & 0.1 & 0.47 ( 17 ) & 2.0 & 83 & 1000.0 & 391 & 4 & 4.1$\times 10^{-5}$ & 5.3$\times 10^{2}$ & 5.5$\times 10^{2}$ \\
$R_{ap}$ = 11 \arcsec
& 0.40 & 480 & 0.1 & 0.51 ( 18 ) & 2.0 & 89 & 1000.0 & 477 & 4 & 4.3$\times 10^{-5}$ & 5.5$\times 10^{2}$ & 5.7$\times 10^{2}$ \\
= 0.30 pc
& 0.45 & 200 & 0.1 & 0.33 ( 12 ) & 4.0 & 13 & 1000.0 & 194 & 10 & 4.8$\times 10^{-5}$ & 1.9$\times 10^{3}$ & 6.7$\times 10^{2}$ \\
& 0.46 & 240 & 0.1 & 0.36 ( 13 ) & 2.0 & 29 & 1000.0 & 233 & 6 & 3.6$\times 10^{-5}$ & 5.2$\times 10^{2}$ & 5.2$\times 10^{2}$ \\
& 0.47 & 320 & 0.1 & 0.42 ( 15 ) & 2.0 & 13 & 775.8 & 315 & 5 & 3.9$\times 10^{-5}$ & 5.9$\times 10^{2}$ & 3.8$\times 10^{2}$ \\
& 0.51 & 200 & 0.3 & 0.19 ( 7 ) & 1.0 & 71 & 1000.0 & 196 & 4 & 5.8$\times 10^{-5}$ & 5.1$\times 10^{2}$ & 5.2$\times 10^{2}$ \\
& 0.51 & 160 & 0.1 & 0.29 ( 11 ) & 4.0 & 22 & 1000.0 & 151 & 12 & 4.5$\times 10^{-5}$ & 9.0$\times 10^{2}$ & 9.1$\times 10^{2}$ \\
& 0.54 & 240 & 0.1 & 0.36 ( 13 ) & 4.0 & 89 & 1000.0 & 229 & 9 & 5.1$\times 10^{-5}$ & 9.0$\times 10^{2}$ & 1.0$\times 10^{3}$ \\
& 0.55 & 100 & 0.3 & 0.13 ( 5 ) & 32.0 & 13 & 1000.0 & 4 & 81 & 5.0$\times 10^{-5}$ & 3.0$\times 10^{5}$ & 1.4$\times 10^{5}$ \\
Averages & 0.45 & 243 & 0.1 & 0.32 ( 12 ) & 3.2 & 46 & 977.6 & 174 & 14 & 4.6$\times 10^{-5}$ & 1.3$\times 10^{3}$ & 1.0$\times 10^{3}$ \\
\hline\noalign{\smallskip}
D6
& 0.03 & 320 & 0.1 & 0.42 ( 15 ) & 2.0 & 13 & 240.2 & 315 & 5 & 3.9$\times 10^{-5}$ & 5.9$\times 10^{2}$ & 3.8$\times 10^{2}$ \\
$d$ = 5.7 kpc
& 0.03 & 200 & 0.1 & 0.33 ( 12 ) & 2.0 & 22 & 200.2 & 194 & 7 & 3.5$\times 10^{-5}$ & 3.6$\times 10^{2}$ & 3.5$\times 10^{2}$ \\
$R_{ap}$ = 15 \arcsec
& 0.03 & 160 & 0.1 & 0.29 ( 11 ) & 2.0 & 13 & 294.3 & 156 & 8 & 3.3$\times 10^{-5}$ & 1.0$\times 10^{3}$ & 4.3$\times 10^{2}$ \\
= 0.41 pc
& 0.03 & 320 & 0.3 & 0.23 ( 8 ) & 96.0 & 83 & 521.5 & 7 & 85 & 6.6$\times 10^{-5}$ & 1.5$\times 10^{4}$ & 1.2$\times 10^{6}$ \\
& 0.03 & 120 & 0.1 & 0.25 ( 9 ) & 2.0 & 22 & 239.2 & 117 & 9 & 3.0$\times 10^{-5}$ & 4.3$\times 10^{2}$ & 4.3$\times 10^{2}$ \\
& 0.04 & 240 & 0.1 & 0.36 ( 13 ) & 1.0 & 22 & 130.1 & 240 & 4 & 2.6$\times 10^{-5}$ & 2.4$\times 10^{2}$ & 2.4$\times 10^{2}$ \\
& 0.04 & 100 & 0.3 & 0.13 ( 5 ) & 32.0 & 55 & 495.5 & 4 & 81 & 5.0$\times 10^{-5}$ & 1.7$\times 10^{5}$ & 1.4$\times 10^{5}$ \\
& 0.04 & 100 & 0.3 & 0.13 ( 5 ) & 0.5 & 13 & 43.0 & 99 & 4 & 3.5$\times 10^{-5}$ & 3.1$\times 10^{2}$ & 2.4$\times 10^{2}$ \\
& 0.04 & 320 & 0.1 & 0.42 ( 15 ) & 1.0 & 22 & 95.1 & 315 & 3 & 2.8$\times 10^{-5}$ & 2.1$\times 10^{2}$ & 2.0$\times 10^{2}$ \\
& 0.04 & 80 & 0.1 & 0.21 ( 8 ) & 4.0 & 13 & 428.4 & 71 & 18 & 3.7$\times 10^{-5}$ & 4.4$\times 10^{3}$ & 8.5$\times 10^{2}$ \\
Averages & 0.03 & 173 & 0.1 & 0.26 ( 9 ) & 3.2 & 28 & 268.8 & 84 & 22 & 3.6$\times 10^{-5}$ & 1.3$\times 10^{3}$ & 1.5$\times 10^{3}$ \\
\hline\noalign{\smallskip}
D8
& 0.07 & 240 & 0.1 & 0.36 ( 13 ) & 2.0 & 80 & 1000.0 & 233 & 6 & 3.6$\times 10^{-5}$ & 4.8$\times 10^{2}$ & 5.2$\times 10^{2}$ \\
$d$ = 5.7 kpc
& 0.07 & 100 & 0.3 & 0.13 ( 5 ) & 1.0 & 89 & 1000.0 & 97 & 6 & 4.9$\times 10^{-5}$ & 4.2$\times 10^{2}$ & 4.4$\times 10^{2}$ \\
$R_{ap}$ = 11 \arcsec
& 0.07 & 320 & 0.1 & 0.42 ( 15 ) & 4.0 & 89 & 1000.0 & 308 & 7 & 5.5$\times 10^{-5}$ & 4.9$\times 10^{2}$ & 5.4$\times 10^{2}$ \\
= 0.30 pc
& 0.08 & 200 & 0.1 & 0.33 ( 12 ) & 4.0 & 89 & 1000.0 & 194 & 10 & 4.8$\times 10^{-5}$ & 5.7$\times 10^{2}$ & 6.7$\times 10^{2}$ \\
& 0.08 & 60 & 0.3 & 0.10 ( 4 ) & 1.0 & 62 & 1000.0 & 58 & 8 & 4.3$\times 10^{-5}$ & 4.2$\times 10^{2}$ & 4.7$\times 10^{2}$ \\
& 0.09 & 30 & 1.0 & 0.04 ( 1 ) & 0.5 & 80 & 1000.0 & 29 & 8 & 6.0$\times 10^{-5}$ & 3.8$\times 10^{2}$ & 4.2$\times 10^{2}$ \\
& 0.09 & 80 & 0.3 & 0.12 ( 4 ) & 1.0 & 34 & 936.9 & 79 & 7 & 4.6$\times 10^{-5}$ & 3.3$\times 10^{2}$ & 3.5$\times 10^{2}$ \\
& 0.10 & 100 & 0.1 & 0.23 ( 8 ) & 4.0 & 77 & 1000.0 & 91 & 15 & 4.0$\times 10^{-5}$ & 6.6$\times 10^{2}$ & 8.8$\times 10^{2}$ \\
& 0.11 & 20 & 1.0 & 0.03 ( 1 ) & 0.5 & 71 & 1000.0 & 19 & 10 & 5.4$\times 10^{-5}$ & 3.7$\times 10^{2}$ & 4.5$\times 10^{2}$ \\
& 0.12 & 40 & 1.0 & 0.05 ( 2 ) & 0.5 & 86 & 1000.0 & 39 & 6 & 6.5$\times 10^{-5}$ & 3.8$\times 10^{2}$ & 4.0$\times 10^{2}$ \\
Averages & 0.09 & 84 & 0.3 & 0.13 ( 5 ) & 1.3 & 76 & 993.7 & 81 & 8 & 4.9$\times 10^{-5}$ & 4.4$\times 10^{2}$ & 4.9$\times 10^{2}$ \\
\hline\noalign{\smallskip}
D9
& 0.24 & 200 & 0.1 & 0.33 ( 12 ) & 0.5 & 13 & 189.2 & 200 & 3 & 1.7$\times 10^{-5}$ & 1.5$\times 10^{2}$ & 1.3$\times 10^{2}$ \\
$d$ = 5.7 kpc
& 0.29 & 100 & 0.1 & 0.23 ( 8 ) & 1.0 & 13 & 453.5 & 98 & 7 & 2.0$\times 10^{-5}$ & 4.4$\times 10^{2}$ & 2.0$\times 10^{2}$ \\
$R_{ap}$ = 12 \arcsec
& 0.30 & 80 & 0.1 & 0.21 ( 8 ) & 1.0 & 13 & 354.4 & 77 & 8 & 1.9$\times 10^{-5}$ & 4.6$\times 10^{2}$ & 1.9$\times 10^{2}$ \\
= 0.33 pc
& 0.33 & 200 & 0.1 & 0.33 ( 12 ) & 1.0 & 89 & 479.5 & 197 & 4 & 2.5$\times 10^{-5}$ & 1.7$\times 10^{2}$ & 1.8$\times 10^{2}$ \\
& 0.35 & 160 & 0.1 & 0.29 ( 11 ) & 1.0 & 89 & 597.6 & 156 & 5 & 2.3$\times 10^{-5}$ & 1.9$\times 10^{2}$ & 2.0$\times 10^{2}$ \\
& 0.36 & 120 & 0.1 & 0.25 ( 9 ) & 1.0 & 13 & 698.7 & 120 & 6 & 2.2$\times 10^{-5}$ & 4.3$\times 10^{2}$ & 2.2$\times 10^{2}$ \\
& 0.36 & 60 & 0.1 & 0.18 ( 7 ) & 1.0 & 13 & 344.3 & 57 & 10 & 1.8$\times 10^{-5}$ & 6.1$\times 10^{2}$ & 2.0$\times 10^{2}$ \\
& 0.38 & 50 & 0.1 & 0.16 ( 6 ) & 2.0 & 13 & 508.5 & 46 & 16 & 2.4$\times 10^{-5}$ & 1.6$\times 10^{3}$ & 3.1$\times 10^{2}$ \\
& 0.42 & 60 & 0.1 & 0.18 ( 7 ) & 2.0 & 13 & 849.8 & 55 & 15 & 2.5$\times 10^{-5}$ & 1.7$\times 10^{3}$ & 3.5$\times 10^{2}$ \\
& 0.45 & 80 & 0.1 & 0.21 ( 8 ) & 2.0 & 89 & 1000.0 & 75 & 12 & 2.7$\times 10^{-5}$ & 2.8$\times 10^{2}$ & 3.5$\times 10^{2}$ \\
Averages & 0.34 & 99 & 0.1 & 0.23 ( 8 ) & 1.1 & 36 & 547.5 & 95 & 9 & 2.2$\times 10^{-5}$ & 4.3$\times 10^{2}$ & 2.2$\times 10^{2}$ \\
\hline\noalign{\smallskip}
E1
& 0.21 & 30 & 0.1 & 0.13 ( 5 ) & 2.0 & 86 & 466.5 & 25 & 23 & 2.0$\times 10^{-5}$ & 1.4$\times 10^{2}$ & 2.4$\times 10^{2}$ \\
$d$ = 5.1 kpc
& 0.21 & 30 & 0.1 & 0.13 ( 5 ) & 1.0 & 86 & 599.6 & 27 & 15 & 1.5$\times 10^{-5}$ & 1.3$\times 10^{2}$ & 1.7$\times 10^{2}$ \\
$R_{ap}$ = 15 \arcsec
& 0.22 & 80 & 1.0 & 0.07 ( 3 ) & 32.0 & 83 & 709.7 & 3 & 79 & 1.4$\times 10^{-4}$ & 2.1$\times 10^{3}$ & 1.6$\times 10^{5}$ \\
= 0.37 pc
& 0.22 & 20 & 0.3 & 0.06 ( 2 ) & 4.0 & 89 & 1000.0 & 11 & 38 & 5.4$\times 10^{-5}$ & 3.1$\times 10^{2}$ & 1.1$\times 10^{3}$ \\
& 0.22 & 30 & 0.1 & 0.13 ( 5 ) & 8.0 & 86 & 1000.0 & 9 & 57 & 2.6$\times 10^{-5}$ & 7.5$\times 10^{2}$ & 6.3$\times 10^{3}$ \\
& 0.23 & 60 & 0.1 & 0.18 ( 7 ) & 16.0 & 83 & 348.3 & 7 & 76 & 2.0$\times 10^{-5}$ & 6.5$\times 10^{2}$ & 2.5$\times 10^{4}$ \\
& 0.23 & 40 & 0.1 & 0.15 ( 6 ) & 0.5 & 58 & 0.0 & 39 & 8 & 1.1$\times 10^{-5}$ & 8.1$\times 10^{1}$ & 8.8$\times 10^{1}$ \\
& 0.23 & 50 & 0.1 & 0.16 ( 7 ) & 0.5 & 77 & 230.2 & 49 & 7 & 1.2$\times 10^{-5}$ & 7.9$\times 10^{1}$ & 8.7$\times 10^{1}$ \\
& 0.24 & 40 & 0.1 & 0.15 ( 6 ) & 1.0 & 89 & 1000.0 & 38 & 12 & 1.6$\times 10^{-5}$ & 1.3$\times 10^{2}$ & 1.7$\times 10^{2}$ \\
& 0.24 & 20 & 0.3 & 0.06 ( 2 ) & 2.0 & 89 & 1000.0 & 15 & 27 & 4.2$\times 10^{-5}$ & 2.2$\times 10^{2}$ & 4.8$\times 10^{2}$ \\
Averages & 0.23 & 36 & 0.2 & 0.11 ( 5 ) & 2.6 & 82 & 635.4 & 16 & 34 & 2.6$\times 10^{-5}$ & 2.5$\times 10^{2}$ & 9.6$\times 10^{2}$ \\
\hline\noalign{\smallskip}
E2
& 0.13 & 30 & 0.3 & 0.07 ( 3 ) & 2.0 & 86 & 1000.0 & 26 & 21 & 4.8$\times 10^{-5}$ & 3.7$\times 10^{2}$ & 6.2$\times 10^{2}$ \\
$d$ = 5.1 kpc
& 0.14 & 30 & 0.3 & 0.07 ( 3 ) & 1.0 & 86 & 1000.0 & 28 & 13 & 3.5$\times 10^{-5}$ & 3.3$\times 10^{2}$ & 4.3$\times 10^{2}$ \\
$R_{ap}$ = 15 \arcsec
& 0.14 & 40 & 0.3 & 0.08 ( 3 ) & 0.5 & 22 & 886.9 & 39 & 7 & 2.7$\times 10^{-5}$ & 1.8$\times 10^{2}$ & 1.9$\times 10^{2}$ \\
= 0.37 pc
& 0.14 & 40 & 0.1 & 0.15 ( 6 ) & 4.0 & 71 & 933.9 & 30 & 27 & 3.0$\times 10^{-5}$ & 4.2$\times 10^{2}$ & 7.5$\times 10^{2}$ \\
& 0.15 & 30 & 0.3 & 0.07 ( 3 ) & 4.0 & 77 & 1000.0 & 21 & 29 & 6.4$\times 10^{-5}$ & 5.4$\times 10^{2}$ & 1.2$\times 10^{3}$ \\
& 0.15 & 50 & 0.3 & 0.09 ( 4 ) & 0.5 & 80 & 1000.0 & 48 & 6 & 2.9$\times 10^{-5}$ & 1.8$\times 10^{2}$ & 1.9$\times 10^{2}$ \\
& 0.15 & 50 & 0.1 & 0.16 ( 7 ) & 2.0 & 77 & 295.3 & 46 & 16 & 2.4$\times 10^{-5}$ & 2.2$\times 10^{2}$ & 3.1$\times 10^{2}$ \\
& 0.16 & 60 & 0.1 & 0.18 ( 7 ) & 2.0 & 86 & 1000.0 & 55 & 15 & 2.5$\times 10^{-5}$ & 2.6$\times 10^{2}$ & 3.5$\times 10^{2}$ \\
& 0.17 & 60 & 0.1 & 0.18 ( 7 ) & 1.0 & 86 & 113.1 & 57 & 10 & 1.8$\times 10^{-5}$ & 1.8$\times 10^{2}$ & 2.0$\times 10^{2}$ \\
& 0.17 & 50 & 0.1 & 0.16 ( 7 ) & 12.0 & 80 & 1000.0 & 15 & 59 & 3.4$\times 10^{-5}$ & 1.5$\times 10^{3}$ & 1.4$\times 10^{4}$ \\
Averages & 0.15 & 43 & 0.2 & 0.11 ( 5 ) & 1.8 & 75 & 822.9 & 34 & 20 & 3.1$\times 10^{-5}$ & 3.3$\times 10^{2}$ & 5.5$\times 10^{2}$ \\
\hline\noalign{\smallskip}
F3
& 0.26 & 10 & 0.3 & 0.04 ( 2 ) & 4.0 & 77 & 270.3 & 1 & 68 & 2.4$\times 10^{-5}$ & 4.9$\times 10^{1}$ & 6.7$\times 10^{2}$ \\
$d$ = 3.7 kpc
& 0.27 & 30 & 0.3 & 0.07 ( 4 ) & 12.0 & 89 & 304.3 & 1 & 81 & 2.2$\times 10^{-5}$ & 7.0$\times 10^{1}$ & 1.2$\times 10^{4}$ \\
$R_{ap}$ = 6 \arcsec
& 0.29 & 10 & 0.1 & 0.07 ( 4 ) & 2.0 & 86 & 156.2 & 4 & 50 & 1.1$\times 10^{-5}$ & 2.0$\times 10^{1}$ & 1.3$\times 10^{2}$ \\
= 0.11 pc
& 0.32 & 40 & 0.1 & 0.15 ( 8 ) & 12.0 & 89 & 276.3 & 2 & 82 & 9.5$\times 10^{-6}$ & 5.7$\times 10^{1}$ & 1.1$\times 10^{4}$ \\
& 0.90 & 10 & 0.1 & 0.07 ( 4 ) & 1.0 & 86 & 239.2 & 7 & 31 & 1.0$\times 10^{-5}$ & 4.4$\times 10^{1}$ & 1.1$\times 10^{2}$ \\
& 1.16 & 10 & 0.1 & 0.07 ( 4 ) & 0.5 & 89 & 235.2 & 9 & 20 & 7.8$\times 10^{-6}$ & 4.6$\times 10^{1}$ & 7.5$\times 10^{1}$ \\
& 1.92 & 10 & 0.3 & 0.04 ( 2 ) & 2.0 & 44 & 499.5 & 5 & 43 & 3.0$\times 10^{-5}$ & 3.3$\times 10^{2}$ & 2.8$\times 10^{2}$ \\
& 3.04 & 20 & 0.1 & 0.10 ( 6 ) & 0.5 & 13 & 479.5 & 19 & 13 & 9.6$\times 10^{-6}$ & 3.9$\times 10^{2}$ & 9.0$\times 10^{1}$ \\
& 3.07 & 20 & 0.1 & 0.10 ( 6 ) & 2.0 & 89 & 456.5 & 15 & 30 & 1.7$\times 10^{-5}$ & 8.7$\times 10^{1}$ & 1.9$\times 10^{2}$ \\
& 3.18 & 10 & 1.0 & 0.02 ( 1 ) & 4.0 & 89 & 629.6 & 1 & 59 & 7.7$\times 10^{-5}$ & 1.1$\times 10^{2}$ & 1.1$\times 10^{3}$ \\
Averages & 0.54 & 14 & 0.2 & 0.07 ( 4 ) & 2.7 & 80 & 283.0 & 3 & 54 & 1.5$\times 10^{-5}$ & 6.0$\times 10^{1}$ & 6.0$\times 10^{2}$ \\
\hline\noalign{\smallskip}
F4
& 0.00 & 40 & 0.1 & 0.15 ( 8 ) & 12.0 & 13 & 955.0 & 2 & 82 & 9.5$\times 10^{-6}$ & 2.1$\times 10^{4}$ & 1.1$\times 10^{4}$ \\
$d$ = 3.7 kpc
& 0.01 & 80 & 0.1 & 0.21 ( 12 ) & 0.5 & 89 & 1000.0 & 79 & 5 & 1.4$\times 10^{-5}$ & 8.6$\times 10^{1}$ & 9.2$\times 10^{1}$ \\
$R_{ap}$ = 15 \arcsec
& 0.01 & 120 & 0.1 & 0.25 ( 14 ) & 0.5 & 89 & 1000.0 & 118 & 4 & 1.5$\times 10^{-5}$ & 8.4$\times 10^{1}$ & 8.8$\times 10^{1}$ \\
= 0.27 pc
& 0.01 & 100 & 0.1 & 0.23 ( 13 ) & 0.5 & 86 & 1000.0 & 99 & 4 & 1.5$\times 10^{-5}$ & 8.7$\times 10^{1}$ & 9.1$\times 10^{1}$ \\
& 0.01 & 60 & 0.1 & 0.18 ( 10 ) & 0.5 & 65 & 1000.0 & 59 & 6 & 1.3$\times 10^{-5}$ & 8.1$\times 10^{1}$ & 8.7$\times 10^{1}$ \\
& 0.01 & 50 & 0.1 & 0.16 ( 9 ) & 0.5 & 34 & 1000.0 & 49 & 7 & 1.2$\times 10^{-5}$ & 8.5$\times 10^{1}$ & 8.7$\times 10^{1}$ \\
& 0.01 & 40 & 0.1 & 0.15 ( 8 ) & 0.5 & 22 & 1000.0 & 39 & 8 & 1.1$\times 10^{-5}$ & 8.7$\times 10^{1}$ & 8.8$\times 10^{1}$ \\
& 0.01 & 30 & 0.3 & 0.07 ( 4 ) & 12.0 & 13 & 789.8 & 1 & 81 & 2.2$\times 10^{-5}$ & 2.7$\times 10^{4}$ & 1.2$\times 10^{4}$ \\
& 0.01 & 30 & 0.1 & 0.13 ( 7 ) & 0.5 & 13 & 960.0 & 29 & 10 & 1.1$\times 10^{-5}$ & 2.9$\times 10^{2}$ & 9.0$\times 10^{1}$ \\
& 0.01 & 30 & 0.1 & 0.13 ( 7 ) & 1.0 & 77 & 1000.0 & 27 & 15 & 1.5$\times 10^{-5}$ & 1.3$\times 10^{2}$ & 1.7$\times 10^{2}$ \\
Averages & 0.01 & 51 & 0.1 & 0.16 ( 9 ) & 1.0 & 50 & 970.5 & 26 & 22 & 1.3$\times 10^{-5}$ & 3.1$\times 10^{2}$ & 2.5$\times 10^{2}$ \\
\hline\noalign{\smallskip}
H1
& 0.05 & 40 & 0.1 & 0.15 ( 10 ) & 12.0 & 71 & 182.2 & 2 & 82 & 9.5$\times 10^{-6}$ & 6.9$\times 10^{3}$ & 1.1$\times 10^{4}$ \\
$d$ = 2.9 kpc
& 0.05 & 20 & 0.1 & 0.10 ( 7 ) & 2.0 & 83 & 1000.0 & 15 & 30 & 1.7$\times 10^{-5}$ & 8.7$\times 10^{1}$ & 1.9$\times 10^{2}$ \\
$R_{ap}$ = 15 \arcsec
& 0.05 & 30 & 0.3 & 0.07 ( 5 ) & 12.0 & 62 & 201.2 & 1 & 81 & 2.2$\times 10^{-5}$ & 1.2$\times 10^{4}$ & 1.2$\times 10^{4}$ \\
= 0.21 pc
& 0.05 & 20 & 0.3 & 0.06 ( 4 ) & 8.0 & 80 & 1000.0 & 2 & 66 & 4.4$\times 10^{-5}$ & 7.7$\times 10^{2}$ & 9.9$\times 10^{3}$ \\
& 0.05 & 10 & 0.3 & 0.04 ( 3 ) & 2.0 & 13 & 160.2 & 5 & 43 & 3.0$\times 10^{-5}$ & 1.1$\times 10^{3}$ & 2.8$\times 10^{2}$ \\
& 0.05 & 20 & 0.1 & 0.10 ( 7 ) & 0.5 & 89 & 1000.0 & 19 & 13 & 9.6$\times 10^{-6}$ & 7.0$\times 10^{1}$ & 9.0$\times 10^{1}$ \\
& 0.05 & 10 & 0.1 & 0.07 ( 5 ) & 0.5 & 13 & 131.1 & 9 & 20 & 7.8$\times 10^{-6}$ & 4.6$\times 10^{2}$ & 7.5$\times 10^{1}$ \\
& 0.06 & 10 & 0.3 & 0.04 ( 3 ) & 1.0 & 89 & 1000.0 & 8 & 28 & 2.5$\times 10^{-5}$ & 1.1$\times 10^{2}$ & 2.6$\times 10^{2}$ \\
& 0.06 & 10 & 1.0 & 0.02 ( 2 ) & 4.0 & 13 & 380.4 & 1 & 59 & 7.7$\times 10^{-5}$ & 3.4$\times 10^{3}$ & 1.1$\times 10^{3}$ \\
& 0.06 & 10 & 3.2 & 0.01 ( 1 ) & 4.0 & 48 & 1000.0 & 2 & 56 & 1.9$\times 10^{-4}$ & 2.3$\times 10^{3}$ & 1.9$\times 10^{3}$ \\
Averages & 0.05 & 16 & 0.3 & 0.06 ( 4 ) & 2.7 & 56 & 605.5 & 4 & 48 & 2.6$\times 10^{-5}$ & 8.3$\times 10^{2}$ & 8.8$\times 10^{2}$ \\
\hline\noalign{\smallskip}
H2
& 1.44 & 10 & 0.1 & 0.07 ( 5 ) & 2.0 & 13 & 269.3 & 4 & 50 & 1.1$\times 10^{-5}$ & 3.9$\times 10^{2}$ & 1.3$\times 10^{2}$ \\
$d$ = 2.9 kpc
& 1.46 & 40 & 0.1 & 0.15 ( 10 ) & 12.0 & 80 & 368.4 & 2 & 82 & 9.5$\times 10^{-6}$ & 6.7$\times 10^{2}$ & 1.1$\times 10^{4}$ \\
$R_{ap}$ = 10 \arcsec
& 1.74 & 10 & 0.3 & 0.04 ( 3 ) & 4.0 & 34 & 353.4 & 1 & 68 & 2.4$\times 10^{-5}$ & 1.4$\times 10^{3}$ & 6.7$\times 10^{2}$ \\
= 0.14 pc
& 1.74 & 10 & 0.1 & 0.07 ( 5 ) & 1.0 & 29 & 354.4 & 7 & 31 & 1.0$\times 10^{-5}$ & 3.9$\times 10^{2}$ & 1.1$\times 10^{2}$ \\
& 1.83 & 30 & 0.3 & 0.07 ( 5 ) & 12.0 & 77 & 389.4 & 1 & 81 & 2.2$\times 10^{-5}$ & 9.0$\times 10^{2}$ & 1.2$\times 10^{4}$ \\
& 1.87 & 10 & 0.1 & 0.07 ( 5 ) & 0.5 & 13 & 380.4 & 9 & 20 & 7.8$\times 10^{-6}$ & 4.6$\times 10^{2}$ & 7.5$\times 10^{1}$ \\
& 2.99 & 10 & 0.3 & 0.04 ( 3 ) & 2.0 & 44 & 396.4 & 5 & 43 & 3.0$\times 10^{-5}$ & 3.3$\times 10^{2}$ & 2.8$\times 10^{2}$ \\
& 3.50 & 20 & 0.1 & 0.10 ( 7 ) & 0.5 & 13 & 411.4 & 19 & 13 & 9.6$\times 10^{-6}$ & 3.9$\times 10^{2}$ & 9.0$\times 10^{1}$ \\
& 3.68 & 20 & 0.1 & 0.10 ( 7 ) & 2.0 & 89 & 329.3 & 15 & 30 & 1.7$\times 10^{-5}$ & 8.7$\times 10^{1}$ & 1.9$\times 10^{2}$ \\
& 4.08 & 10 & 1.0 & 0.02 ( 2 ) & 4.0 & 58 & 497.5 & 1 & 59 & 7.7$\times 10^{-5}$ & 8.2$\times 10^{2}$ & 1.1$\times 10^{3}$ \\
Averages & 1.67 & 15 & 0.1 & 0.07 ( 5 ) & 2.9 & 41 & 352.5 & 3 & 55 & 1.3$\times 10^{-5}$ & 6.3$\times 10^{2}$ & 6.8$\times 10^{2}$ \\
\hline\noalign{\smallskip}
H3
& 1.19 & 10 & 0.1 & 0.07 ( 5 ) & 2.0 & 51 & 181.2 & 4 & 50 & 1.1$\times 10^{-5}$ & 8.1$\times 10^{1}$ & 1.3$\times 10^{2}$ \\
$d$ = 2.9 kpc
& 1.26 & 40 & 0.1 & 0.15 ( 10 ) & 12.0 & 89 & 196.2 & 2 & 82 & 9.5$\times 10^{-6}$ & 5.7$\times 10^{1}$ & 1.1$\times 10^{4}$ \\
$R_{ap}$ = 11 \arcsec
& 1.47 & 30 & 0.3 & 0.07 ( 5 ) & 12.0 & 89 & 225.2 & 1 & 81 & 2.2$\times 10^{-5}$ & 7.0$\times 10^{1}$ & 1.2$\times 10^{4}$ \\
= 0.15 pc
& 1.56 & 10 & 0.3 & 0.04 ( 3 ) & 4.0 & 80 & 179.2 & 1 & 68 & 2.4$\times 10^{-5}$ & 4.0$\times 10^{1}$ & 6.7$\times 10^{2}$ \\
& 2.52 & 10 & 0.1 & 0.07 ( 5 ) & 1.0 & 29 & 286.3 & 7 & 31 & 1.0$\times 10^{-5}$ & 3.9$\times 10^{2}$ & 1.1$\times 10^{2}$ \\
& 2.76 & 10 & 0.1 & 0.07 ( 5 ) & 0.5 & 13 & 308.3 & 9 & 20 & 7.8$\times 10^{-6}$ & 4.6$\times 10^{2}$ & 7.5$\times 10^{1}$ \\
& 4.52 & 10 & 0.3 & 0.04 ( 3 ) & 2.0 & 44 & 310.3 & 5 & 43 & 3.0$\times 10^{-5}$ & 3.3$\times 10^{2}$ & 2.8$\times 10^{2}$ \\
& 4.64 & 20 & 0.1 & 0.10 ( 7 ) & 0.5 & 13 & 321.3 & 19 & 13 & 9.6$\times 10^{-6}$ & 3.9$\times 10^{2}$ & 9.0$\times 10^{1}$ \\
& 5.15 & 20 & 0.1 & 0.10 ( 7 ) & 2.0 & 89 & 236.2 & 15 & 30 & 1.7$\times 10^{-5}$ & 8.7$\times 10^{1}$ & 1.9$\times 10^{2}$ \\
& 5.46 & 30 & 0.1 & 0.13 ( 9 ) & 0.5 & 13 & 301.3 & 29 & 10 & 1.1$\times 10^{-5}$ & 2.9$\times 10^{2}$ & 9.0$\times 10^{1}$ \\
Averages & 1.36 & 19 & 0.2 & 0.08 ( 5 ) & 5.8 & 77 & 195.4 & 1 & 70 & 1.6$\times 10^{-5}$ & 6.0$\times 10^{1}$ & 1.8$\times 10^{3}$ \\
\hline\noalign{\smallskip}
H4
& 0.09 & 40 & 0.1 & 0.15 ( 10 ) & 12.0 & 71 & 929.9 & 2 & 82 & 9.5$\times 10^{-6}$ & 6.9$\times 10^{3}$ & 1.1$\times 10^{4}$ \\
$d$ = 2.9 kpc
& 0.11 & 30 & 0.3 & 0.07 ( 5 ) & 12.0 & 65 & 1000.0 & 1 & 81 & 2.2$\times 10^{-5}$ & 1.0$\times 10^{4}$ & 1.2$\times 10^{4}$ \\
$R_{ap}$ = 10 \arcsec
& 0.12 & 10 & 0.1 & 0.07 ( 5 ) & 0.5 & 29 & 1000.0 & 9 & 20 & 7.8$\times 10^{-6}$ & 6.1$\times 10^{1}$ & 7.5$\times 10^{1}$ \\
= 0.14 pc
& 0.14 & 10 & 0.1 & 0.07 ( 5 ) & 1.0 & 13 & 981.0 & 7 & 31 & 1.0$\times 10^{-5}$ & 5.0$\times 10^{2}$ & 1.1$\times 10^{2}$ \\
& 0.20 & 10 & 0.3 & 0.04 ( 3 ) & 4.0 & 13 & 334.3 & 1 & 68 & 2.4$\times 10^{-5}$ & 1.6$\times 10^{3}$ & 6.7$\times 10^{2}$ \\
& 0.30 & 10 & 0.3 & 0.04 ( 3 ) & 2.0 & 89 & 1000.0 & 5 & 43 & 3.0$\times 10^{-5}$ & 5.8$\times 10^{1}$ & 2.8$\times 10^{2}$ \\
& 0.43 & 10 & 0.1 & 0.07 ( 5 ) & 2.0 & 13 & 138.1 & 4 & 50 & 1.1$\times 10^{-5}$ & 3.9$\times 10^{2}$ & 1.3$\times 10^{2}$ \\
& 0.50 & 10 & 1.0 & 0.02 ( 2 ) & 4.0 & 71 & 1000.0 & 1 & 59 & 7.7$\times 10^{-5}$ & 1.4$\times 10^{2}$ & 1.1$\times 10^{3}$ \\
& 0.66 & 20 & 0.1 & 0.10 ( 7 ) & 0.5 & 89 & 1000.0 & 19 & 13 & 9.6$\times 10^{-6}$ & 7.0$\times 10^{1}$ & 9.0$\times 10^{1}$ \\
& 0.68 & 20 & 0.1 & 0.10 ( 7 ) & 2.0 & 89 & 1000.0 & 15 & 30 & 1.7$\times 10^{-5}$ & 8.7$\times 10^{1}$ & 1.9$\times 10^{2}$ \\
Averages & 0.25 & 15 & 0.2 & 0.07 ( 5 ) & 2.3 & 54 & 838.3 & 4 & 48 & 1.7$\times 10^{-5}$ & 3.8$\times 10^{2}$ & 4.7$\times 10^{2}$ \\
\hline\noalign{\smallskip}
H5
& 1.19 & 40 & 0.1 & 0.15 ( 10 ) & 12.0 & 65 & 292.3 & 2 & 82 & 9.5$\times 10^{-6}$ & 9.5$\times 10^{3}$ & 1.1$\times 10^{4}$ \\
$d$ = 2.9 kpc
& 1.37 & 10 & 0.1 & 0.07 ( 5 ) & 0.5 & 13 & 241.2 & 9 & 20 & 7.8$\times 10^{-6}$ & 4.6$\times 10^{2}$ & 7.5$\times 10^{1}$ \\
$R_{ap}$ = 11 \arcsec
& 1.51 & 10 & 0.1 & 0.07 ( 5 ) & 1.0 & 29 & 218.2 & 7 & 31 & 1.0$\times 10^{-5}$ & 3.9$\times 10^{2}$ & 1.1$\times 10^{2}$ \\
= 0.15 pc
& 1.53 & 30 & 0.3 & 0.07 ( 5 ) & 12.0 & 62 & 303.3 & 1 & 81 & 2.2$\times 10^{-5}$ & 1.2$\times 10^{4}$ & 1.2$\times 10^{4}$ \\
& 2.36 & 10 & 0.3 & 0.04 ( 3 ) & 4.0 & 13 & 235.2 & 1 & 68 & 2.4$\times 10^{-5}$ & 1.6$\times 10^{3}$ & 6.7$\times 10^{2}$ \\
& 2.37 & 10 & 0.3 & 0.04 ( 3 ) & 2.0 & 44 & 240.2 & 5 & 43 & 3.0$\times 10^{-5}$ & 3.3$\times 10^{2}$ & 2.8$\times 10^{2}$ \\
& 3.10 & 20 & 0.1 & 0.10 ( 7 ) & 0.5 & 13 & 256.3 & 19 & 13 & 9.6$\times 10^{-6}$ & 3.9$\times 10^{2}$ & 9.0$\times 10^{1}$ \\
& 3.24 & 20 & 0.1 & 0.10 ( 7 ) & 2.0 & 89 & 168.2 & 15 & 30 & 1.7$\times 10^{-5}$ & 8.7$\times 10^{1}$ & 1.9$\times 10^{2}$ \\
& 3.37 & 10 & 1.0 & 0.02 ( 2 ) & 4.0 & 58 & 322.3 & 1 & 59 & 7.7$\times 10^{-5}$ & 8.2$\times 10^{2}$ & 1.1$\times 10^{3}$ \\
& 3.92 & 10 & 0.1 & 0.07 ( 5 ) & 2.0 & 13 & 155.2 & 4 & 50 & 1.1$\times 10^{-5}$ & 3.9$\times 10^{2}$ & 1.3$\times 10^{2}$ \\
Averages & 1.66 & 15 & 0.2 & 0.07 ( 5 ) & 2.9 & 37 & 255.1 & 3 & 54 & 1.5$\times 10^{-5}$ & 1.5$\times 10^{3}$ & 7.7$\times 10^{2}$ \\
\enddata
\end{deluxetable}
\clearpage

\clearpage

\begin{thebibliography}{}
\bibitem[Anglada(1996)]{1996ASPC...93....3A} Anglada, G.\ 1996, Radio Emission from the Stars and the Sun, 3.
\bibitem[Anglada et al.(2015)]{2015aska.confE.121A} Anglada, G., Rodr{\'\i}guez, L.~F., \& Carrasco-Gonzalez, C.\ 2015, Advancing Astrophysics with the Square Kilometre Array (AASKA14), 121.
\bibitem[Anglada et al.(2018)]{2018A&ARv..26....3A} Anglada, G., Rodr{\'\i}guez, L.~F., \& Carrasco-Gonz{\'a}lez, C.\ 2018, Astronomy and Astrophysics Review, 26, 3.
\bibitem[Arce, \& Goodman(2001)]{2001ApJ...551L.171A} Arce, H.~G., \& Goodman, A.~A.\ 2001, \apj, 551, L171.
\bibitem[Arce et al.(2007)]{2007prpl.conf..245A} Arce, H.~G., Shepherd, D., Gueth, F., et al.\ 2007, Protostars and Planets V, 245.
\bibitem[Bally \& Zinnecker(2005)]{2005AJ....129.2281B} Bally, J. \& Zinnecker, H.\ 2005, \aj, 129, 2281. doi:10.1086/429098
\bibitem[Beltran \& de Wit(2016)]{} Beltr\'an, M. T. \& de Wit, W. J. 2016, A\&ARv, 24, 6
\bibitem[Beuther et al.(2002)]{2002ASPC..267..341B} Beuther, H., Schilke, P., Menten, K.~M., et al.\ 2002, Hot Star Workshop III: The Earliest Phases of Massive Star Birth, 341.
\bibitem[Beuther et al.(2018)]{2018A&A...617A.100B} Beuther, H., Mottram, J.~C., Ahmadi, A., et al.\ 2018, \aap, 617, A100. doi:10.1051/0004-6361/201833021
\bibitem[Beuther et al.(2021)]{2021arXiv210402420B} Beuther, H., Gieser, C., Suri, S., et al.\ 2021, arXiv:2104.02420
\bibitem[Bonnell et al.(1998)]{1998MNRAS.298...93B} Bonnell, I.~A., Bate, M.~R., \& Zinnecker, H.\ 1998, \mnras, 298, 93.
\bibitem[Bonnell et al.(2001)]{2001MNRAS.323..785B} Bonnell, I.~A., Bate, M.~R., Clarke, C.~J., \& Pringle, J.~E.\ 2001, \mnras, 323, 785.
\bibitem[Brogan et al.(2018)]{2018ApJ...866...87B} Brogan, C.~L., Hunter, T.~R., Cyganowski, C.~J., et al.\ 2018, \apj, 866, 87.
\bibitem[Butler, \& Tan(2012)]{2012ApJ...754....5B} Butler, M.~J., \& Tan, J.~C.\ 2012, \apj, 754, 5.
\bibitem[Carrasco-Gonz{\'a}lez et al.(2010)]{2010Sci...330.1209C} Carrasco-Gonz{\'a}lez, C., Rodr{\'\i}guez, L.~F., Anglada, G., et al.\ 2010, Science, 330, 1209. doi:10.1126/science.1195589
\bibitem[Caselli et al.(1997)]{1997A&A...322..296C} Caselli, P., Hartquist, T.~W., \& Havnes, O.\ 1997, \aap, 322, 296
\bibitem[Chen et al.(2016)]{2016ApJ...824...72C} Chen, X., Arce, H.~G., Zhang, Q., et al.\ 2016, \apj, 824, 72.
\bibitem[Choi et al.(2017)]{2017ApJS..232...24C} Choi, M., Kang, M., Lee, J.-E., et al.\ 2017, The Astrophysical Journal Supplement Series, 232, 24.
\bibitem[Churchwell et al.(1990)]{1990A&AS...83..119C} Churchwell, E., Walmsley, C.~M., \& Cesaroni, R.\ 1990, Astronomy and Astrophysics Supplement Series, 83, 119.
\bibitem[Claussen et al.(1996)]{1996ApJS..106..111C} Claussen, M.~J., Wilking, B.~A., Benson, P.~J., et al.\ 1996, The Astrophysical Journal Supplement Series, 106, 111.
\bibitem[Codella et al.(1999)]{1999A&A...343..585C} Codella, C., Bachiller, R., \& Reipurth, B.\ 1999, \aap, 343, 585.
\bibitem[Codella et al.(2007)]{2007A&A...462L..53C} Codella, C., Cabrit, S., Gueth, F., et al.\ 2007, \aap, 462, L53.
\bibitem[Cosentino et al.(2018)]{2018MNRAS.474.3760C} Cosentino, G., Jim{\'e}nez-Serra, I., Henshaw, J.~D., et al.\ 2018, \mnras, 474, 3760
\bibitem[Cosentino et al.(2020)]{2020MNRAS.tmp.2764C} Cosentino, G., Jim{\'e}nez-Serra, I., Henshaw, J.~D., et al.\ 2020, \mnras, doi:10.1093/mnras/staa2942
\bibitem[Csengeri et al.(2016)]{2016A&A...586A.149C} Csengeri, T., Leurini, S., Wyrowski, F., et al.\ 2016, \aap, 586, A149.
\bibitem[Cunningham et al.(2009)]{2009ApJ...692..943C} Cunningham, N.~J., Moeckel, N., \& Bally, J.\ 2009, \apj, 692, 943.
\bibitem[Curiel et al.(1987)]{1987RMxAA..14..595C} Curiel, S., Canto, J., \& Rodriguez, L.~F.\ 1987, Revista Mexicana de Astronomia y Astrofisica, vol. 14, 14, 595.
\bibitem[De Buizer et al.(2017)]{2017ApJ...843...33D} De Buizer, J.~M., Liu, M., Tan, J.~C., et al.\ 2017, \apj, 843, 33.
\bibitem[Draine(2011)]{2011piim.book.....D} Draine, B.~T.\ 2011, Physics of the Interstellar and Intergalactic Medium by Bruce T. Draine. Princeton University Press.
\bibitem[Duarte-Cabral et al.(2014)]{2014A&A...570A...1D} Duarte-Cabral, A., Bontemps, S., Motte, F., et al.\ 2014, \aap, 570, A1 
\bibitem[Federrath et al.(2014)]{2014ApJ...790..128F} Federrath, C., Schr{\"o}n, M., Banerjee, R., et al.\ 2014, \apj, 790, 128.
\bibitem[Frank et al.(2014)]{2014prpl.conf..451F} Frank, A., Ray, T.~P., Cabrit, S., et al.\ 2014, Protostars and Planets VI, 451.
\bibitem[Gibb et al.(2007)]{2007MNRAS.382.1213G} Gibb, A.~G., Davis, C.~J., \& Moore, T.~J.~T.\ 2007, \mnras, 382, 1213.
\bibitem[Ginsburg et al.(2013)]{2013ApJS..208...14G} Ginsburg, A., Glenn, J., Rosolowsky, E., et al.\ 2013, \apjs, 208, 14. doi:10.1088/0067-0049/208/2/14
\bibitem[Goddi et al.(2018)]{2018arXiv180505364G} Goddi, C., Ginsburg, A., Maud, L., et al.\ 2018, arXiv e-prints , arXiv:1805.05364.
\bibitem[Goldsmith, \& Langer(1999)]{1999ApJ...517..209G} Goldsmith, P.~F., \& Langer, W.~D.\ 1999, \apj, 517, 209.
\bibitem[Gonz{\'a}lez, \& Cant{\'o}(2002)]{2002ApJ...580..459G} Gonz{\'a}lez, R.~F., \& Cant{\'o}, J.\ 2002, \apj, 580, 459.
\bibitem[Guilloteau et al.(1992)]{1992A&A...265L..49G} Guilloteau, S., Bachiller, R., Fuente, A., et al.\ 1992, \aap, 265, L49.
\bibitem[Harju et al.(1998)]{1998A&AS..132..211H} Harju, J., Lehtinen, K., Booth, R.~S., et al.\ 1998, Astronomy and Astrophysics Supplement Series, 132, 211.
\bibitem[Hirota et al.(2017)]{2017NatAs...1E.146H} Hirota, T., Machida, M.~N., Matsushita, Y., et al.\ 2017, Nature Astronomy, 1, 146.
\bibitem[Jim{\'e}nez-Serra et al.(2010)]{2010MNRAS.406..187J} Jim{\'e}nez-Serra, I., Caselli, P., Tan, J.~C., et al.\ 2010, \mnras, 406, 187.
\bibitem[K\"onigl, \& Pudritz(2000)]{2000prpl.conf..759K} Konigl, A., \& Pudritz, R.~E.\ 2000, Protostars and Planets IV, 759.
\bibitem[Kong et al.(2017)]{2017ApJ...834..193K} Kong, S., Tan, J.~C., Caselli, P., et al.\ 2017, \apj, 834, 193 
\bibitem[Kong et al.(2019)]{2019ApJ...874..104K} Kong, S., Arce, H.~G., Maureira, M.~J., et al.\ 2019, \apj, 874, 104
\bibitem[Kurtz et al.(1994)]{1994ApJS...91..659K} Kurtz, S., Churchwell, E., \& Wood, D.~O.~S.\ 1994, The Astrophysical Journal Supplement Series, 91, 659.
\bibitem[Lada, \& Fich(1996)]{1996ApJ...459..638L} Lada, C.~J., \& Fich, M.\ 1996, \apj, 459, 638.
\bibitem[Lee et al.(2000)]{2000ApJ...542..925L} Lee, C.-F., Mundy, L.~G., Reipurth, B., et al.\ 2000, \apj, 542, 925.
\bibitem[Lee, \& Sahai(2004)]{2004ApJ...606..483L} Lee, C.-F., \& Sahai, R.\ 2004, \apj, 606, 483.
\bibitem[Lee et al.(2010)]{2010ApJ...713..731L} Lee, C.-F., Hasegawa, T.~I., Hirano, N., et al.\ 2010, \apj, 713, 731.
\bibitem[Lee et al.(2015)]{2015ApJ...805..186L} Lee, C.-F., Hirano, N., Zhang, Q., et al.\ 2015, \apj, 805, 186.
\bibitem[Leurini et al.(2014)]{2014A&A...570A..49L} Leurini, S., Codella, C., L{\'o}pez-Sepulcre, A., et al.\ 2014, \aap, 570, A49.
\bibitem[Li et al.(2019a)]{2019ApJ...878...29L} Li, S., Wang, J., Fang, M., et al.\ 2019, \apj, 878, 29. doi:10.3847/1538-4357/ab1e4c
\bibitem[Li et al.(2019b)]{2019ApJ...886..130L} Li, S., Zhang, Q., Pillai, T., et al.\ 2019, \apj, 886, 130. doi:10.3847/1538-4357/ab464e
\bibitem[Lim \& De Buizer(2019)]{2019ApJ...873...51L} Lim, W. \& De Buizer, J.~M.\ 2019, \apj, 873, 51. doi:10.3847/1538-4357/ab0288
\bibitem[Liu et al.(2018)]{2018ApJ...862..105L} Liu, M., Tan, J.~C., Cheng, Y., et al.\ 2018, \apj, 862, 105.
\bibitem[Liu et al.(2019)]{2019ApJ...874...16L} Liu, M., Tan, J.~C., De Buizer, J.~M., et al.\ 2019, \apj, 874, 16.
\bibitem[Liu et al.(2020)]{2020ApJ...904...75L} Liu, M., Tan, J.~C., De Buizer, J.~M., et al.\ 2020, \apj, 904, 75. doi:10.3847/1538-4357/abbefb
\bibitem[Liu et al.(2020)]{2020MNRAS.496.2790L} Liu, T., Evans, N.~J., Kim, K.-T., et al.\ 2020, \mnras, 496, 2790. doi:10.1093/mnras/staa1577
\bibitem[L{\'o}pez-Sepulcre et al.(2011)]{2011A&A...526L...2L} L{\'o}pez-Sepulcre, A., Walmsley, C.~M., Cesaroni, R., et al.\ 2011, \aap, 526, L2.
\bibitem[Ma et al.(2013)]{2013ApJ...779...79M} Ma, B., Tan, J.~C., \& Barnes, P.~J.\ 2013, \apj, 779, 79.
\bibitem[Maud et al.(2015)]{2015MNRAS.453..645M} Maud, L.~T., Moore, T.~J.~T., Lumsden, S.~L., et al.\ 2015, \mnras, 453, 645.
\bibitem[Matsushita et al.(2017)]{2017MNRAS.470.1026M} Matsushita, Y., Machida, M.~N., Sakurai, Y., et al.\ 2017, \mnras, 470, 1026.
\bibitem[McKee \& Tan(2003)]{MT03} McKee, C.~F., \& Tan, J.~C.\ 2003, \apj, 585, 850.
\bibitem[McMullin et al.(2007)]{2007ASPC..376..127M} McMullin, J.~P., Waters, B., Schiebel, D., et al.\ 2007, Astronomical Data Analysis Software and Systems XVI, 376, 127
\bibitem[Molinari et al.(2008)]{2008A&A...481..345M} Molinari, S., Pezzuto, S., Cesaroni, R., et al.\ 2008, \aap, 481, 345.
\bibitem[Moscadelli et al.(2016)]{2016A&A...585A..71M} Moscadelli, L., S{\'a}nchez-Monge, {\'A}., Goddi, C., et al.\ 2016, \aap, 585, A71. doi:10.1051/0004-6361/201526238
\bibitem[Moser et al.(2020)]{2020ApJ...897..136M} Moser, E., Liu, M., Tan, J.~C., et al.\ 2020, \apj, 897, 136.
\bibitem[Motte et al.(2007)]{2007A&A...476.1243M} Motte, F., Bontemps, S., Schilke, P., et al.\ 2007, \aap, 476, 1243.
\bibitem[Motte et al.(2018)]{2018NatAs...2..478M} Motte, F., Nony, T., Louvet, F., et al.\ 2018, Nature Astronomy, 2, 478. doi:10.1038/s41550-018-0452-x
\bibitem[Offner, \& Arce(2014)]{2014ApJ...784...61O} Offner, S.~S.~R., \& Arce, H.~G.\ 2014, \apj, 784, 61.
\bibitem[Offner, \& Chaban(2017)]{2017ApJ...847..104O} Offner, S.~S.~R., \& Chaban, J.\ 2017, \apj, 847, 104.
\bibitem[Ossenkopf, \& Henning(1994)]{1994A&A...291..943O} Ossenkopf, V., \& Henning, T.\ 1994, \aap, 291, 943.
\bibitem[Palla et al.(1993)]{1993A&A...280..599P} Palla, F., Cesaroni, R., Brand, J., et al.\ 1993, \aap, 280, 599.
\bibitem[Plambeck et al.(2009)]{2009ApJ...704L..25P} Plambeck, R.~L., Wright, M.~C.~H., Friedel, D.~N., et al.\ 2009, \apj, 704, L25.
\bibitem[Plunkett et al.(2015)]{2015Natur.527...70P} Plunkett, A.~L., Arce, H.~G., Mardones, D., et al.\ 2015, \nat, 527, 70.
\bibitem[Podio et al.(2020)]{2020arXiv201215379P} Podio, L., Tabone, B., Codella, C., et al.\ 2020, arXiv:2012.15379
\bibitem[Principe et al.(2018)]{2018MNRAS.473..879P} Principe, D.~A., Cieza, L., Hales, A., et al.\ 2018, \mnras, 473, 879.
\bibitem[Purser et al.(2016)]{2016MNRAS.460.1039P} Purser, S.~J.~D., Lumsden, S.~L., Hoare, M.~G., et al.\ 2016, \mnras, 460, 1039.
\bibitem[Qiu, \& Zhang(2009)]{2009ApJ...702L..66Q} Qiu, K., \& Zhang, Q.\ 2009, \apj, 702, L66.
\bibitem[Reipurth et al.(1992)]{1992ApJ...392..145R} Reipurth, B., Raga, A.~C., \& Heathcote, S.\ 1992, \apj, 392, 145.
\bibitem[Rodr{\'\i}guez et al.(2008)]{2008AJ....135.2370R} Rodr{\'\i}guez, L.~F., Moran, J.~M., Franco-Hern{\'a}ndez, R., et al.\ 2008, \aj, 135, 2370. doi:10.1088/0004-6256/135/6/2370
\bibitem[Rohde et al.(2019)]{2019MNRAS.483.2563R} Rohde, P.~F., Walch, S., Seifried, D., et al.\ 2019, \mnras, 483, 2563.
\bibitem[Rosero et al.(2016)]{2016ApJS..227...25R} Rosero, V., Hofner, P., Claussen, M., et al.\ 2016, The Astrophysical Journal Supplement Series, 227, 25.
\bibitem[Rosero et al.(2019a)]{2019ApJ...873...20R} Rosero, V., Tanaka, K.~E.~I., Tan, J.~C., et al.\ 2019, \apj, 873, 20.
\bibitem[Rosero et al.(2019b)]{2019ApJ...880...99R} Rosero, V., Hofner, P., Kurtz, S., et al.\ 2019, \apj, 880, 99. doi:10.3847/1538-4357/ab2595
\bibitem[Rosolowsky et al.(2008)]{2008ApJ...679.1338R} Rosolowsky, E.~W., Pineda, J.~E., Kauffmann, J., et al.\ 2008, \apj, 679, 1338.
\bibitem[Sakai et al.(2010)]{2010ApJ...714.1658S} Sakai, T., Sakai, N., Hirota, T., et al.\ 2010, \apj, 714, 1658.
\bibitem[S{\'a}nchez-Monge et al.(2013a)]{2013A&A...550A..21S} S{\'a}nchez-Monge, {\'A}., Beltr{\'a}n, M.~T., Cesaroni, R., et al.\ 2013, \aap, 550, A21.
\bibitem[S{\'a}nchez-Monge et al.(2013b)]{2013A&A...557A..94S} S{\'a}nchez-Monge, {\'A}., L{\'o}pez-Sepulcre, A., Cesaroni, R., et al.\ 2013, \aap, 557, A94.
\bibitem[Sanhueza et al.(2012)]{2012ApJ...756...60S} Sanhueza, P., Jackson, J.~M., Foster, J.~B., et al.\ 2012, \apj, 756, 60.
\bibitem[Sanhueza et al.(2019)]{2019ApJ...886..102S} Sanhueza, P., Contreras, Y., Wu, B., et al.\ 2019, \apj, 886, 102. doi:10.3847/1538-4357/ab45e9
\bibitem[Sanna et al.(2018)]{2018A&A...619A.107S} Sanna, A., Moscadelli, L., Goddi, C., et al.\ 2018, \aap, 619, A107.
\bibitem[Schilke et al.(1997)]{1997A&A...321..293S} Schilke, P., Walmsley, C.~M., Pineau des Forets, G., et al.\ 1997, \aap, 321, 293.
\bibitem[Shu et al.(2000)]{2000prpl.conf..789S} Shu, F.~H., Najita, J.~R., Shang, H., et al.\ 2000, Protostars and Planets IV, 789.
\bibitem[Stone, \& Norman(1993)]{1993ApJ...413..210S} Stone, J.~M., \& Norman, M.~L.\ 1993, \apj, 413, 210.
\bibitem[Tanaka et al.(2016)]{2016ApJ...818...52T} Tanaka, K.~E.~I., Tan, J.~C., \& Zhang, Y.\ 2016, \apj, 818, 52. doi:10.3847/0004-637X/818/1/52
\bibitem[Urquhart et al.(2014)]{2014MNRAS.443.1555U} Urquhart, J.~S., Moore, T.~J.~T., Csengeri, T., et al.\ 2014, \mnras, 443, 1555.
\bibitem[Wang et al.(2006)]{2006ApJ...651L.125W} Wang, Y., Zhang, Q., Rathborne, J.~M., et al.\ 2006, \apj, 651, L125.
\bibitem[Wang et al.(2010)]{2010ApJ...709...27W} Wang, P., Li, Z.-Y., Abel, T., \& Nakamura, F.\ 2010, \apj, 709, 27.
\bibitem[Wouterloot, \& Walmsley(1986)]{1986A&A...168..237W} Wouterloot, J.~G.~A., \& Walmsley, C.~M.\ 1986, \aap, 168, 237.
\bibitem[Zhang et al.(2015)]{2015ApJ...804..141Z} Zhang, Q., Wang, K., Lu, X., et al.\ 2015, \apj, 804, 141.
\bibitem[Zhang, \& Tan(2018)]{2018ApJ...853...18Z} Zhang, Y., \& Tan, J.~C.\ 2018, \apj, 853, 18.
\bibitem[Zhang et al.(2019)]{2019ApJ...873...73Z} Zhang, Y., Tan, J.~C., Sakai, N., et al.\ 2019, \apj, 873, 73.
\end{thebibliography}
\end{document}